\documentclass[lettersize,journal]{IEEEtran}
\usepackage{graphicx}
\usepackage{latexsym}
\usepackage{tabularx}
\usepackage{cite}
\usepackage[labelformat=simple]{subcaption}

\usepackage{caption} 
\usepackage{color}
\usepackage{wrapfig}
\usepackage{amsfonts,amssymb}
\usepackage{amsthm,amsmath}
\usepackage{url}
\usepackage[encapsulated]{CJK} 
\usepackage{indentfirst}
\usepackage{soul}
\usepackage{comment}
\usepackage{setspace}
\usepackage{balance}
\usepackage{array}
\usepackage{orcidlink}
\usepackage{hyperref}
\hypersetup{colorlinks,
      linkcolor=blue,
      citecolor=blue,
      urlcolor=black,
      filecolor=blue}

\usepackage{cleveref}
\usepackage{makecell}

\usepackage{xcolor}

\newcommand{\revised}[1]{{\color{black}#1}}

\definecolor{dv}{RGB}{148, 0, 211}

\makeatletter
\newcommand\footnoteref[1]{\protected@xdef\@thefnmark{\ref{#1}}\@footnotemark}
\makeatother

\usepackage{enumitem}

\usepackage{algorithmicx}
\usepackage{algorithm}
\usepackage[noend]{algpseudocode}

\usepackage{booktabs}
\usepackage{multirow}
\usepackage{adjustbox}

\def\BibTeX{{\rm B\kern-.05em{\sc i\kern-.025em b}\kern-.08em
    T\kern-.1667em\lower.7ex\hbox{E}\kern-.125emX}}



\makeatletter

\newcommand{\algmargin}{\the\ALG@thistlm}
\newlength{\whilewidth}
\algnewcommand{\parState}[1]{\State%
  \parbox[t]{\dimexpr\linewidth-\algmargin}{\strut #1\strut}}

\algrenewtext{If}[1]
    {\parbox[t]{\dimexpr\linewidth-\algmargin}
    {\hangindent\whilewidth\algorithmicif\ #1\ \algorithmicthen}}

\algrenewtext{ElsIf}[1]
    {\parbox[t]{\dimexpr\linewidth-\algmargin}
    {\hangindent\whilewidth\algorithmicelsif\ #1\ \algorithmicthen}}

\algnewcommand\algorithmicinput{\textbf{Input:}}
\algnewcommand\INPUT{\item[\algorithmicinput]}

\algnewcommand\algorithmicoutput{\textbf{Output:}}
\algnewcommand\OUTPUT{\item[\algorithmicoutput]}

    \newcounter{phase}[algorithm]
    \newlength{\phaserulewidth}
    \newcommand{\setphaserulewidth}{\setlength{\phaserulewidth}}
    
    \setphaserulewidth{.7pt}
\makeatother

\usepackage{thmtools}

\theoremstyle{definition}

\theoremstyle{plain}

\DeclareMathOperator*{\argmin}{arg\,min}

\graphicspath{ {images/} }

\usepackage{etoolbox}

\renewcommand{\baselinestretch}{0.985}
\normalsize
\allowdisplaybreaks

\begin{document}

\title{AgileDART: An Agile and Scalable Edge Stream Processing Engine}

\author{
        Cheng-Wei~Ching\,\orcidlink{0000-0001-6621-4907},
        Xin~Chen\,\orcidlink{0009-0008-8188-8211},
        Chaeeun~Kim\,\orcidlink{0000-0002-4715-7836},
        Tongze~Wang\,\orcidlink{0009-0001-0033-9114},
        Dong~Chen\,\orcidlink{0000-0002-1052-5658},
        Dilma Da~Silva\,\orcidlink{0000-0001-6538-2888},
        and     
        Liting~Hu\,\orcidlink{0009-0007-7222-5507}

\vspace{-0.1in}
        
\IEEEcompsocitemizethanks{
    \IEEEcompsocthanksitem This work was supported by the National Science Foundation under Grants
    NSF-OAC-2313738, NSF-CAREER-2313737, NSF-SPX-1919181, NSF-SPX-2202859, and NSF-CNS-2322919. A preliminary version of this paper appeared at the 2021 USENIX Annual Technical Conference (USENIX ATC’ 21)~\cite{dart}. \textit{(Corresponding author: Liting Hu.)}	
	\IEEEcompsocthanksitem Cheng-Wei Ching is with the Department of Computer Science and Engineering, University of California, Santa Cruz, Santa Cruz, CA 95064 USA (e-mail: cching1@ucsc.edu). 
    \IEEEcompsocthanksitem Xin Chen is with the Department of Computer Science and Engineering, Georgia Institute of Technology, Atlanta, GA 30332 USA (e-mail: xchen384@gatech.edu). 
	\IEEEcompsocthanksitem Chaeeun Kim is with the Department of Computer Science and Engineering, University of California, Santa Cruz, Santa Cruz, CA 95064 USA (e-mail: ckim151@ucsc.edu).
    \IEEEcompsocthanksitem Tongze Wang is with the Department of Computer Science and Engineering, University of California, Santa Cruz, Santa Cruz, CA 95064 USA (e-mail: twang141@ucsc.edu).
    \IEEEcompsocthanksitem Dong Chen is with Colorado School of Mines, Golden, CO 80401 USA (e-mail: dongchen@mines.edu).
    \IEEEcompsocthanksitem Dilma Da Silva is with Texas A\&M University, College Station, TX 77843 USA (e-mail: dilma@cse.tamu.edu).
    \IEEEcompsocthanksitem Liting Hu is with the Department of Computer Science and Engineering, University of California, Santa Cruz, Santa Cruz, CA 95064 USA (e-mail: liting@ucsc.edu).

}
}
 
\markboth{Journal of \LaTeX\ Class Files,~Vol.~18, No.~9, September~2020}%
{Submitted paper}

\IEEEtitleabstractindextext{
\begin{abstract}

\revised{
Edge applications generate a large influx of sensor data on massive scales, and these massive data streams must be processed shortly to derive actionable intelligence. However, traditional data processing systems are not well-suited for these edge applications as they often do not scale well with a large number of concurrent stream queries, do not support low-latency processing under limited edge computing resources, and do not adapt to the level of heterogeneity and dynamicity commonly present in edge computing environments.
As such, we present AgileDart, an agile and scalable edge stream processing engine that enables fast stream processing of many concurrently running low-latency edge applications' queries at scale in dynamic, heterogeneous edge environments. The novelty of our work lies in a dynamic dataflow abstraction that leverages distributed hash table-based peer-to-peer overlay networks to autonomously place, chain, and scale stream operators to reduce query latencies, adapt to workload variations, and recover from failures and a bandit-based path planning model that re-plans the data shuffling paths to adapt to unreliable and heterogeneous edge networks. We show that AgileDart outperforms Storm and EdgeWise on query latency and significantly improves scalability and adaptability when processing many real-world edge stream applications' queries.
}

\end{abstract}

\begin{IEEEkeywords}
Edge stream processing, distributed hash table, edge networks, bandit algorithm.
\end{IEEEkeywords}

}

\maketitle

\IEEEdisplaynontitleabstractindextext

\IEEEpeerreviewmaketitle

\section{Introduction}
\IEEEPARstart{T}{oday}, we have an unprecedented opportunity to gain insights from a steady stream of real-time data --- for example, clickstreams from web servers, application and infrastructure log data, and Internet of Things (IoTs) data. This almost continuous flow of data can derive decision-making and operational intelligence to new heights of agility and responsiveness. For example, by connecting cars to the internet and critical infrastructure, we can quickly find empty parking slots or get safety warnings such as forward collision, intersection movement assist, emergency vehicle approaching, and road work. By connecting all the light bulbs on the freeway to the internet and Long-Term Evolution (LTE) modems, if any anomaly like a car crash happens, it can detect it right away and communicate back to us. 

Under many time-critical scenarios, these massive data streams must be processed in a very short time to derive actionable intelligence. Regrettably, traditional cloud-based data processing systems~\cite{ApacheFlink, ApacheSamza, ApacheSpark, storm, MillWheel, Heron, S4} and their edge support~\cite{Edgent, EdgeWise, azureiot, greengrass, shen2015} predominantly adopt a server-client architecture, where the front-end sensors send time-series observations of the physical or human system to the back-end cloud for analysis, and the analytics drive recommendations that are enacted on, or notified to, the system to improve it. The long-distance communications between the sensor and the cloud give rise to two major issues, making it \emph{\textbf{not}} well-suited for these time-critical edge applications: 

\begin{enumerate}
    \item \textbf{High latency:} The substantial latency inherent in this approach can render the results obsolete. 
    \item \textbf{Network infrastructure constraints:} The network infrastructure often struggles to handle substantial data streams. As an example, the extensive video streams produced by the increasingly deployed cameras put significant pressure on today's high-end Metropolitan Area Networks (MANs), which typically offer bandwidth capacities of only 100 Gbps~\cite{yangfog}.
\end{enumerate}

Therefore, these time-critical edge applications require a distributed solution that analyzes the data on the fly before the data arrives in the cloud. As such, it motivates the emergence of ``\emph{edge stream processing}'', which is a new means to distribute processing from the centralized cloud servers towards decentralized processing units close to the data sources at the edge nodes, where unbounded streaming data are continuously generated and analyzed in near real-time.

\begin{figure}[t]
  \centering
  \includegraphics[width=0.425\textwidth]{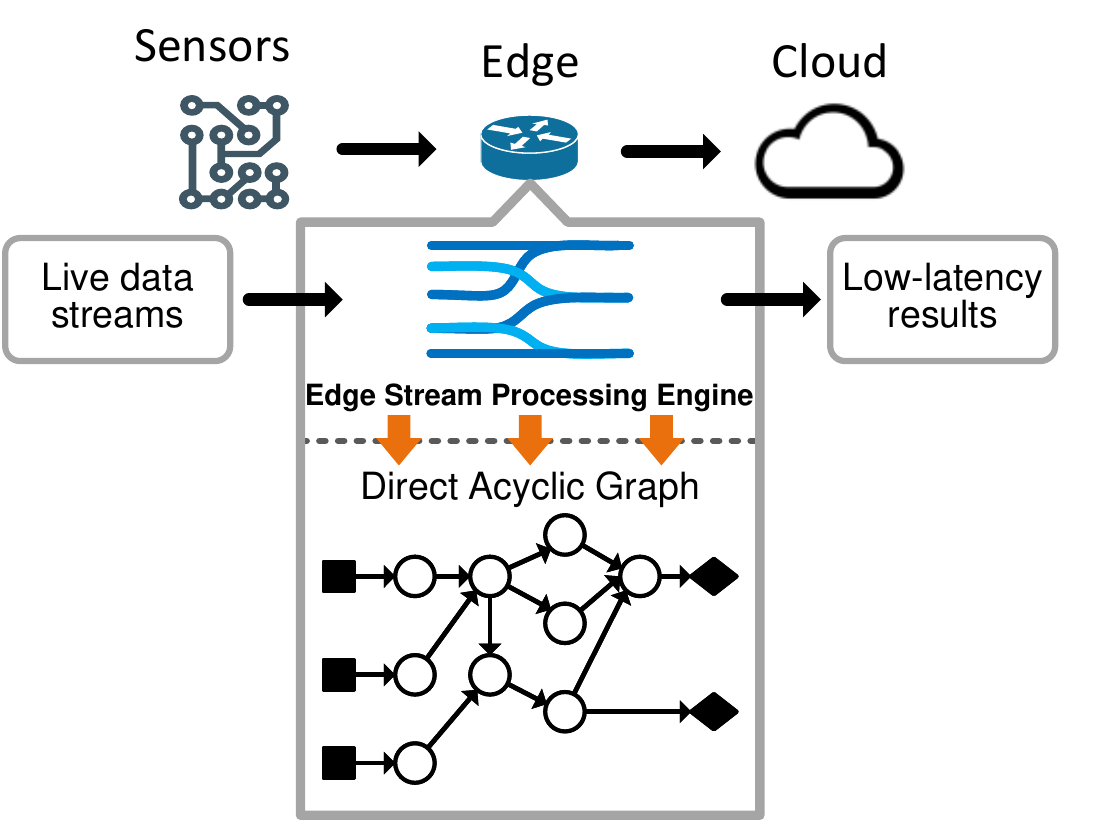}
  \caption{Edge stream processing overview.}
  \label{fig:esp}
\end{figure}

\textbf{\textit{What is edge stream processing?}} To put it simply, edge stream processing is to apply the stream processing paradigm to the edge computing architecture. Broadly speaking, edge systems consist of Things, Gateways, and the Cloud. Things are sensors (e.g., smart wearables, self-driving car sensors) that ``read" from the world and actuators that "write" to it. Gateways orchestrate Things and bridge them with the cloud~\cite{EdgeWise}. Instead of relying on the cloud to process sensor data and trigger actuators, the edge stream processing engine relies on distributed edge compute nodes including gateways, edge routers, and powerful sensors to process sensor data and trigger actuators.

Figure~\ref{fig:esp} shows the overview of the edge stream processing engine. Things generate streams of data continuously---unbounded sequences of data points (tuples) that carry timestamps. These data streams are consumed by the edge stream processing engine, which creates a logical topology of stream processing nodes connected in a Directed Acyclic Graph (DAG), processes the tuples of streams as they flow through the DAG from sources to sinks, and outputs the results in a very short time. Each \textit{\textbf{source}} node is a sensor. Each \textit{\textbf{sink}} node is an actuator or a message queue to the Cloud. Each \textit{\textbf{inner}} node is an operator that performs arbitrary computation on the data points, ranging from simple computation such as \textit{map}, \textit{reduce}, \textit{join}, \textit{filter} to complex computation such as \textit{ML-based classification algorithms}. DAGs can be implemented via many patterns, such as the partition/aggregate pattern which scales out by partitioning tasks into many sub-tasks (e.g., Dryad~\cite{dryad}), sequential/dependent pattern in which streams are processed sequentially and subsequent streams depend on the results of previous ones (e.g., Storm~\cite{storm}), and hybrid pattern with sequential/dependent and partition/aggregate (e.g., Spark Streaming~\cite{sparkstreaming}, Naiad~\cite{naiad}).

\begin{figure*}[t] 
 \centering 
 \includegraphics[scale=0.465]{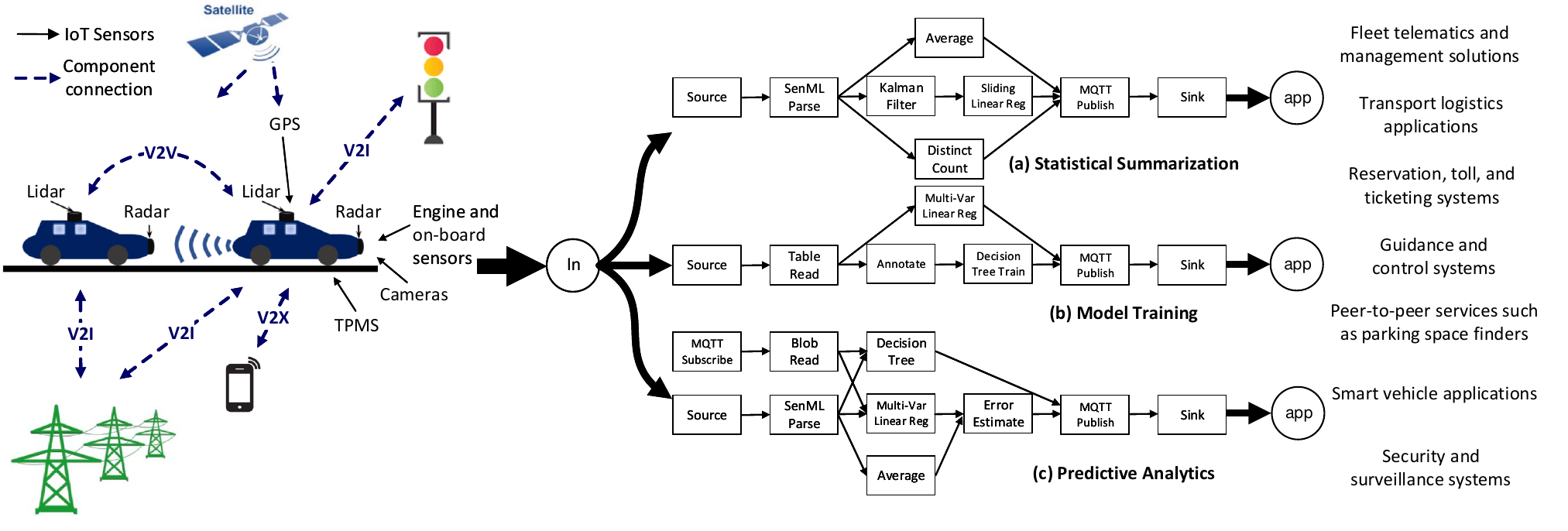}
 \caption{Motivating time-critical edge applications in Intelligent Transportation Systems illustrate the need for the Edge stream processing engine (V2V: Vehicle-to-Vehicle, V2I: Vehicle-to-Infrastructure, V2X: Vehicle-to-Everything).}
 \label{fig:app}
\end{figure*}

\textbf{\textit{Why does edge stream processing matter?}}
Edge stream processing brings an effective and practical solution that benefits many time-critical edge applications in the areas of factory automation, tactile internet, connected cars, smart grid, healthcare monitoring, remote control, automated guide vehicle, and process automation. Figure~\ref{fig:app} illustrates a concrete use case scenario~\cite{Dilmagrant}. In future Intelligent Transportation (ITS) Systems, such as the efforts currently funded by the US Department of Transportation~\cite{JPOplan}, all cars are connected to each other and equipped with Internet access. Even at a low level of autonomy, each car generates at least 3 Gbit/s of data per second from sensors including 4-6 Radars (0.1$\sim$15 Mbit/s per sensor), 1-5 LIDARs (20$\sim$100 Mbit/s), 6-12 Cameras (500$\sim$3500 Mbit/s), 8-16 Ultrasonics ($<$ 0.01 Mbit/s), GPS, and IMU~\cite{carnews}.  All cars' sensor data, along with road and weather sensor data, are continuously logged by the ITS system. These live data streams are massive and distributed. On the backend, many stream processing applications will run concurrently, consuming these live data streams to quickly derive insights and make decisions. Examples include intelligently routing vehicles to avoid traffic congestion, parking space finders, protecting against safety hazards and terrorist threats, and enabling automated payment and ticketing. These applications usually have different topologies. We list three representative topologies extracted from RIoTBench benchmark suite~\cite{riotbench} that are used for statistical summarization, model training, and predictive analytics. For example, the predictive analytics topology forks data streams to the decision tree classifier and the multivariate regression to predict future trends and decide if any action is required.

To facilitate fast stream processing for a vast number of concurrently running time-critical edge applications' jobs at scale in a dynamic (varying workloads and bandwidths) and heterogeneous (different network layouts) edge environment, this paper proposes AgileDart, a next-generation scalable and adaptive edge stream processing engine.
\revised{In sharp contrast to existing stream processing systems relying on a single monolithic master, the key innovation of AgileDart lies in \textit{a decentralized architecture} in which all peer nodes \textit{autonomously} place operators, plan dataflow paths for data shuffling, and scale operators, thereby dramatically improving scalability and adaptivity.}

We make the following contributions in this paper. 

\revised{First, we highlight the challenges when designing an efficient edge stream processing engine (Section~\ref{sec:challenges}) and then study existing stream processing systems' query model and software architecture and their limitations in supporting dynamic and heterogeneous edge environments. To the best of our knowledge, we are the \emph{first} to observe the lack of scalability and adaptivity in existing stream processing systems for handling massive edge applications in practical edge settings (Section~\ref{sec:background}).}

\revised{Second, we design a novel dynamic dataflow abstraction that \textit{autonomously} places, chains, and parallelizes stream operators using the distributed hash table (DHT) based peer-to-peer (P2P) overlay networks. The main advantage of using DHT is that it avoids the original monolithic master, and P2P overlay networks enable all peer nodes to make operator-mapping decisions jointly. Moreover, nodes can join or leave the system with minimal work by redistributing keys. Hence, the design allows our system to scale to vast numbers of nodes, applications, and jobs simultaneously. To the best of our knowledge, we are the \emph{first} to explore DHT-based P2P technologies to pursue extreme scalability in edge stream processing systems~(Section~\ref{sec:design}). 

Third, we introduce a novel bandit-based exploitation-exploration path planning model to reinforce the DHT-based routing protocols in the dynamic dataflow abstraction. Instead of relying only on the DHT-based routing tables with fixed prefixes, the model dynamically replans the stream data shuffling paths to adapt to unreliable and heterogeneous edge networks~(Section~\ref{sec:model}).

Fourth, using DHTs, we decompose the stream processing system architecture from one master to $N$ workers to $M$ masters to $N$ workers, which removes a monolithic centralized master and ensures that each edge zone can have an independent master for handling applications and operating individually without any centralized state. As a result of our distributed management, AgileDart improves overall query latencies for concurrently running applications and significantly reduces application deployment times. To the best of our knowledge, we offer the \emph{first} stream processing engine to enable operating edge applications in a DevOps fashion~(Section~\ref{sec:impl}).}

Finally, we demonstrate AgileDart’s scalability and latency gains over Apache Storm~\cite{storm} and EdgeWise~\cite{EdgeWise} on IoT stream benchmarks~(Section~\ref{sec:eval}).

\section{Challenges}
\label{sec:challenges}

\subsection{Requirements}
\label{subsec:requirements}

We outline the following specific requirements for an efficient edge stream processing engine:

\begin{itemize}
\item \textbf{R1: Low latency.} It must continuously analyze incoming data in real-time and offer low-latency results, otherwise users/scientists should simply transmit raw data to the cloud for batch analysis. 

\item \textbf{R2: Parallelism.} It must support arbitrary topologies on combined edge resources of multiple devices. Many edge applications are data-intensive. Running these ``big data'' applications cannot be accomplished using a single edge device. Instead, it requires a set of multiple edge nodes to jointly provide the service in order to exploit the parallelism in which each edge node processes distributed incoming data in parallel.  

\item \textbf{R3: Scalability.} As edge applications generate numerous stream queries tailored to the unique requirements of individual users, the volume of concurrent stream queries sent to the edge can grow substantially. Therefore, it is critical that the edge stream processing engine can scale gracefully to a vast number of concurrent edge stream queries.

\item \textbf{R4: Adaptability to the edge network dynamics.} Modern stream processing engines have mainly focused on addressing computational bottlenecks in the cloud and, hence, are WAN-agnostic~\cite{StreamCloud, kuo2021energy, timestream, ApacheFlink, streamscope, chronostream, ching2023dual}. Considering the fundamental differences in the environments, the edge stream processing engine must be WAN-aware, i.e., adapt to the bandwidth and workload variations.

\item \textbf{R5: Fault-tolerance.} The edge stream processing engine must offer strong fault tolerance guarantees, i.e., be able to quickly recover from any transient network failure or any edge device failure. For example, the query processing can continue even if multiple edge devices fail or become unreachable.

\end{itemize}

\begin{figure*}[t!]
  \centering
  \begin{minipage}[b]{0.3\textwidth}
    \includegraphics[width=\textwidth]{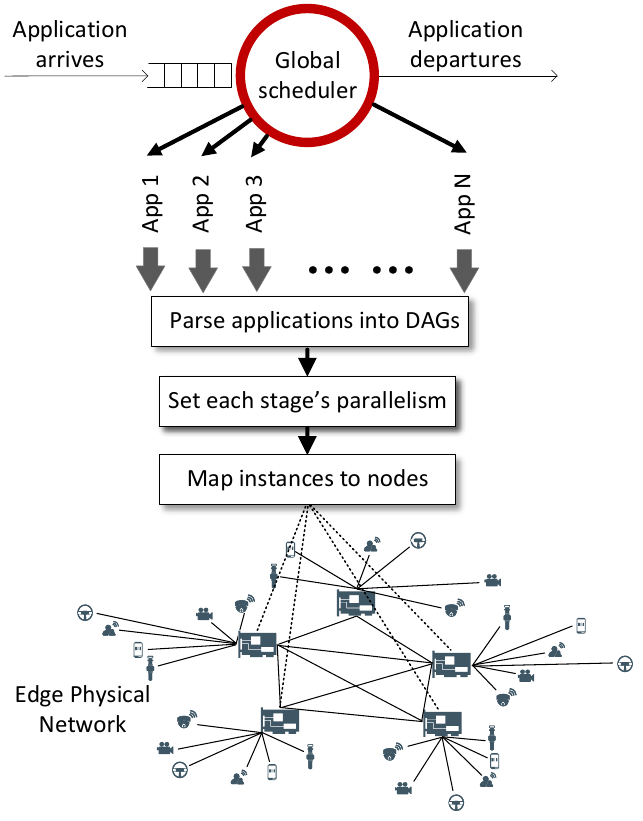}
    \caption{The global scheduler.}
    \label{fig:globalscheduler}
  \end{minipage}
  \begin{minipage}[b]{0.6\textwidth}
    \includegraphics[width=\textwidth]{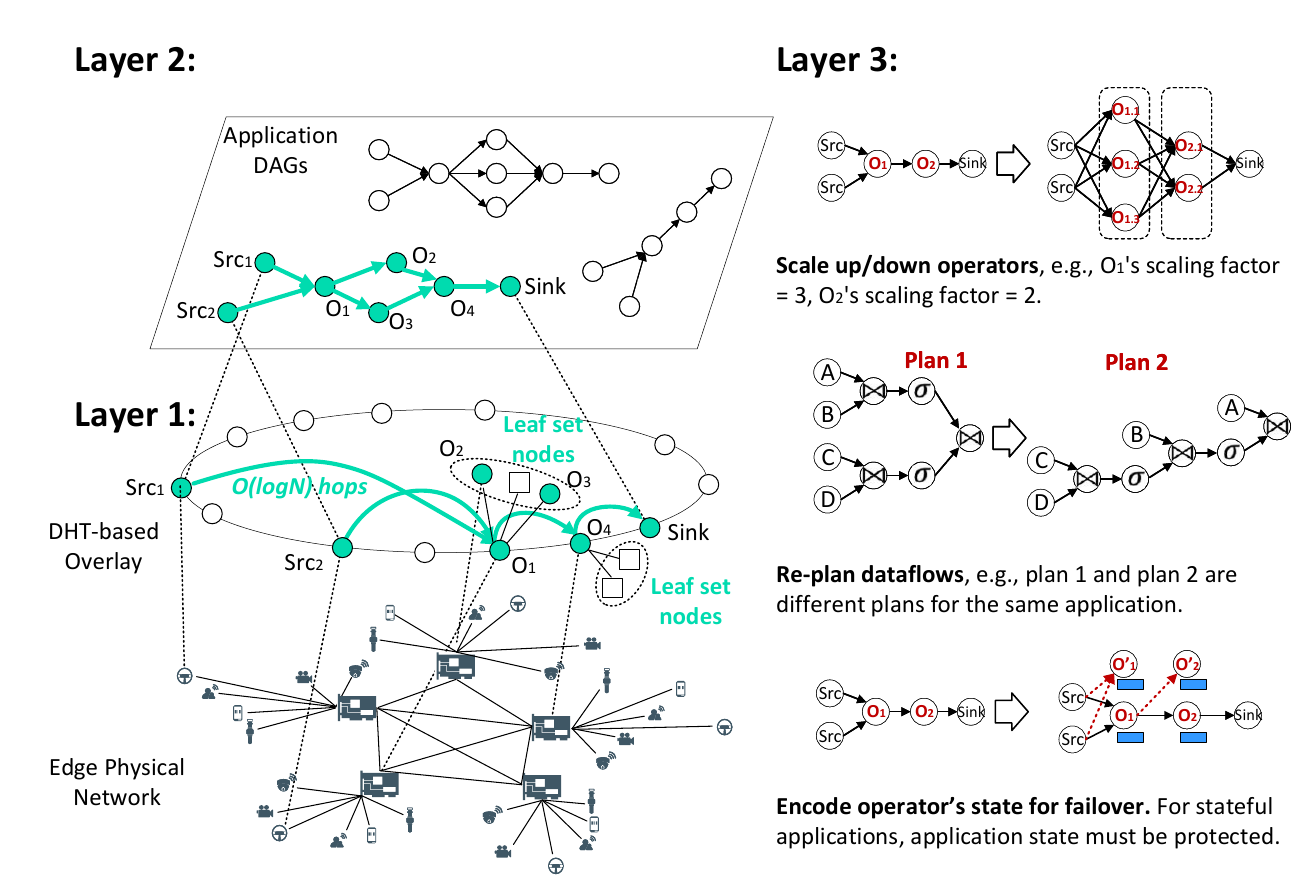}
    \caption{Dynamic dataflow graph abstraction for operator placement.}
      \label{fig:dynamicdataflow}
  \end{minipage}
\end{figure*}

\subsection{Challenges in Edge Stream Processing}

Unfortunately, as IoT systems grow in number and complexity, we face significant challenges in building edge stream processing engines that can meet all of the above requirements.

\textbf{\textit{Challenge \#1: it is very challenging to offer low latency and high throughput in resource-constrained edge environments.}} Unlike the cloud, edge nodes have limited computing resources: few-core processors, little memory, and little permanent storage~\cite{armbrust2010view, shi2016edge, ching2024totoro}. For example, Cisco’s IoT gateways have a quad-core processor and 1 GB of RAM. Unlike in cloud-based solutions, which can assume a persistent data source (e.g., Kafka~\cite{ApacheKafka}), there is no backpressure mechanism at the edge. If any data source has a burst of data, there is no place that can buffer this data. The limited hardware resources, together with the lack of a ``stop the world'' scheme, make it particularly challenging to handle ``big data'' needs. 

\textbf{\textit{Challenge \#2: it is very challenging to scale to numerous concurrently running edge stream queries that have diverse topologies.}} Modern stream engines (e.g., Apache Storm~\cite{storm}, Apache Flink~\cite{flink}, Twitter’s Heron~\cite{Heron}) are based on the “one master/many workers” architecture, in which the only master is responsible for all scheduling activities, e.g., distributing query tasks to different machines, tracking each query’s progress, and handling failures and stragglers. 
They use a first-come, first-serve method, making deployment times accumulate and leading to long-tail latencies. As such, they easily suffer a central bottleneck when handling a large number of edge stream queries. 

\textbf{\textit{Challenge \#3: it is very challenging to adapt to the dynamic edge networks and handle failures or stragglers to ensure system reliability.}} Edge devices and gateways are interconnected using a wireless network based on WiFi, Zigbee, or BlueTooth. Unlike wired networks, wireless networks experience dynamically fluctuating bandwidth and connectivity between nodes due to factors such as signal attenuation, interference, and wireless channel contention~\cite{2018frontier}. Transient network failures are also very common. Given the substantial data flow through wireless networks, any noticeable delays or lags resulting from wireless instability can easily violate the low-latency requirement for edge stream queries. Existing studies on the adaptability in stream processing systems~\cite{seep,dhalion,StreamCloud,threestep,drizzle} mainly focus on the cloud environment, where the primary sources of dynamics come from workload variability, failures, and stragglers. In this case, a solution typically allocates additional computational resources or re-distributes the workload of the bottleneck execution across multiple nodes within a data center. Unfortunately, unlike the cloud servers, edge nodes have limited computing resources and do not support backpressure. Consequently, the previous adaptability techniques by reallocating resources or buffering data at data sources cannot be applied in edge stream processing systems.

\section{Background}
\label{sec:background}

As shown in Figure~\ref{fig:globalscheduler}, existing stream processing systems ~\cite{flutter,hung2,hung,pixida,iridium,jetstream,SAGE,CLARINET,geode,EdgeWise,2018frontier} mostly rely on a master-slave architecture in which a ``single'' monolithic master administers many applications (if any). The responsibilities include accepting new applications, parsing each application’s DAG into stages, determining the number of parallel execution instances (tasks) under each stage, mapping these instances onto edge nodes, and tracking their progress. 

This centralized architecture may run well when handling a small number of applications in the cloud. However, when it comes to a practical edge environment where new users join and exit more frequently and launch a large number of edge applications to run simultaneously, the architecture easily becomes a scalability bottleneck and jeopardizes the performance of latency-sensitive applications. This deficiency is due to two factors:  (1) \textbf{high deployment latency}. These systems use a first-come, first-served approach to deploying applications, causing applications to wait in a long queue and thus leading to long query latencies; and (2) \textbf{lack of flexibility for dataflow path planning}. They limit themselves to a fixed execution model and lack the flexibility to design different dataflow paths for different applications to adapt to the dynamic conditions of heterogeneous edge environments. The edge stream processing systems' query model and execution pipeline are presented in 
Appendix~\ref{subsec:query} and Appendix~\ref{subsec:pipeline}, respectively.

\section{Design}

\label{sec:design}
This section introduces AgileDart's dynamic dataflow abstraction and shows how to scale operators up and down and perform failure recovery on top of this abstraction. 

\subsection{Overview}
\label{subsec:d_overview}

As shown in Figure~\ref{fig:dynamicdataflow}, AgileDart consists of three layers: \emph{the DHT-based consistent ring overlay}, \emph{the dynamic dataflow abstraction}, and \emph{the scaling and failure recovery mechanisms}.

\textbf{\textit{Layer 1: DHT-based consistent ring overlay.}} All distributed edge ``nodes'' (e.g., routers, gateways, or powerful sensors) are self-organized into a DHT-based overlay, which has been commonly used in Bitcoin~\cite{Nakamoto2008} and BitTorrent~\cite{bittorrent}. Each node is randomly assigned a unique ``NodeId'' in a large circular NodeId space. NodeIds are used to identify the nodes and route stream data. \emph{No matter where the data is generated, it is guaranteed that the data can be routed to any destination node within $O(\log N)$ hops.} To do that, each node needs to maintain two data structures: a routing table and a leaf set. The routing table is used for building dynamic dataflows. The leaf set is used for scaling and failure recovery.

\textbf{\textit{Layer 2: Dynamic dataflow abstraction.}} Built upon the overlay, we introduce a novel dynamic dataflow abstraction. The key innovation is to leverage DHT-based routing protocols to approximate the optimal routes between source nodes and sink nodes, which can automatically place and chain operators to form a dataflow graph for each application. 

\textbf{\textit{Layer 3: Scaling and failure recovery mechanisms.}} Every node has a leaf set that contains the physically ``closest'' nodes to this node. The leaf set provides the elasticity for (1) scaling up and down operators to adapt to workload variations; (2) replanning the data shuffling paths to adapt to the edge network dynamics. As stream data moves along the dataflow graph, the system makes dynamic decisions about the downstream node to send streams to, which increases network path diversity and becomes more resilient to changes in network conditions; and (3) replicating operators to handle failures and stragglers. If any node fails or becomes a straggler, the system can automatically switch over to a replica. Then, the system can continue processing with very little or no disruption. 

\subsection{Dynamic Dataflow Abstraction}
\label{subsec:d_abstraction}

In the peer-to-peer (P2P) model (e.g., Pastry~\cite{pastry}, Chord~\cite{chord}), each node is equal to the other nodes in that they have the same rights and duties. The primary purpose of the P2P model is to enable all nodes to work collaboratively to deliver a specific service. For example, in BitTorrent~\cite{bittorrent}, if someone downloads some file, the file is downloaded to her computer in bits and parts that come from many other computers in the system that already have that file. At the same time, the file is also sent (uploaded) from her computer to others who ask for it. Similarly to BitTorrent, in which many machines work collaboratively to undertake the duties of downloading and uploading files, we enable all distributed edge nodes to work collaboratively to undertake the responsibilities of the original monolithic master’s.

\begin{figure}[t]
  \centering
  \includegraphics[width=0.4\textwidth]{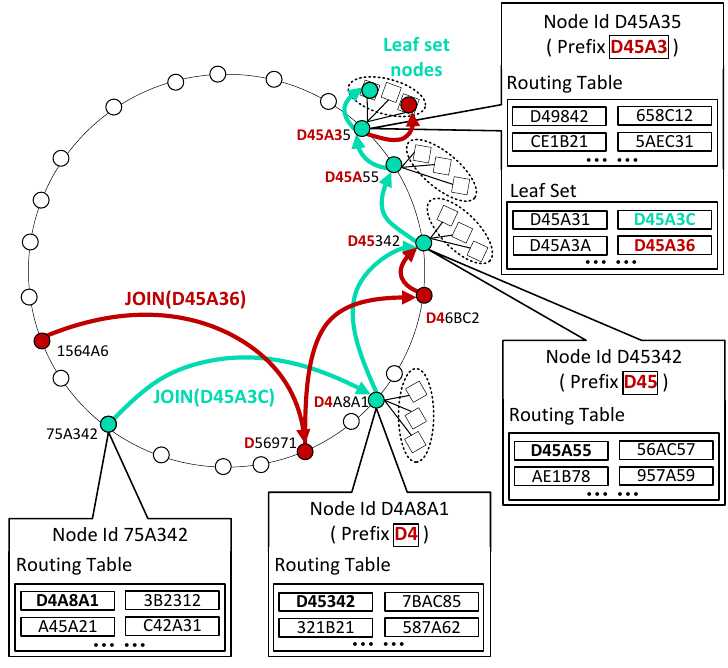}
  \caption{The process of building a dynamic dataflow graph.}
  \label{fig:dht}
\end{figure}

Figure~\ref{fig:dht} shows the process of building the dynamic dataflow graph for an edge stream application. \emph{First}, we organize distributed edge nodes into a P2P overlay network, which is similar to the way BitTorrent nodes use the Kademila DHT~\cite{maymounkov2002kademlia} for ``trackerless'' torrents. Each node is randomly assigned a unique identifier known as the ``NodeId'' in a large circular node ID space (e.g., $0 \sim 2^{128}$). \emph{Second}, given an IoT stream application, we map the source operators to the sensors that generate the data streams. We map the sink operators to IoT actuators or message queues to the cloud service. \emph{Third}, every source node sends a \texttt{JOIN} message towards a key, where the key is the hash of the sink node's NodeId. Because all source nodes belonging to the same application have the same key, their messages will be routed to a rendezvous point---the sink node(s). Then, we keep a record of the nodes that these messages pass through during routing and link them together to form the dataflow graph for this application. 

To achieve low latency, the overlay guarantees that the stream data can be routed from source nodes to sink nodes within $O(\log N)$ hops, thus ensuring the query latency upper bound. To achieve locality, the dynamic dataflow graph covers a set of nodes from sources to sinks. The first hop is always the node closer to the data source (data locality). Each node in the path has many leaf set nodes, which provides enough heterogeneous candidate nodes with different capacities and increases network path diversity. For example, if there are more operators than nodes, extra operators can map onto leaf set nodes. For that purpose, each node maintains two data structures: a routing table and a leaf set.

\begin{itemize}[topsep=0pt,itemsep=-1ex,partopsep=1ex,parsep=1ex]
    \item \emph{Routing table}: it consists of node characteristics organized in rows by the length of the common prefix. The routing works based on prefix-based matching. Every node knows \emph{m} other nodes in the ring and the distance of the nodes it knows increases exponentially. It jumps closer and closer to the destination, like a greedy algorithm, within $\lceil \log_{2^b}N \rceil$ hops. We add extra entries in the routing table to incorporate proximity metrics (e.g., hop count, RTT, cross-site link congestion level) in the routing process so as to handle bandwidth variations. 
    \item \emph{Leaf set}: it contains a fixed number of nodes whose NodeIds are ``physically'' closest to that node, which assists in rebuilding routing tables and reconstructing the operator’s state when any node fails.
\end{itemize}

As shown in Figure~\ref{fig:dht}, node \texttt{75A342} and node \texttt{1564A6} are two source nodes and node \texttt{D45A3C} is the sink node.  The source nodes route \texttt{JOIN} messages towards the sink node, and their messages are routed to a rendezvous point (s) — the sink node(s). We choose the forwarder nodes along routing paths based on RTT and node capacity. Afterward, we keep a record of the nodes that their messages pass through during routings (e.g., node \texttt{D4A8A1}, node \texttt{D45342}, node \texttt{D45A55}, node \texttt{D45A35}, node \texttt{D45A3C}), and reversely link them together to build the dataflow graph. 

The efficiency comes from several factors. \emph{First}, the application's instances can be instantly placed without the intervention of any centralized master, which benefits the time-critical deadline-based edge application's queries. \emph{Second}, because keys are different, the paths and the rendezvous nodes of all application's dataflow graphs will also be different, distributing operators evenly over the overlay, which significantly improves the scalability. \emph{Third}, the DHT-based leaf set increases elasticity for handling failures and adapting to the bandwidth and workload variations.

\subsection{Elastic Scaling Mechanism}
\label{subsec:d_scale}
After an application's operators are mapped onto the nodes along this application's dataflow graph, \emph{how to auto-scale them to adapt to bandwidth and workload variations?} We need to consider various factors. Scaling up/down is to increase/decrease the parallelism (the number of instances) of the operator within a node. Scaling out means instantiating new instances on another node by re-distributing the data streams across extra network links. In general, scaling up or down usually affects an individual stage at a time, but scaling out may affect many stages at a time and thus incur more overhead. However, scaling out can solve the bandwidth bottleneck by increasing network path diversity, while scaling up/down may not. 

We design a heuristic approach that adapts execution based on various factors: bottleneck (compute or bandwidth), operator (stateless or stateful), dynamics (long-term or short-term), dataflow structures, and noises. If there are computational bottlenecks, we scale up the problematic operators. The intuition is that when data queuing increases, automatically adding more instances to the system will avoid the bottleneck. We leverage the Secant root-finding method~\cite{avriel2003nonlinear} to automatically calculate the optimal instance number based on the current system’s health value. The policy is pluggable. \revised{Suppose each instance can increase input rate and queue size by $r$ and $q$, respectively. In phase $p_n$, the rate at which new tasks or requests arrive at the system is $R_{p_n}$, and the number of tasks waiting to be processed is $Q_{p_n}$. Let $x_n$ and $x_{n-1}$ be the number of instances during phases $p_n$ and $p_{n-1}$. We define health score $f(x_n)$ as
\begin{align}
    f(x_n) = \alpha\frac{x_n\times r}{R_{p_n}} + (1-\alpha)\frac{x_n\times q}{Q_{p_n}},
    \nonumber
\end{align}
where $\alpha$ is the weight tuning the importance of two terms. Then the number of instances required for the next phase $p_{n+1}$ such that $f\cong1$ can be given by:
\begin{align}
x_{n+1} = x_n + (1- f(x_n)) \times \frac{x_n-x_{n-1}}{f(x_n)-f(x_{n-1})}.
\nonumber
\end{align}
}

For bandwidth bottlenecks, we further consider whether the operator is stateless or stateful. In the case of stateless operators, we simply scale out operators across nodes. For stateful operators, we migrate the operator with its state to a new node in the leaf set that increases the network path diversity. Intuitively, when the original path only achieves low throughput, an operator may achieve higher throughput by sending the data over another network path.

\subsection{Failure Recovery Mechanism}
\label{subsec:d_failure}

Since the overlay is self-organizing and self-repairing, the dataflow graph for each edge application can be automatically recovered by restarting the failed operator on another node. Here, the challenge is, \emph{how to resume the processing without losing intermediate data (i.e., operator state)}? Examples of operator states include keeping some aggregation or summary of the received tuples in memory or keeping a state machine for detecting patterns (for example, for fraudulent financial transactions) in memory. 
A conventional approach is replication~\cite{shah2004highly, borealis}, which uses backup nodes to process the same stream in parallel with the primary set of nodes, and the inputs are sent to both. The system then automatically switches over to the secondary set of nodes upon failures. This approach, however, costs 2$\times$ the hardware resources, therefore unsuitable for edge environments due to their limited hardware resources. Another approach is checkpointing~\cite{storm, Trident, timestream, drizzle}, which periodically checkpoints all operators’ states to a persistent storage system (e.g., HDFS or HBase) and the failover node retrieves the checkpointed state upon failures. It avoids the 2$\times$ hardware cost. However, this approach is slow because it must transfer state over the edge networks, which typically have very limited bandwidth.

To address this challenge, we propose a heuristic approach that adapts failure recovery mechanisms based on various factors: application characteristic (stateless or stateful), lifetime (short-term or long-term), state size (small or large), and DAG structure (simple or complex). If the application is stateless, we do not need state recovery. If the application is stateful but short-lived, the cost of data recovery may outweigh the cost of state unavailability and, thus, we also do not need state recovery. We can simply restart the failed operator on a new edge node. For stateful edge applications that run long-term with large-sized states, we propose a parallel recovery approach by leveraging erasure coding~\cite{reed1960polynomial}. 

\begin{figure}[t]
  \centering
  \includegraphics[width=0.5\textwidth]{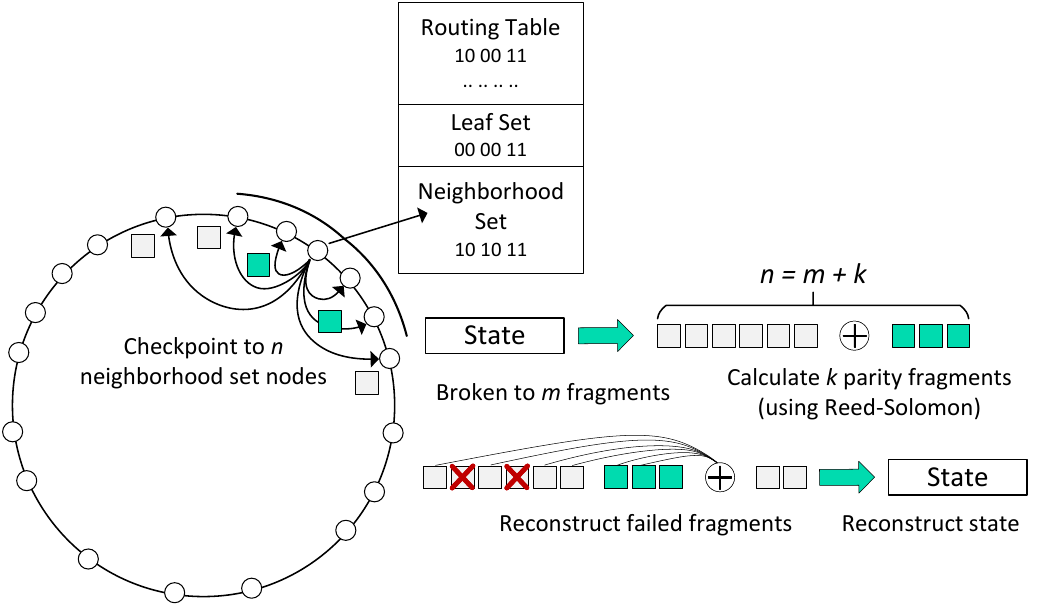}
  \caption{The erasure-coding-based parallel recovery approach.}
  \label{fig:failure_recovery}
\end{figure}

As shown in Figure~\ref{fig:failure_recovery}, each operator's larger-than-memory state is divided into $m$ fragments and then encoded into $n$ fragments and checkpointed to $n$ neighborhood set nodes in the DHT-based overlay in parallel ($n>m$). Erasure coding guarantees that the original state can be reconstructed from any $m$ fragments and it can tolerate a maximum of ($n-m$) failures at any time. By doing that, we do not need any central coordinator. The failure recovery process is fast because many nodes can leverage the application's dataflow graph to recompute the lost state in parallel upon failures. The replica number, the checkpointing frequency, the number of raw fragments $m$, the number of parity fragments $k$, and the number of encoded fragments $n$ are tunable parameters. They are determined based on state size, running environment, and the application’s service-level agreements (SLAs).

\section{Data Shuffling Model}
\label{sec:model}
Building the dynamic dataflow abstraction for operator placement is not efficient. Although we try to place connected operators on the same edge zone (or even the same edge node) to minimize the data transfer overhead between operators, common operators such as \texttt{union} and \texttt{join} may require a large volume of data to be transmitted across sites since their inputs may be generated at different locations. Thus, data shuffling, if planned poorly, may become a significant performance bottleneck.

In this section, we introduce a novel bandit-based exploitation-exploration path-planning model that can
dynamically replan the data shuffling paths to adapt to the edge network dynamics.

\subsection{Problem Formulation}

The challenge is that, in many cases, we don't know the quality of the edge network links in advance, such as the probability of successfully transmitting data shuffling packets in wireless sensor networks. This information can only be obtained by actually sending packets and observing the outcomes. Therefore, when we design the data shuffling paths for edge stream application’s queries, we face a dilemma: whether to explore new or unknown network links or exploit well-known ones. If we only rely on the known paths, we may miss out on finding a better path with faster communication or higher success rates. However, exploring too many new paths may result in more packet losses and higher communication delays.

Inspired by the Multi-Armed Bandit (MAB) models~\cite{stochastic_online_shortest_path,regret_analysis, adaptive_allocation_rules,combinatorial_mab}, we can formulate the data shuffling path-planning problem as a combinatorial Multi-Armed Bandit (MAB) optimization problem: we explore different paths to learn about their rewards (i.e., link quality), and exploit the paths that offer the highest rewards. By doing so, we can gradually improve our knowledge of the edge networks to find optimal paths.

The edge network can be modeled as a directed graph $G= (V, E)$, where $V$ is the set of nodes and $E$ is the set of links. Given a single source-sink pair $(s,d)\in V^2$, let $\mathcal{P}\subseteq \{0,1\}^{|E|}$ denote the set of all possible loop-free paths from node $s$ to node $d$ in $G$, where each path $p \in \mathcal{P}$ is a $|E|$-dimensional binary vector; for any $i\in E$, $p_i=1$ if and only if $i$ belongs to $p$. In practical edge networks, each link $i\in E$ can be unreliable. At any given time $t$, we can use a binary variable $X_i(t)$ to represent whether a transmission on link $i$ is successful. $X_i(t)$ is a sequence of independent random events, where each event is either a success or a failure with some unknown probability $\theta_i$. If we repeatedly attempt to send a packet on link $i$ until it succeeds, the time it takes (called the packet delay) follows a geometric distribution with mean $1/\theta_i$. Among the set of paths $\mathcal{P}$, there must exist a path $p^*$ that yields the minimal packet delay. Formally, $p^* \in \argmin_{p\in \mathcal{P}}D_\theta(p)$, where $D_\theta(p)$ is the total packet delay of path $p$.

\textbf{\textit{Optimization objective.}} Our goal is to efficiently route $K$ packets, such as data shuffling tuples, from a source operator to a sink operator, minimizing the overall time taken. This can be measured in terms of \emph{regret}, defined as the cumulative difference of expected delay between the path chosen by a policy and the unknown optimal path.
Therefore, the regret $R^\pi(K)$ of policy $\pi$ up to the $K$-th packet is the expected difference in delays between policy $\pi$ and the optimal policy that selects the best path $p^*$ for transmission:
\begin{align}
    R^\pi(K) =\mathbb{E}\Bigg[\sum_{k=1}^K D^\pi(k)\Bigg]-K D_\theta(p^*),
    \nonumber
\end{align}
where $D^\pi(k)$ denotes the end-to-end delay of the $k$-th packet under policy $\pi$, the expectation $\mathbb{E}[\cdot]$ is taken with respect to the random transmission outcomes and possible randomization in the policy $\pi$, and $D_\theta(p^*)$ is the total packet delay of the best path $p^*$. Our optimization objective is to find a policy $\pi$ among all policies $\Pi$ such that 
\begin{align}
    \min_{\pi \in \Pi} R^\pi(K). \nonumber
\end{align}

\begin{algorithm}[t]

\setstretch{1.25}
    \caption{AgileDart's distributed data shuffling path planning algorithm}
    \label{pseudo:our_algo}
    \begin{algorithmic}[1]
        \For{time slot $\tau \geq 1$ }
            \State Select link $(v,v')\in E$, where
            \label{state:select_link}
            \State \quad $v'\in \argmin_{w\in V\setminus\{v\}:(v,w)\in E}C_\tau(v,w)$, where \newline
                    \hspace*{5em}$C_\tau(v,w) = \omega_\tau(v,w)+J_\tau(w)$
            \label{state:cost_function}
            \State Collect feedback on link $(v,v')$, and update $\omega_\tau(v,v')$ by updating 
        empirical success rate $\hat{\theta}_\tau(v,v')$, 
        number of packets routed $s_\tau(v,v')$,
        total number of transmission attempts $t'_\tau(v,v')$,
        and long-term routing cost $J_\tau(v')$.
        \EndFor
        \label{state:update_feedback}  
    \end{algorithmic}
\end{algorithm}

\begin{table}[t]
\setlength{\belowcaptionskip}{10pt} 
\renewcommand\arraystretch{1.25}
 \small
\begin{tabular}{cp{6.7cm}}
    \Xhline{2\arrayrulewidth}
    \rule{0pt}{1.1\normalbaselineskip}\textbf{Variable} & \textbf{Description} \vspace{0.05in}\\ \hline 
    \rule{0pt}{1.1\normalbaselineskip}    
    
    $s$    &  Source node.\\
    $d$    &  Sink node.   \\
    $V$    &  Set of nodes in $G$.\\
    $E$    &  Set of links in $G$.\\
    $G$    &  Graph with a set of nodes $V$ and a set of links $E$.    \vspace{0.05in}      \\\hline
    \rule{0pt}{1.1\normalbaselineskip}$(v,v')$    &  Link from node $v$ to node $v'$.         \\  
    $C_\tau(v,v')$  &    Cost function of link $(v,v')$.         \\
    $J_\tau(w)$     &  Long-term routing cost of node $w$.  \\
    $n(\tau)$     &     The packet number that is about to be sent or
    is already in the network at time slot $\tau$.        \\
    $s_\tau(v,v')$  &    The number of packets routed through link $(v,v')$ before the $n(\tau)$-th packet is sent.         \\
    $t'_\tau(v,v')$  &   Total number of transmission attempts on link $(v,v')$ up to time slot $\tau$.    \\
    $\hat{\theta}_\tau(v,v')$     &    Empirical success rate of link $(v,v')$ up to time slot $\tau$.         \\
    $\omega_\tau(v,v')$     &  Empirical transmission cost of link $(v,v')$ with exploration adjustment up to time slot $\tau$.     \vspace{0.05in}     \\
    \Xhline{2\arrayrulewidth}
\end{tabular}
\caption{Key variables in AgileDart's distributed data shuffling path-planning algorithm.}
\label{tab:our_notation}
\vspace{-0.2in}
\end{table}

\subsection{Bandit-Based Data Shuffling Algorithm}

We propose a distributed data shuffling path-planning algorithm based on the semi-feedback bandit model~\cite{stochastic_online_shortest_path, bandit_algo_book,online_influence}.

\textbf{\textit{Basic operation.}} The pseudo-code of the distributed data shuffling path-planning algorithm is given in Algorithm~\ref{pseudo:our_algo}. 
Table~\ref{tab:our_notation} summarizes the key variables in the proposed algorithm. 
Whenever node $v$ receives a packet from a previous node at time slot $\tau$, it first runs line~\ref{state:select_link} and line~\ref{state:cost_function} to find the next hop $v' \in V\setminus\{v\}$ and link $(v,v') \in E$, where link $(v,v')$ has the minimum transmission cost in cost function $C_\tau(v,v')$. The cost function $C_{\tau}(v,v')$ contains two terms: the \textit{empirical transmission cost with exploration adjustment} $\omega_\tau(v,v')$ and the \textit{long-term routing cost} $J_t(v')$. After sending the packet to the next hop $v'$ through link $(v,v')$, node $v$ receives feedback from node $v'$ and updates the empirical success rate of link $(v,v')$, number of packets routed through link $(v,v')$, total transmission attempts of link $(v,v')$, and the long-term routing cost of the next hop $v'$ (line~\ref{state:update_feedback}).

\textbf{\textit{Design rationale.}}
The design rationale for the cost function $C_\tau(v,v')$ in Algorithm~\ref{pseudo:our_algo} is as follows:
\begin{enumerate}

\item The total delay of a path depends on the packet delays of all the links that make up the path and the selection of
the first link will affect subsequent link choices.

\item To address this, the cost function is divided into two terms: $\omega_\tau(v,v')$ and $J_\tau(v')$. The first term $\omega_\tau(v,v')$ estimates the packet delay of the link $(v,v')$ based on its empirical packet delay and visit frequency. The second term $J_\tau(v')$ estimates all packet delays of subsequent paths of node $v'$ based on the empirical transmission delays and visit frequencies of the links in the subsequent paths. 

\item In terms of packet delay, the first term tells us how good one link is so far (e.g., node $v_1$ and node $v_2$) and the second term tells us how good a subsequent path of the outgoing node of the link is so far (e.g., one path from $v_1$ to $d$ or the other path from $v_2$ to $d$).

\item By utilizing the two terms in the cost function, we can avoid the potential stragglers in subsequent paths.

\end{enumerate}
 
Now, let's clarify the terms introduced in this rationale:

We first introduce the term $\omega_\tau(v,v')$. For any time slot $\tau$, let $n(\tau)$ denote the packet number that is about to be sent or is already in the network. For any link, let $\hat{\theta}_\tau(v,v')$ denote the empirical success rate of link $(v,v')$ up to time slot $\tau$, that is $\hat{\theta}_\tau(v,v') = s_\tau(v,v')/t'_\tau(v,v')$, where $s_\tau(v,v')$ is the number of packets routed through link $(v,v')$ before the $n(\tau)$-th packet is sent, and $t'_\tau(v,v')$ is the total number of transmission attempts on link $(v,v')$ up to time slot $\tau$. Thus the empirical transmission cost of link $(v,v')$ with exploration adjustment up to time slot $\tau$ is 
\begin{align*}
    \omega_\tau(v,v') = \min \big\{&\frac{1}{u}:u\in [\hat{\theta}_\tau(v,v'), 1],  \\
    &t'_\tau(v,v')\cdot KL\big(\hat{\theta}_\tau(v,v'), u\big) \leq C\log(\tau)
    \big\}, \nonumber
\end{align*}
where $KL(\hat{\theta}_\tau(v,v'), u)$ is the Kullback–Leibler (KL) divergence number~\cite{kl_divergence} between two Bernoulli distributions with respective means $\hat{\theta}_\tau(v,v')$ and $u$, i.e.,
\begin{align*}
    KL(\hat{\theta}_\tau(v,v'), u) =  &\hat{\theta}_\tau(v,v')\cdot\log\Big(\frac{\hat{\theta}_\tau(v,v')}{u}\Big) \\
    & + \big(1-\hat{\theta}_\tau(v,v')\big)\cdot\log\Big(\frac{1-\hat{\theta}_\tau(v,v')}{1-u}\Big),
\end{align*}
where $C \in (0,1]$ is an exploration factor that controls the degree of exploration. When the value of $C$ is close to 1, it tends to select those links that were not frequently visited in the previous trials, which is a good fit for some edge networks that have high communication variations.

We next introduce the term $J_\tau(w)$. Let $\mathcal{P}_w$ denote the set of loop-free paths from node $w$ to the destination of a packet. Then, $J_\tau(w)$ is the minimum total empirical transmission cost with exploration adjustment $\omega_\tau(i)$ of a path $p$ in $\mathcal{P}_{w}$:
\begin{align*}
    J_\tau(w) = \min_{p\in \mathcal{P}_{w}}\sum_{i\in p}\omega_\tau(i),
\end{align*}
where $i\in p$ denotes link $i$ in path $p$.

In summary, this algorithm can efficiently identify the optimal path for data shuffling in unreliable edge networks, taking into account link heterogeneity and minimizing regret in the path selection process. \revised{We conduct complexity analysis in 
Appendix~\ref{sec:time_complexity}.
}

\section{Implementation}
\label{sec:impl}

\begin{figure}[t]
  \centering
  \includegraphics[width=0.45\textwidth]{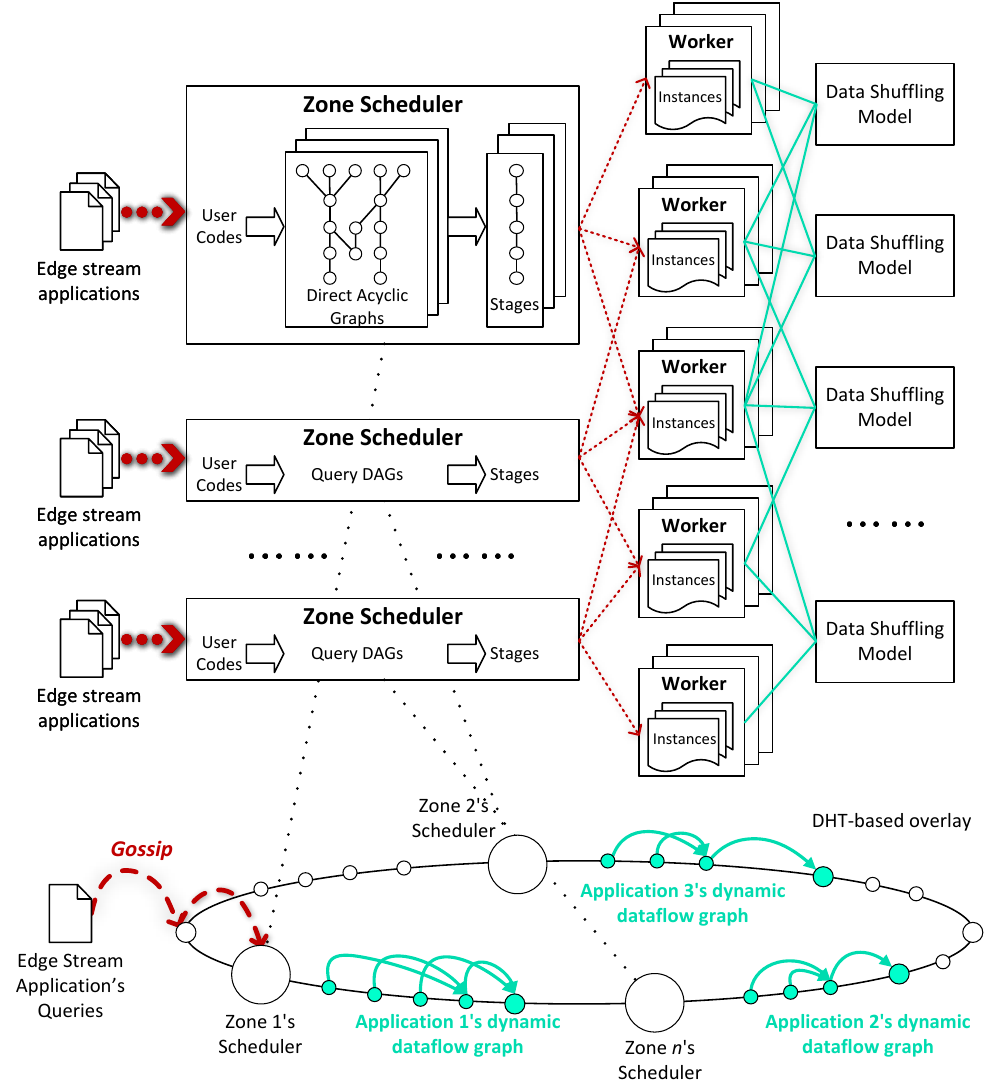}
  \caption{The AgileDart system architecture.}
  \label{fig:impl}
\end{figure}

Instead of implementing another distributed system core, we implement AgileDart on top of Apache Flume~\cite{flume} (v.1.9.0) and Pastry~\cite{freepastry} (v.2.1) software stacks. Flume is a distributed service for collecting and aggregating large amounts of streaming event data, which is widely used with Kafka~\cite{ApacheKafka} and the Spark ecosystem. Pastry is an overlay network and routing network for the implementation of a distributed hash table (DHT) similar to Chord~\cite{chord}, which is widely used in applications such as Bitcoin~\cite{Nakamoto2008}, BitTorrent~\cite{bittorrent}, FAROO~\cite{faroo}, Kazaa~\cite{kazaa}, Limewire~\cite{limewire}, and Acquisition~\cite{acquisition}. A unique feature of Pastry is proximity-aware routing. We leverage Flume’s excellent runtime system (e.g., basic API, code interpreter, transportation layer) and Pastry’s routing substrate and event transport layer to implement the AgileDart system.

We made three major modifications to Flume and Pastry: (1) we implemented the dynamic dataflow abstraction for operator placement and path planning, which includes a list of operations to track the DHT routing paths for chaining operators and a list of operations to capture the performance metrics of nodes for placing operators; (2) we implemented the scaling mechanism and the failure recovery mechanism by introducing queuing-related metrics (queue length, input rate, and output rate), buffering operator’s in-memory state, encoding and replicating state to leaf set nodes; (3) we implemented the distributed schedulers by using Scribe~\cite{scribe} topic-based trees on top of Pastry; and (4) we implemented the bandit-based path-planning model that dynamically replans the stream data shuffling paths to adapt to the unreliable, heterogeneous edge networks. 

Figure~\ref{fig:impl} shows the high-level architecture of the AgileDart system. The system has two components: a set of distributed schedulers that span geographical edge zones and a set of workers. Unlike traditional stream processing systems that manually assign nodes as ``scheduler’’ or ``workers’’, AgileDart dynamically assigns nodes as ``schedulers'' or ``workers''. For the first step, when any new edge stream application is launched, it looks for a nearby scheduler by using the gossip protocol~\cite{gossip}, which is a procedure of P2P communication based on the way that epidemics spread. If it successfully finds a scheduler within $O(\log N)$ hops, the application registers itself to this scheduler. If not, it means there is no scheduler in this zone. This application will vote any random nearby node to be the scheduler and register itself to that scheduler. For the second step, the scheduler processes this application’s queries by parsing the application's user code into a directed acyclic graph (DAG) and dividing this DAG into stages. The vertices represent stream operators and the edges represent data flows between operators. Then, the scheduler automatically parallelizes, chains operators, and places the instances on edge nodes using \emph{the proposed dynamic dataflow abstraction}. These nodes are then set as this application's workers. The dynamic dataflow graph can automatically scale up and out operators and replicate operators to adapt to the workload variations and recover from failures using \emph{the proposed scaling mechanism and failure recovery mechanism}.
For the third step, the dynamic dataflow graph replans the data shuffling paths to adapt to the edge network dynamics using \emph{the proposed bandit-based path-planning model}.

\section{Evaluation}
\label{sec:eval}

We evaluate AgileDart on a real distributed network environment that has 100 Linux virtual machines (VMs) and Raspberry Pis. We use a full-stack standard IoT stream processing benchmark and real-world IoT stream processing applications with real-world datasets. Our evaluation answers these questions:

\begin{itemize}
    \item Does AgileDart enhance latency during the processing of a vast number of edge stream applications?

    \item Does AgileDart scale effectively as the number of concurrently running edge stream applications increases or as the system size (number of edge nodes) grows?

    \item Does AgileDart improve adaptivity in the presence of workload changes, transient failures, and mobility?
    
    \item What is the runtime overhead of AgileDart?
\end{itemize}

\begin{figure*}[t]
 \centering
\captionsetup[subfigure]{width=5cm}
 \subfloat[DAG queue waiting time comparison of AgileDart and EdgeWise.]{
 \includegraphics[width=.26\linewidth]{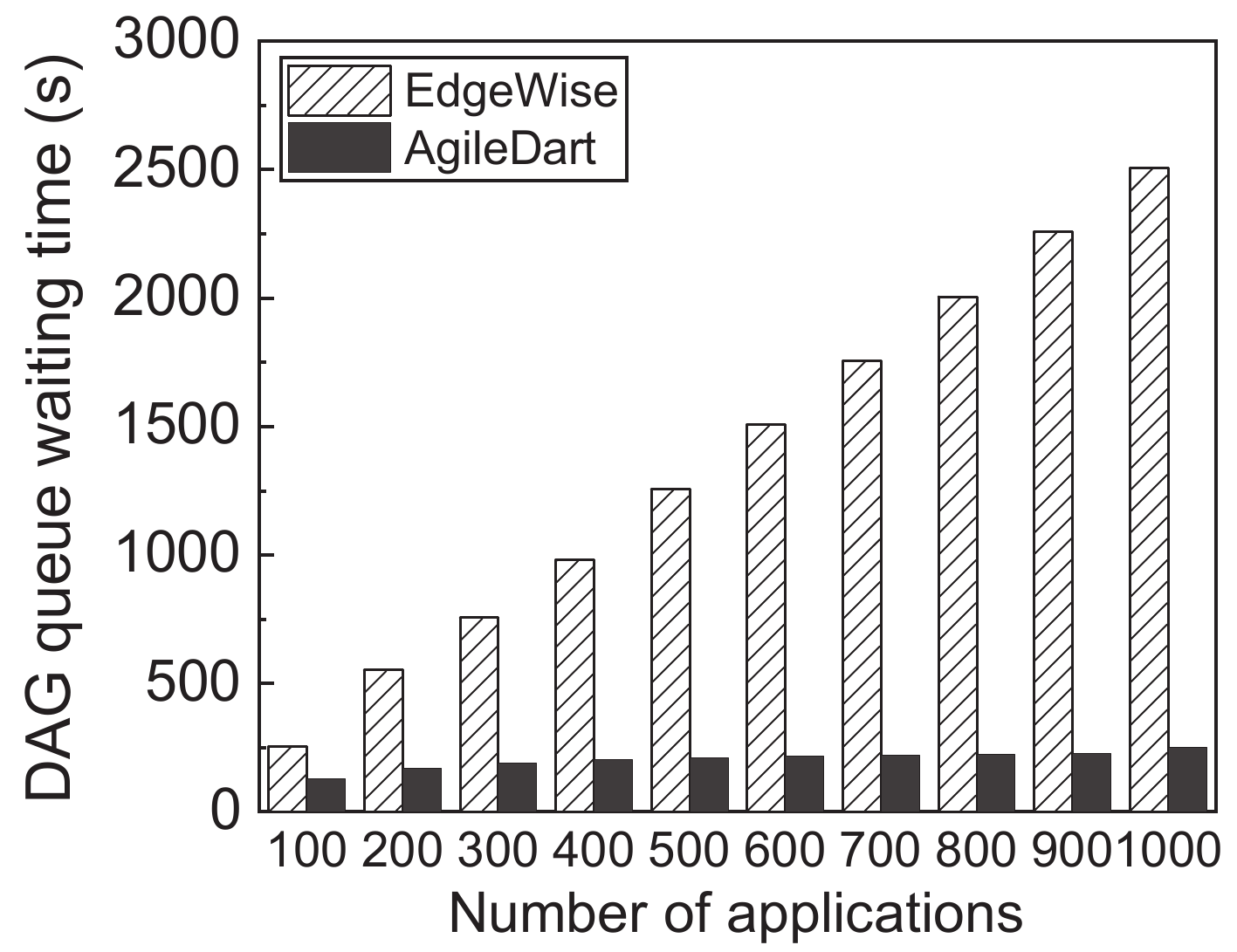}
 \label{subfig:latency_queue}
 }\hspace{1.5em}
 \subfloat[DAG deployment time comparison of AgileDart and EdgeWise.]{
 \includegraphics[width=.26\linewidth]{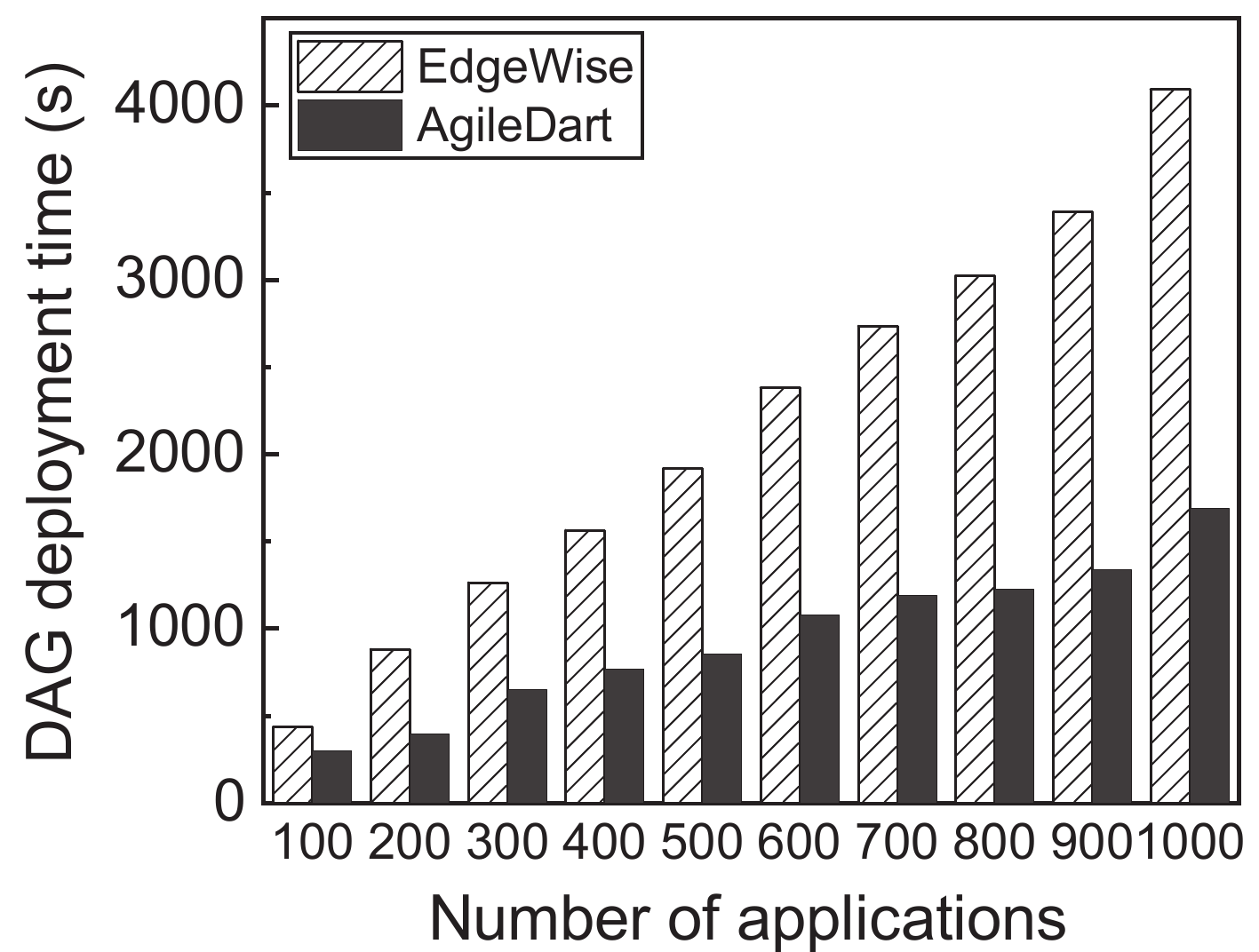}
 \label{subfig:latency_deploy}
 }\hspace{1.5em}
 \subfloat[Query processing time.]{
 \includegraphics[width=.24\linewidth]{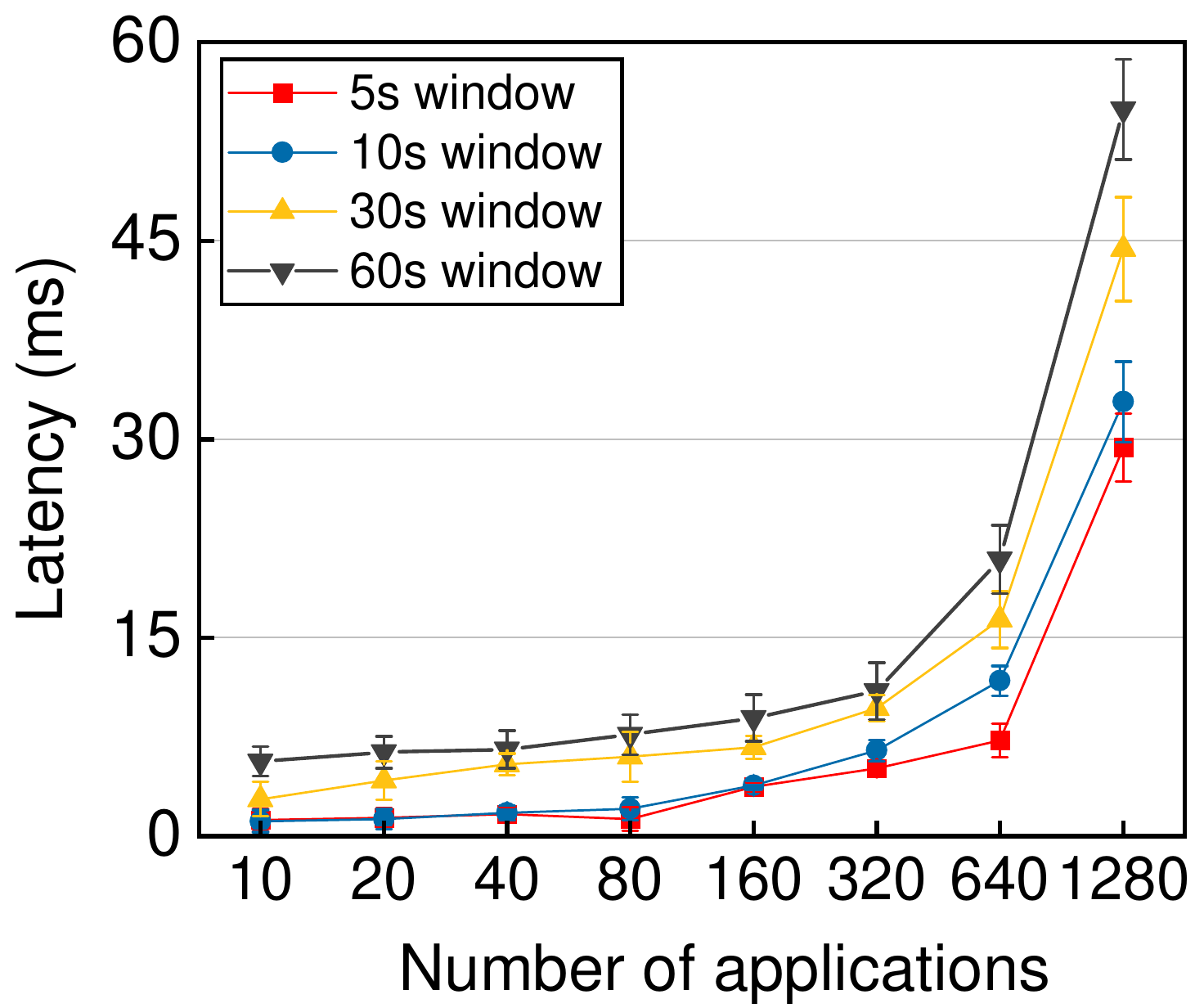}
 \label{subfig:latency_window}
 }
 \vspace{0.15in}
 \caption{The latency comparison of AgileDart and EdgeWise for (a) DAG queue waiting time, (b) DAG deployment time, and (c) query processing time for an increasing number of concurrently running applications.} 
 \vspace{0.1in}
\end{figure*} 

\begin{figure*}[t]
 \centering
\captionsetup[subfigure]{width=5cm}
 \subfloat[Calculating the frequent route in the taxi application.]{
 \includegraphics[width=.24\linewidth]{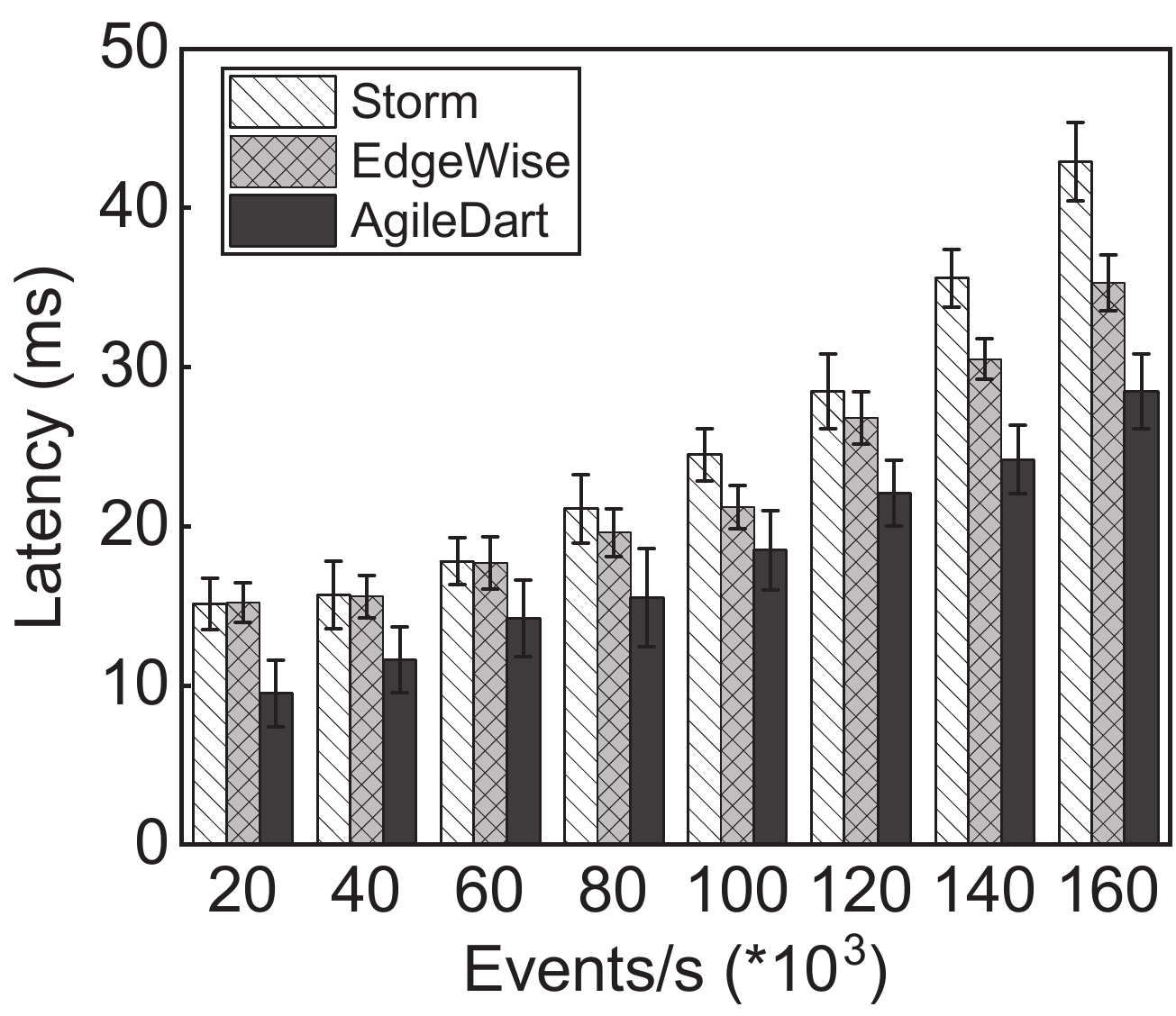}
 \label{subfig:app_taxi_route}
 }\hspace{1.5em}
 \subfloat[Calculating the most profit area in the
taxi application.]{
 \includegraphics[width=.24\linewidth]{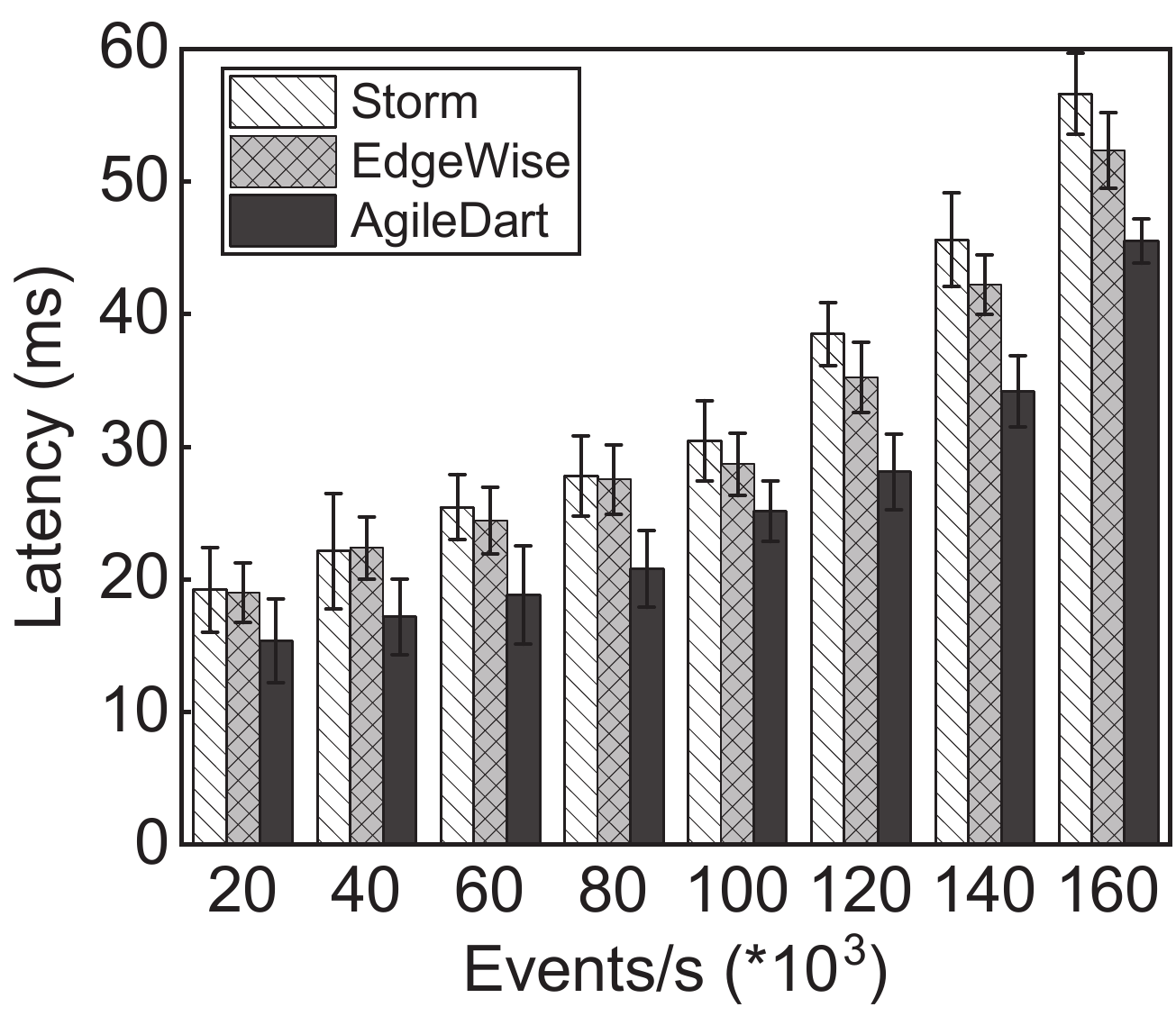}
 \label{subfig:app_taxi_profit}
 }\hspace{1.5em}
 \subfloat[Visualizing environmental changes in the urban sensing application.]{
 \includegraphics[width=.24\linewidth]{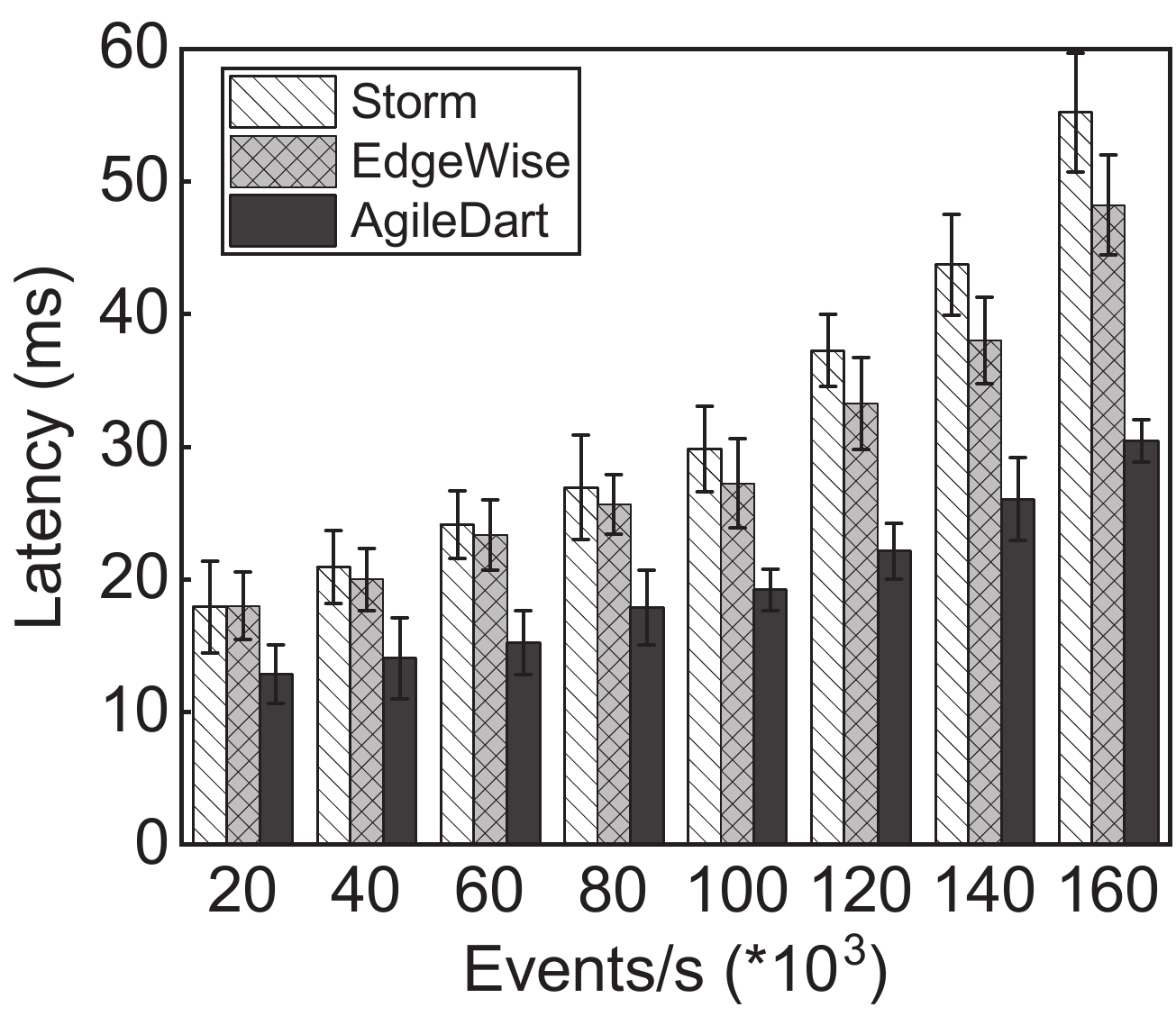}
 \label{subfig:app_urban}
 }
 \vspace{0.15in}
 \caption{The latency comparison of AgileDart, Storm, and EdgeWise for (a) frequent route application, (b) profitable area application, and (c) urban sensing application.} 
\end{figure*} 

\subsection{Setup}
\textbf{\textit{Real hardware.}} In our real hardware experiments, we employ an intermediate class of computing devices representative of IoT edge devices. Specifically, we use 10 Raspberry Pi 4 Model B devices for hosting source operators, each of which has a 1.5GHz 64-bit quad-core ARMv8 CPU with 4GB of RAM and runs Linux raspberrypi 4.19.57. Raspberry Pis are equipped with Gigabit Ethernet Dual-band Wi-Fi. We use 100 Linux virtual machines (VMs) to represent the gateways and routers for hosting internal and sink operators, each of which has a quad-core processor and 1GB of RAM (equivalent to Cisco's IoT gateway ~\cite{ciscofog}). These VMs are connected through a local-area network. In order to make our experiments closer to real edge network scenarios, we used the TC tool~\cite{linux-tc} to control link bandwidth differences.

\textbf{\textit{Distributed deployment.}} In our real distributed network experiments, we use a testbed consisting of 100 VMs running Linux 3.10.0, all interconnected through Gigabit Ethernet.  Each VM has 4 cores and 8GB of RAM, and 60GB disk. Specifically, to evaluate AgileDart's scalability, we launch up to 10,000 edge nodes in our testbed.

\textbf{\textit{Baseline.}} As a baseline, we use Storm~\cite{storm} and EdgeWise~\cite{EdgeWise} for comparison. Apache Storm is version 2.0.0~\cite{storm}, and EdgeWise~\cite{EdgeWise} is downloaded from GitHub~\cite{edgewisegit}. Both engines are configured with 10 TaskManagers, each with 4 slots (maximum parallelism per operator = 36). We run Nimbus and ZooKeeper~\cite{zookeeper} on the VMs and run supervisors on the Raspberry Pis. We use Pastry 2.1~\cite{pastry} configured with leaf set size of 24, max open sockets of 5000 and transport buffer size of 6 MB.

\textbf{\textit{IoT stream benchmark and real-world IoT stream processing applications.}} We simultaneously deploy a large number of applications (topologies) to demonstrate the scalability of our system. These applications are selected from a full-stack standard IoT stream processing benchmark~\cite{riotbench}. Additionally, we implement four IoT stream processing applications that use real-world datasets\cite{sensecity, debs2015, soil, predictingtaxi}. These applications incorporate a broad range of techniques, including \emph{predictive analysis}, \emph{model training}, \emph{data preprocessing}, and \emph{statistical summarization}. The operators in these applications execute a variety of functions, such as \texttt{transform}, \texttt{filter}, \texttt{flatmap}, \texttt{aggregate}, \texttt{duplicate}, and \texttt{hash}. 

We implement the DEBS 2015 application~\cite{debs2015} to process spatio-temporal data streams and calculate real-time indicators of the most frequent routes and most profitable areas in New York City. The sensor data consists of taxi trip reports that include start and drop-off points, corresponding timestamps, and payment information. Data are reported at the end of the trip, i.e., upon arrival in the order of the drop-off timestamps. While the prediction tasks in this application do not demand immediate real-time responses, they do capture the data distribution and query patterns typical of more complex upcoming transportation engines. An application that integrates additional data sources such as bus, subway, car-for-hire (e.g., Uber), ride-sharing, traffic, and weather conditions, would exhibit the same structural topology and query rates that we use in our experiments while offering decision-making support in the scale of seconds.

We implement the Urban Sensing application~\cite{sensecity} to aggregate pollution, dust, light, sound, temperature, and humidity data from seven cities, enabling real-time understanding of urban environmental changes. Given the potential for a practical deployment of environmental sensing with thousands of such sensors per city, we've scaled the input rate by a factor of 1000$\times$ to simulate a larger deployment of 90,000 sensors.

\begin{figure*}[t]
 \centering
\captionsetup[subfigure]{width=5cm}
 \subfloat[The distribution of AgileDart's operators over different edge nodes.]{
 \includegraphics[width=.24\linewidth]{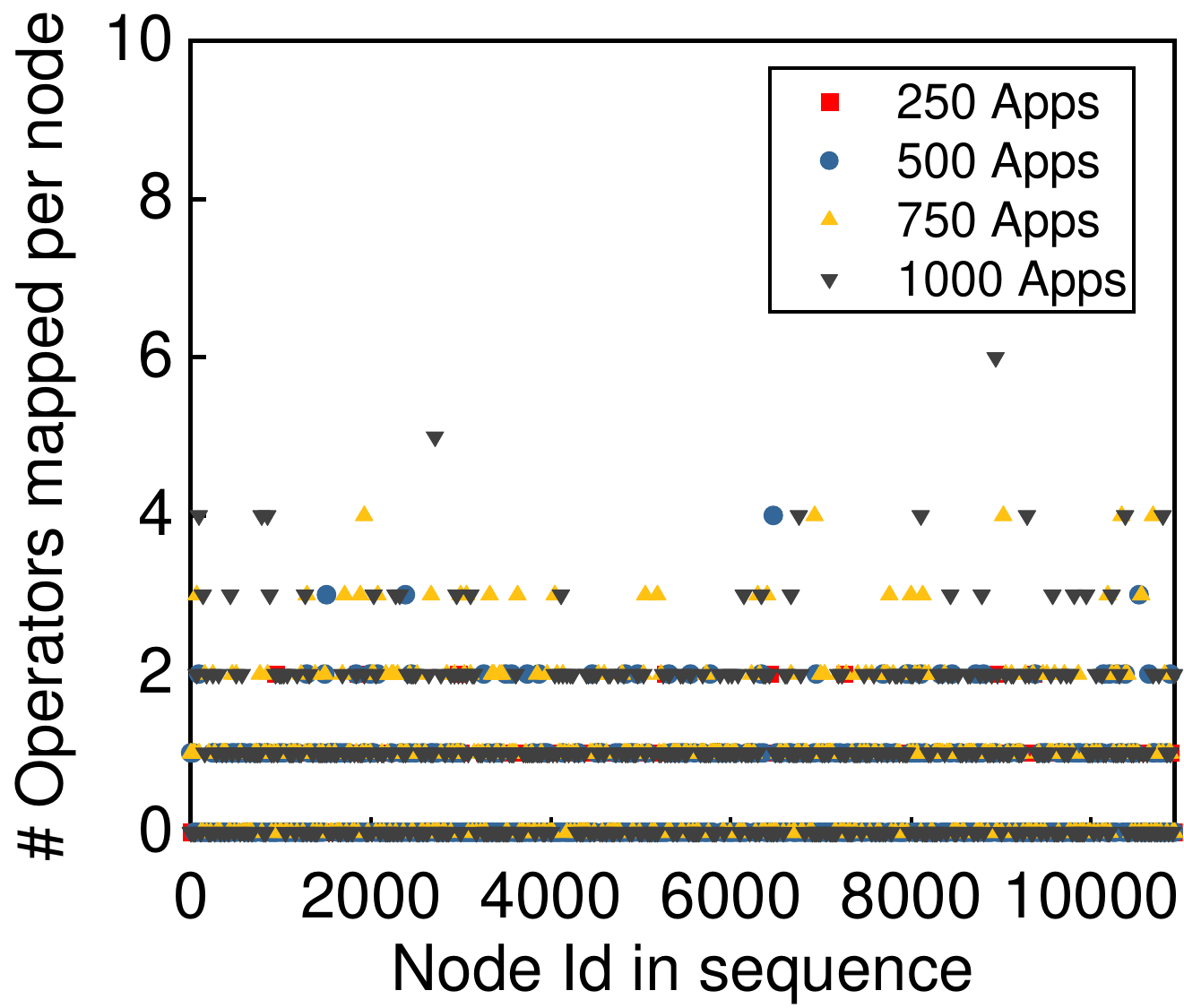}
 \label{subfig:scalability_operators}
 }\hspace{1.5em}
 \subfloat[Normal probability plot of the number of operators mapped per node.]{
 \includegraphics[width=.26\linewidth]{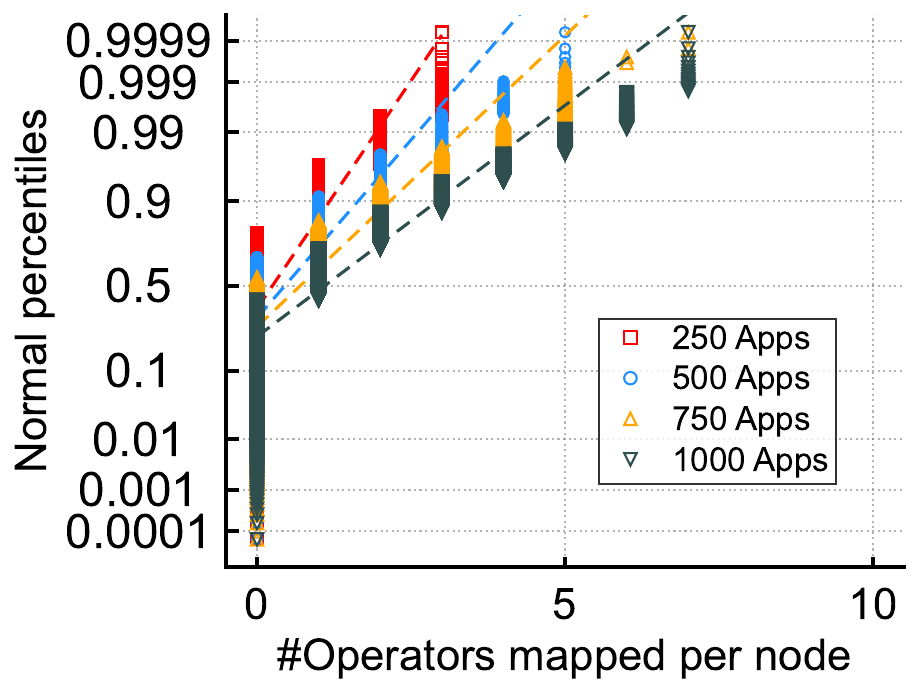}
 \label{subfig:scalability_cdf}
 }\hspace{1.5em}
 \subfloat[The distribution of AgileDart's schedulers over different edge zones.]{
 \includegraphics[width=.24\linewidth]{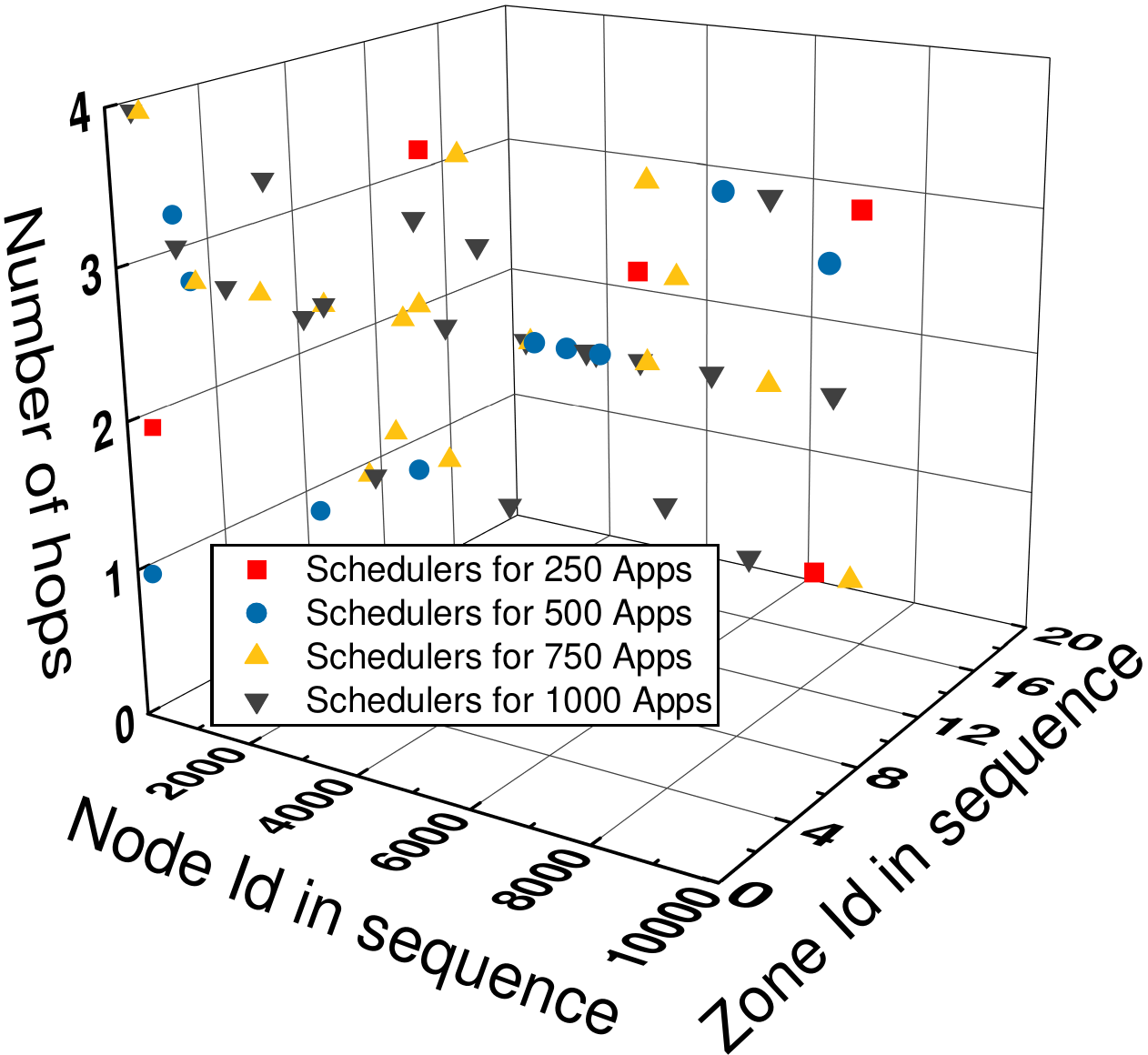}
 \label{subfig:scalability_3d}
 }
 \vspace{0.15in}
 \caption{Scalability study of AgileDart for the distribution of operators and schedulers over large-scale edge topologies.} 
 \vspace{0.1in}
\end{figure*} 

\begin{figure*}[t]
 \centering
\captionsetup[subfigure]{width=5cm}
 \subfloat[Overlay and dataflow recovery time.]{
 \includegraphics[width=.26\linewidth]{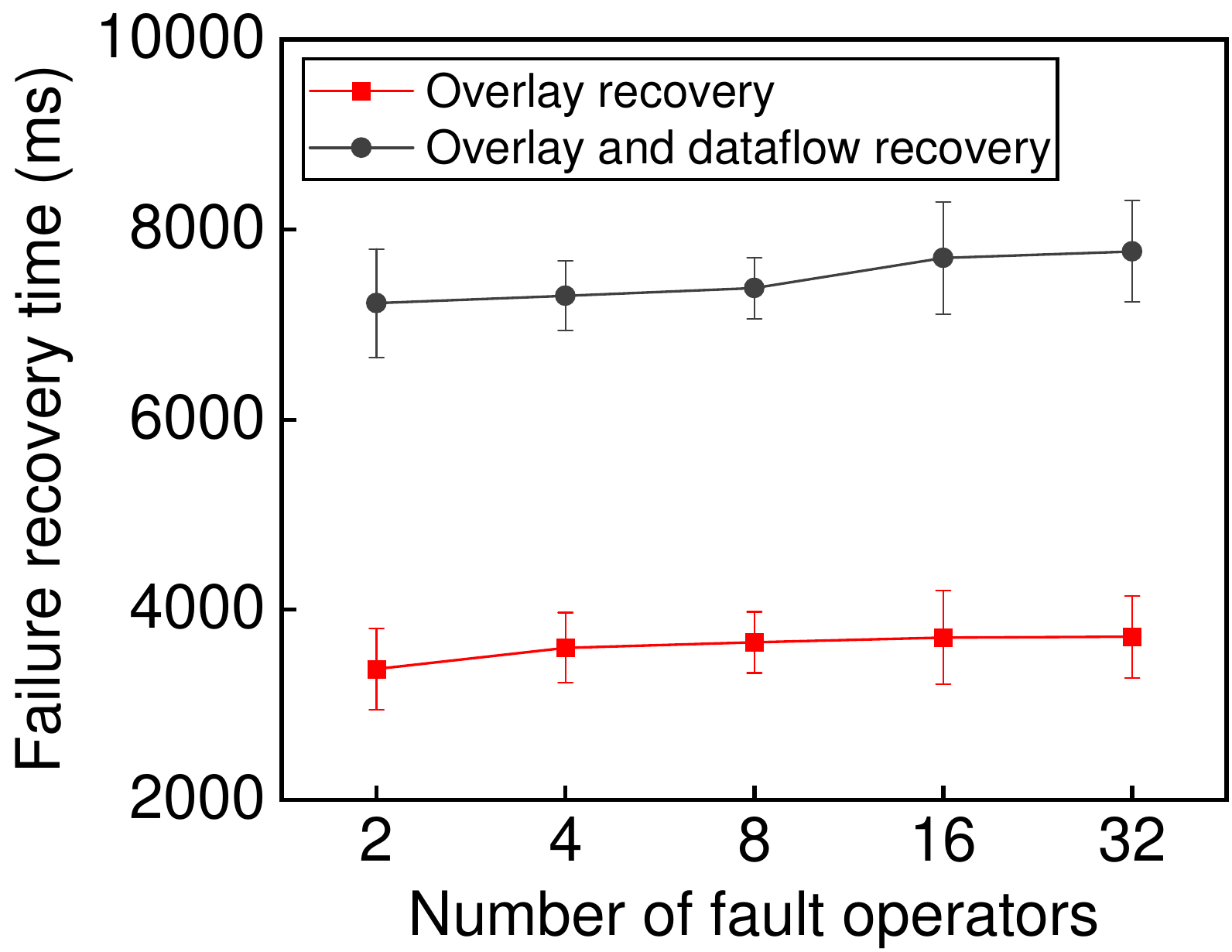}
 \label{subfig:recovery_overlay}
 }\hspace{1.5em}
 \subfloat[State recovery time.]{
 \includegraphics[width=.26\linewidth]{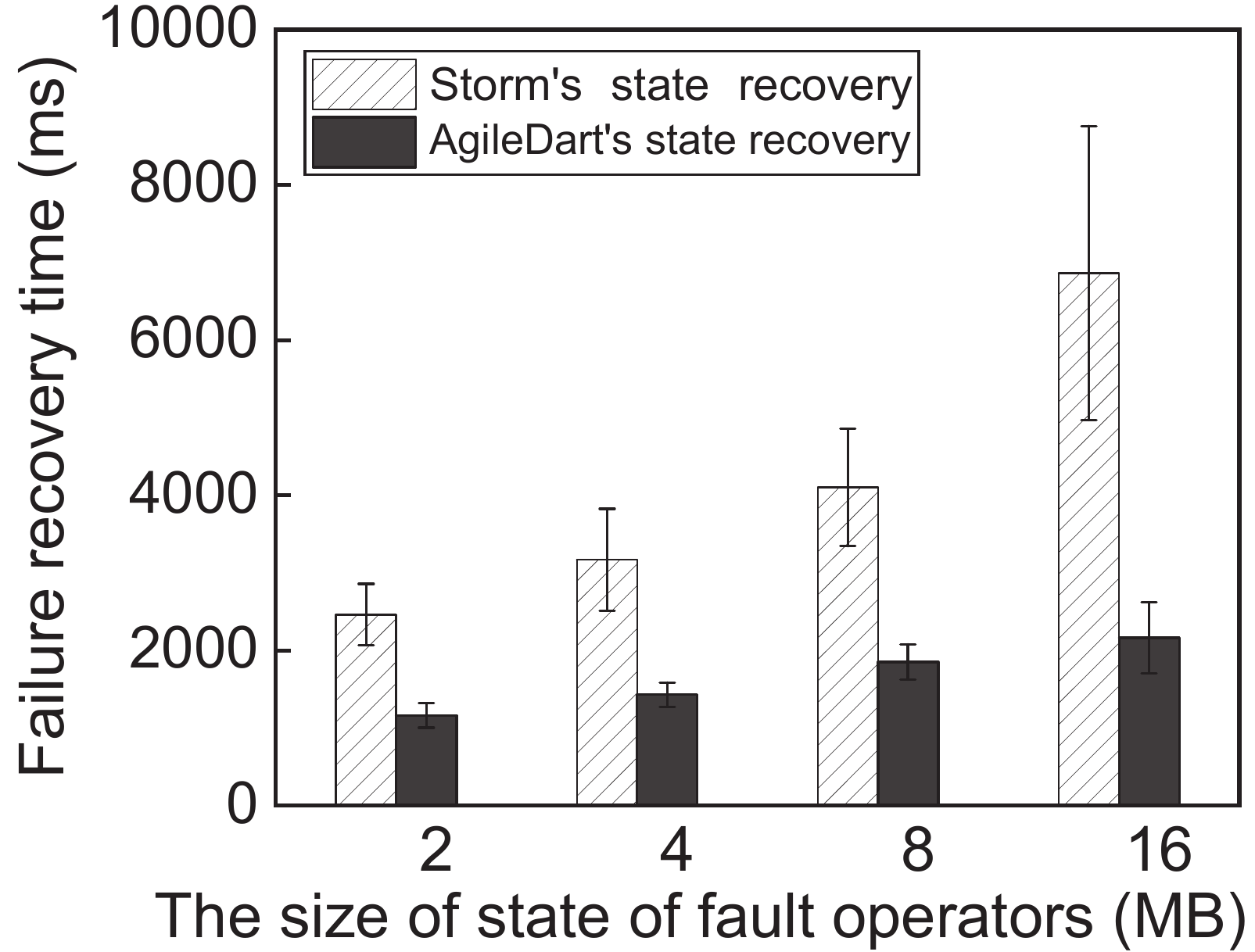}
 \label{subfig:recovery_state}
 }\hspace{1.5em}
 \subfloat[State recovery time of AgileDart for different \emph{m} and \emph{k} (state size = 16 MB).]{
 \includegraphics[width=.26\linewidth]{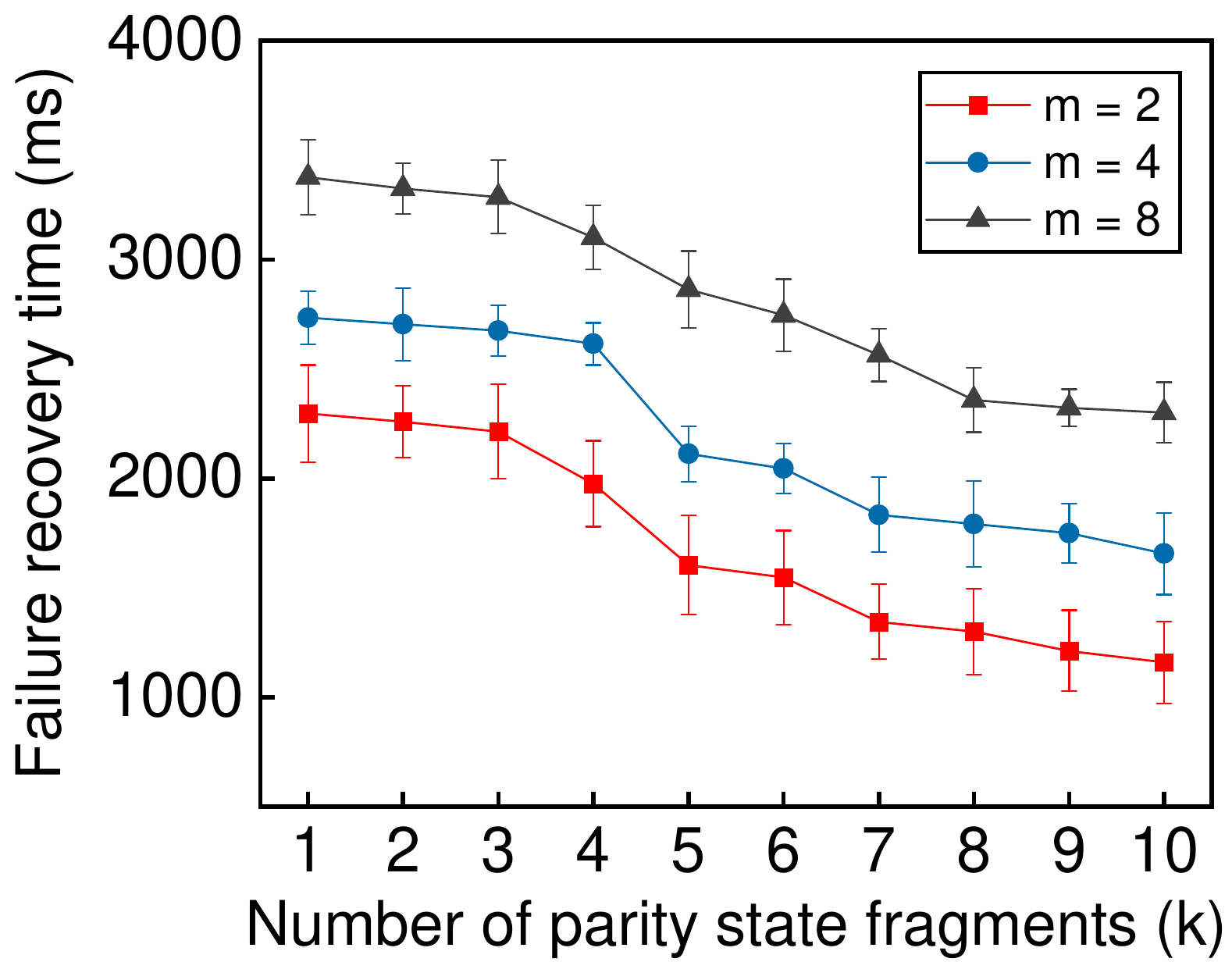}
 \label{subfig:recovery_km}
 }
 \vspace{0.15in}
 \caption{Fault tolerance study of AgileDart for the overlay recovery, dataflow topology recovery, and operator's state recovery.} 
\end{figure*} 

\textbf{\textit{Metrics.}} We focus on the performance metrics of query latency. Query latency is measured by sampling 5\% of the tuples, assigning each tuple a unique ID, and comparing timestamps at the source and the sink node. To evaluate the scalability of AgileDart, we measure the distribution of operators across edge nodes and the allocation of distributed schedulers across edge zones. To evaluate the adaptivity of AgileDart, we intentionally introduce bottlenecks through resource contention and deliberately disable nodes with human intervention.


\subsection{Query Latency}
We measure the query latencies for running real-world IoT stream applications on the Raspberry Pis and VMs across a wide range of input rates.

Figure~\ref{subfig:latency_queue} and Figure~\ref{subfig:latency_deploy} show the latency comparison of AgileDart and EdgeWise for (a) DAG queue waiting time and (b) DAG deployment time as the number of concurrently running applications increases. Applications are selected from a pool containing dataflow topologies (DAGs), including \texttt{ExclamationTopology}, \texttt{JoinBoltExample}, \texttt{LambdaTopology}, \texttt{Prefix}, \texttt{SingleJoinExample}, \texttt{SlidingTupleTsTopology}, \texttt{SlidingWindowTopology} and \texttt{WordCountTopology}. EdgeWise is built on top of Storm. Both Storm and EdgeWise rely on a centralized master (Nimbus) to deploy the application's DAGs, and then process them one by one on a first-come, first-served basis. Therefore, we can see that EdgeWise's DAG queue waiting time and deployment time increase linearly as the number of applications increases. This centralized approach can easily lead to scalability bottlenecks. In contrast, AgileDart mitigates these scalability issues due to its fully decentralized architecture, which doesn't depend on a centralized master for DAG analysis and deployment.

Figure~\ref{subfig:latency_window} shows the query latency of AgileDart as the number of concurrently running applications increases. The results demonstrate that AgileDart exhibits strong scalability when dealing with a large number of concurrently running applications. First, AgileDart's distributed schedulers are capable of processing these applications' queries independently, preventing them from queuing on a single central scheduler, which would result in significant queuing delays. This is analogous to the practice of supermarkets adding more cashiers to reduce waiting queues during peak hours. Second, AgileDart's P2P model ensures that every available node in the system can participate in operator mapping, auto-scaling, and failure recovery. This approach helps avoid central bottlenecks, balances workload, and accelerates the processing.

The performance comparison results for running the frequent route application, the profitable areas application, and the urban sensing application are shown in Figure~\ref{subfig:app_taxi_route}, Figure~\ref{subfig:app_taxi_profit}, and Figure~\ref{subfig:app_urban}. In general, AgileDart, Storm~\cite{storm}, and EdgeWise~\cite{EdgeWise} exhibit similar performance when the system is underutilized, particularly with low input. However, as the system reaches average utilization levels, characterized by relatively high input rates, AgileDart outperforms both Storm and EdgeWise. AgileDart achieves approximately 16.7\% to 52.7\% lower query latency compared to Storm and 9.8\% to 45.6\% lower query latency compared to EdgeWise.

This improved performance is attributed to AgileDart's ability to limit the number of hops between source operators and sink operators to within $O(\log N)$ hops.  
Additionally, AgileDart can dynamically scale operators in response to changes in the input rate. Notably, AgileDart performs exceptionally well in the Urban Sensing application, as this application involves data splitting and aggregation, resulting in numerous I/O operations and data transfers that benefit from AgileDart's dynamic dataflow abstraction. We expect further latency improvement under a limited bandwidth environment since AgileDart selects the path with less delay by using the bandit-based path planning model.

    \begin{figure*}[t]
        \centering
        \subfloat[Process of scaling up.]{%
            \includegraphics[width=.235\linewidth]{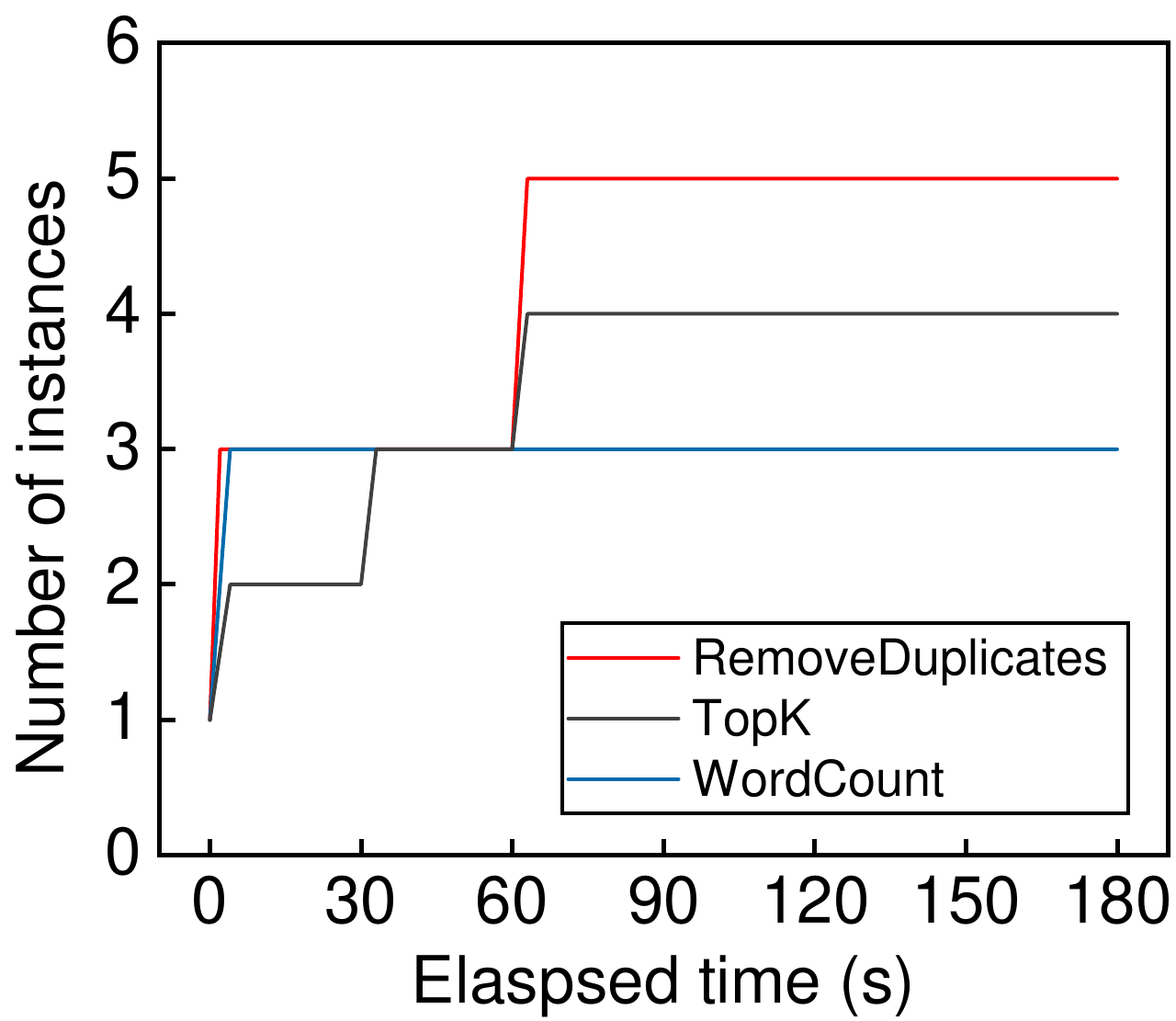}%
            \label{subfig:scale_up}%
        }\hfill
        \subfloat[Process of scaling up and scaling out.]{%
            \includegraphics[width=.235\linewidth]{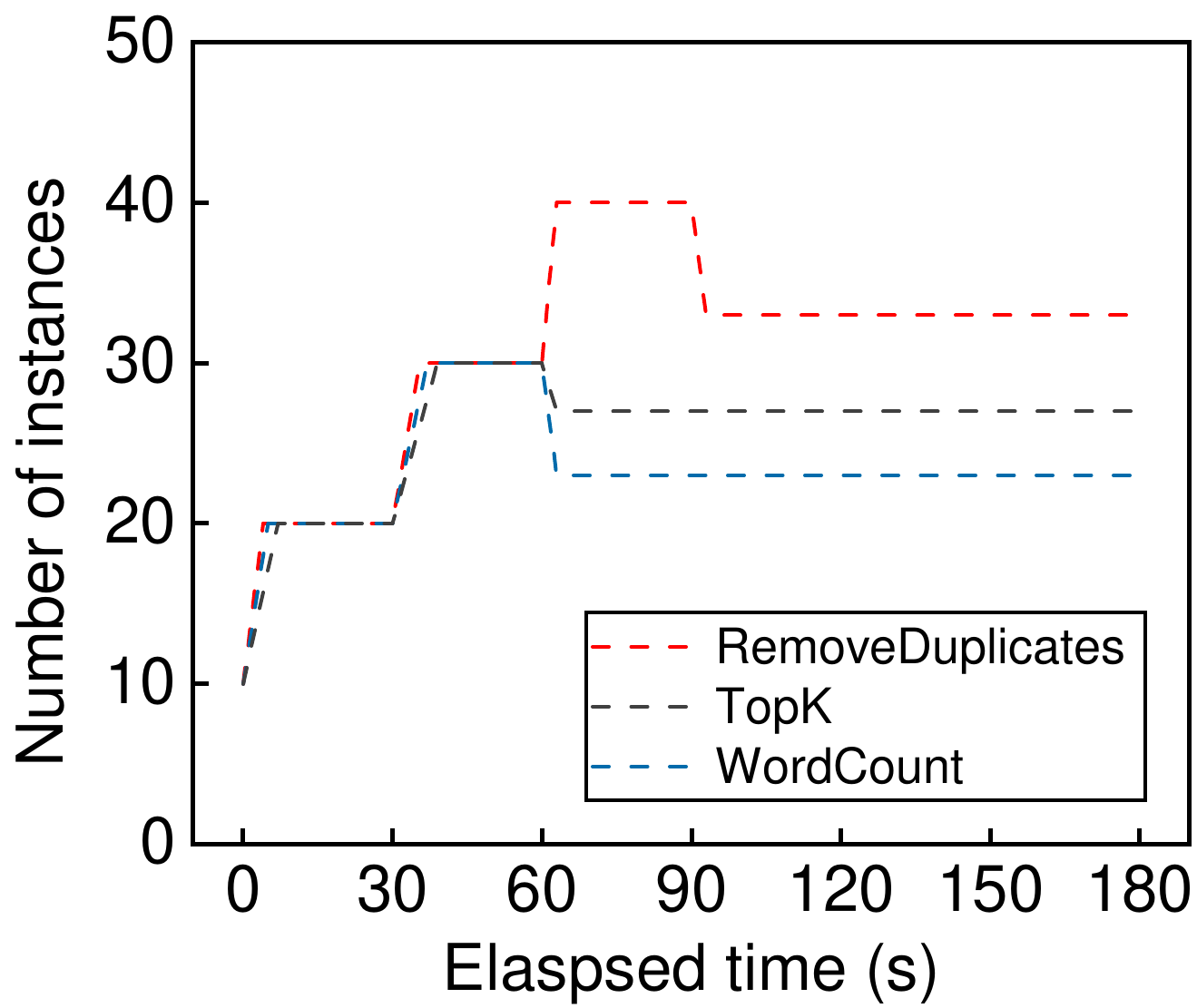}%
            \label{subfig:scale_out}%
        }\hfill
        \subfloat[Health score changes corresponding to Figure~\ref{subfig:scale_up}.]{%
            \includegraphics[width=.235\linewidth]{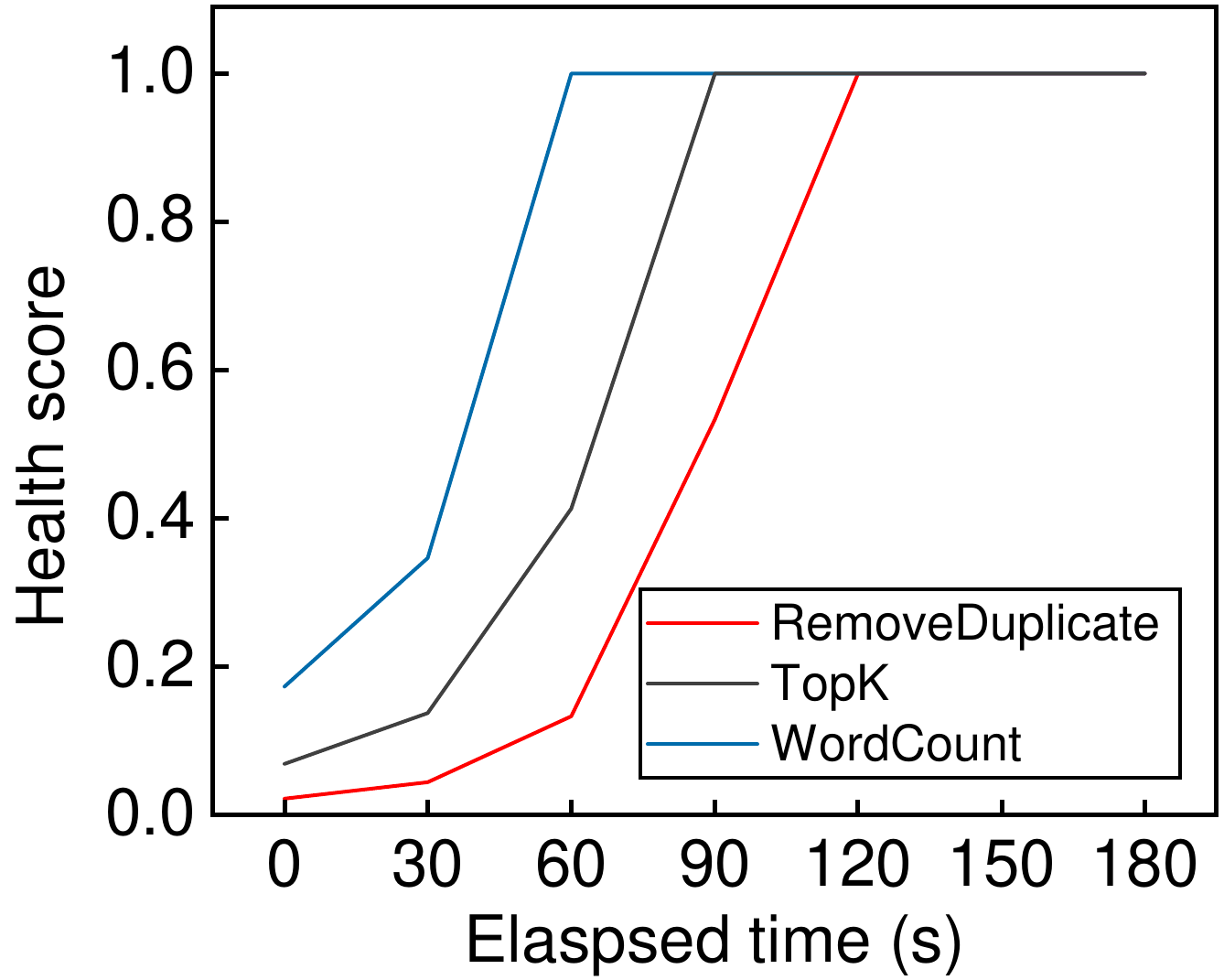}%
            \label{subfig:scale_health_up}%
        }\hfill
        \subfloat[Health score changes corresponding to Figure~\ref{subfig:scale_out}.]{%
            \includegraphics[width=.235\linewidth]{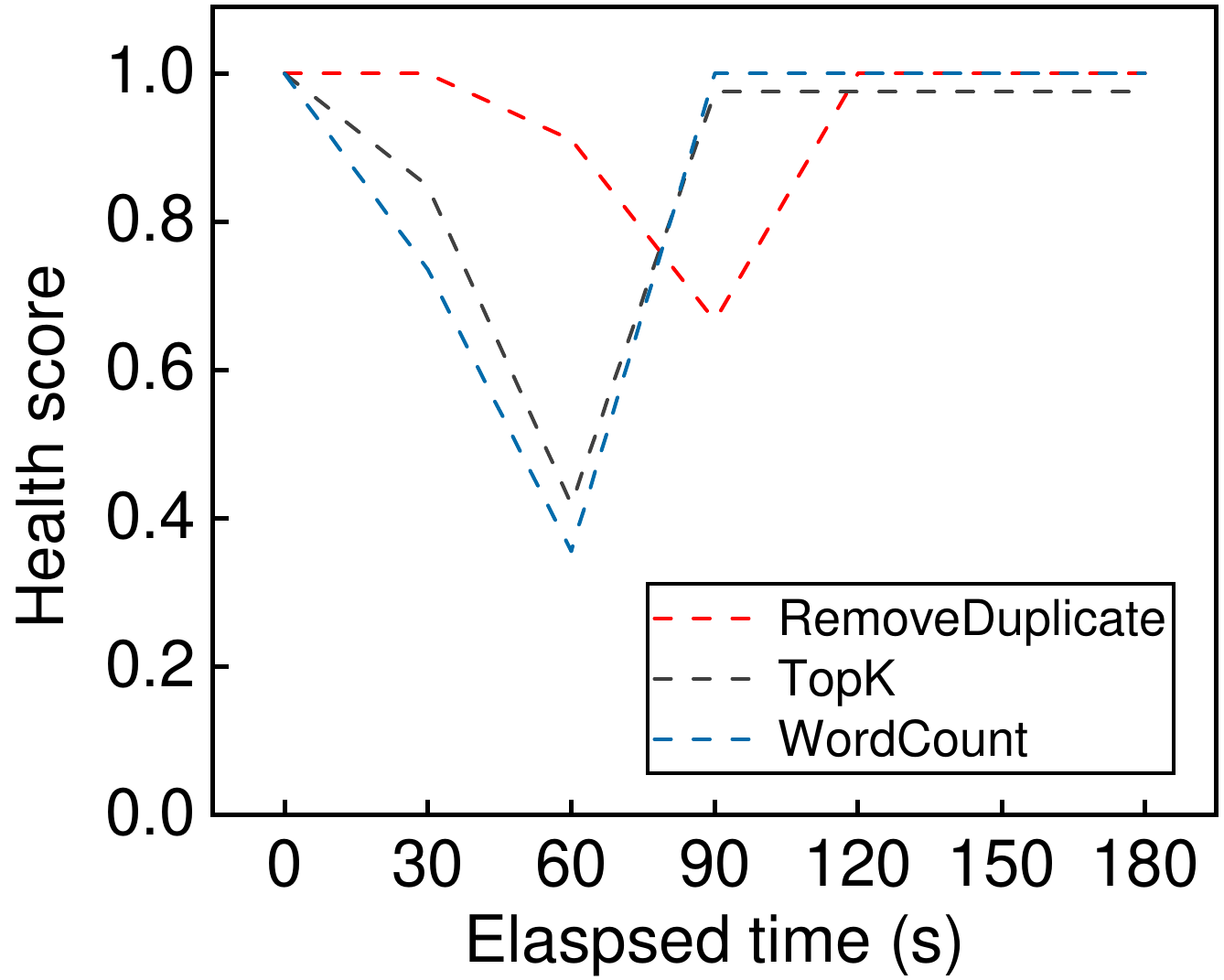}%
            \label{subfig:scale_health_out}%
        }
        \vspace{0.15in}
        \caption{Adaptivity study of AgileDart for the scaling up and the scaling out.}
    \end{figure*}


\subsection{Scalability Analysis}
\revised{We now show scalability: AgileDart decomposes the traditional centralized architecture of stream processing engines into a novel decentralized architecture for operator mapping and query scheduling. This structural change notably enhances the system's capacity to scale effectively with massive concurrently running applications, application operators, and edge zones. We deploy operators from up to 1,000 applications across 10,000 edge nodes over 20 edge zones to study whether AgileDart achieves an even distribution of operators or overburdens some nodes.}

\textbf{\textit{Distribution of operators.}}
Figure~\ref{subfig:scalability_operators} shows the mappings of AgileDart’s operators on edge nodes for 250, 500, 750, and 1,000 concurrently running applications. These applications run a mix of topologies with varying numbers of operators (with an average of 10). 
\revised{Figure~\ref{subfig:scalability_cdf} shows the normal probability plot of the number of operators per node. The results show that when deploying 250 and 500 applications, around 97.85\% nodes host fewer than 3 operators; and when deploying 750 and 1,000 applications, around 99.89\% nodes host fewer than 4 operators.} From Figure~\ref{subfig:scalability_operators} and Figure~\ref{subfig:scalability_cdf}, we can see that these applications' operators are evenly distributed across all edge nodes. This is because AgileDart essentially leverages the DHT routing to map operators on edge nodes. Since the application’s dataflow topologies are different, their routing paths and the rendezvous points will also be different, resulting in operators well balanced across all edge nodes.

\textbf{\textit{Distribution of schedulers.}} Figure~\ref{subfig:scalability_3d} shows the mappings of AgileDart’s distributed schedulers on edge nodes and zones for 250, 500, 750, and 1,000 concurrently running applications, along with the average number of hops for these applications to look for a scheduler.  For AgileDart, it adds a scheduler for every new 50 applications. Following the P2P's gossip protocol, each application searches for a scheduler within the zone using a maximum of $\lceil \log_{2^b}N \rceil$ hops, where $b = 4$. If there is no scheduler in the zone or the number of applications in the zone exceeds a specific threshold, a peer node (usually with powerful computing resources) will be elected as a new scheduler. The results demonstrate that as the number of concurrently running applications increases, the number of schedulers across zones increases proportionally. These schedulers are evenly distributed among different zones, and most of them can be found within 4 hops.

\subsection{Failure Recovery Analysis}
We next show fault tolerance: in the case of stateless IoT stream applications, AgileDart simply resumes the entire execution pipeline since there is no need for state recovery. In the case of stateful IoT stream applications, the distributed states within operators are continually checkpointed to leaf set nodes in parallel and are reconstructed in the event of failures. The results show that even in scenarios where many nodes fail or leave the system, AgileDart can achieve a relatively stable time for both the overlay and dataflow topology. Furthermore, we demonstrate that for stateful applications with varying state sizes, AgileDart exhibits faster state recovery than Storm. Lastly, we investigate how the partition factor ($m$) and redundancy factor ($k$) influence the failure recovery time.

\textbf{\textit{Overlay recovery and dataflow recovery.}} Figure~\ref{subfig:recovery_overlay} shows the overlay recovery time and the dataflow topology recovery time for an increasing number of simultaneous operator failures. To induce simultaneous failures, we intentionally remove several active nodes from the overlay and evaluate the time it takes for AgileDart to recover.  The recovery time includes recomputing the routing table entries, re-planning the dataflow path, synchronizing operators, and resuming the computation. The results show that AgileDart achieves a stable recovery time for an increasing number of simultaneous failures. This is because, in AgileDart, each failed node can be quickly detected and recovered by its neighbors through heartbeat messages without having to talk to a central coordinator, so many simultaneous failures can be repaired in parallel.

\textbf{\textit{State recovery.}} Figure~\ref{subfig:recovery_state} shows a comparison of state recovery times between AgileDart and Storm for stateful applications with varying state sizes. AgileDart’s state-saving time is comparable to Storm’s state-saving time. AgileDart achieves a 34\% to 63\% reduction in state recovery time compared to Storm, and this gap widens as the state size increases. This highlights the advantage of AgileDart's decentralized design, which allows many nodes to participate in state recovery in parallel. In contrast, Storm relies on a single node to retrieve the state, constrained by I/O rate and network bandwidth.

Figure~\ref{subfig:recovery_km} shows the failure recovery time of AgileDart with different partition factors \emph{m} (the number of raw fragments) and different redundancy factors \emph{k} (the number of parity fragments). As shown in Figure~\ref{subfig:recovery_km}, when \emph{m} is fixed, the failure recovery time decreases as \emph{k} increases. When \emph{k} is fixed, the failure recovery time decreases as \emph{m} decreases. This is due to the fact that the recovery time is mainly determined by $mB/(m+k-1)$, where \emph{B} represents the amount of data that any providing peer uploads. The value of $mB/(m+k-1)$ increases with an increase of \emph{m} when the values of \emph{k} and \emph{B} are fixed. The value of $mB/(m+k-1)$ decreases with the increases of \emph{k} when the values of \emph{m} and \emph{B} are fixed. In practice, different IoT stream applications can have different settings of \emph{m} and \emph{k} to best suit their use case scenarios.

\revised{Computation and communication overhead comparison of AgileDart, Replication, and Checkpointing is presented in 
Appendix~\ref{sec:recovery_overhead}.
}


    \begin{figure*}[t]
    \centering
        \subfloat[Four applications and corresponding optimal paths.]{%
            \includegraphics[width=.4\linewidth]{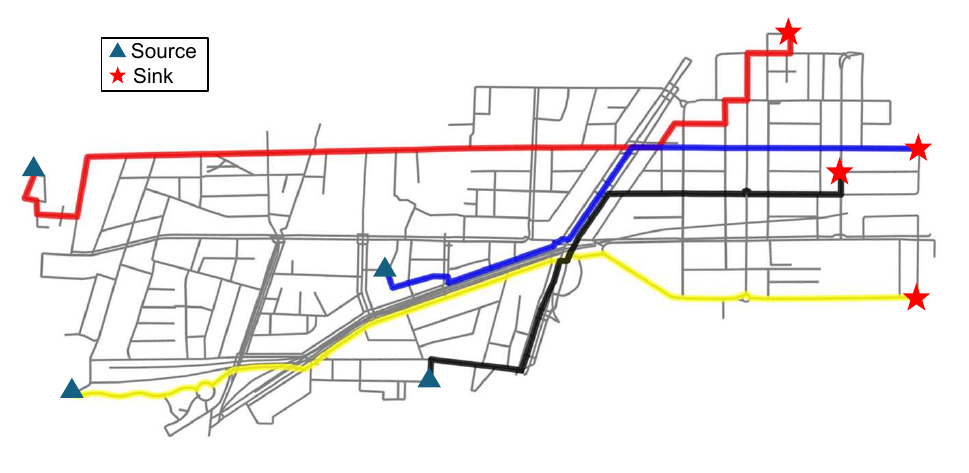}%
            \label{subfig:different_source_sink_paths}%
        }\hspace{1.5em}
        \subfloat[Paths selected by different algorithms.]{%
            \includegraphics[width=.4\linewidth]{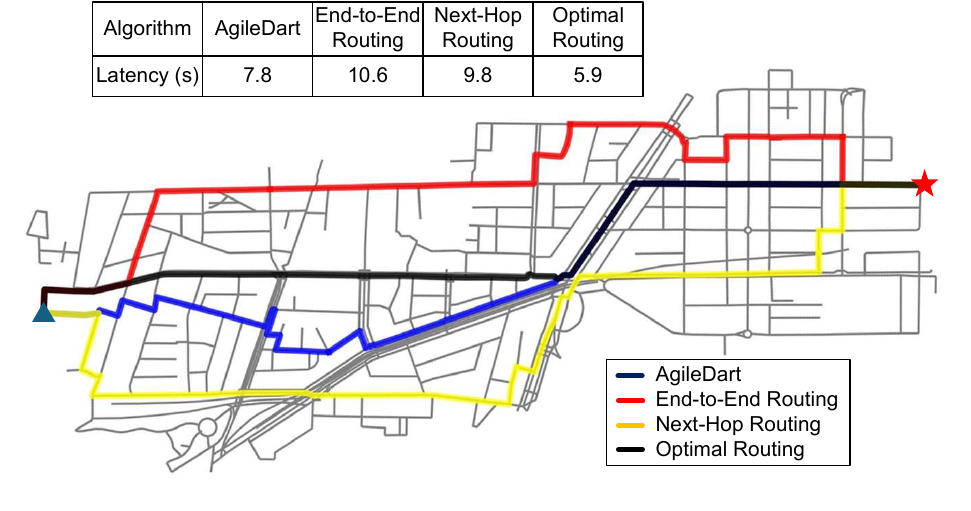}%
            \label{subfig:same_source_sink_paths}%
        }
        \vspace{0.15in}
        \caption{Real-world road map in Sydney, Australia.} 
        \vspace{0.15in}
    \end{figure*}

\begin{figure*}[t!]
    \centering
    \begin{minipage}{.235\linewidth}
        \centering
                \captionsetup{width=\textwidth} %
        \includegraphics[width=\linewidth]{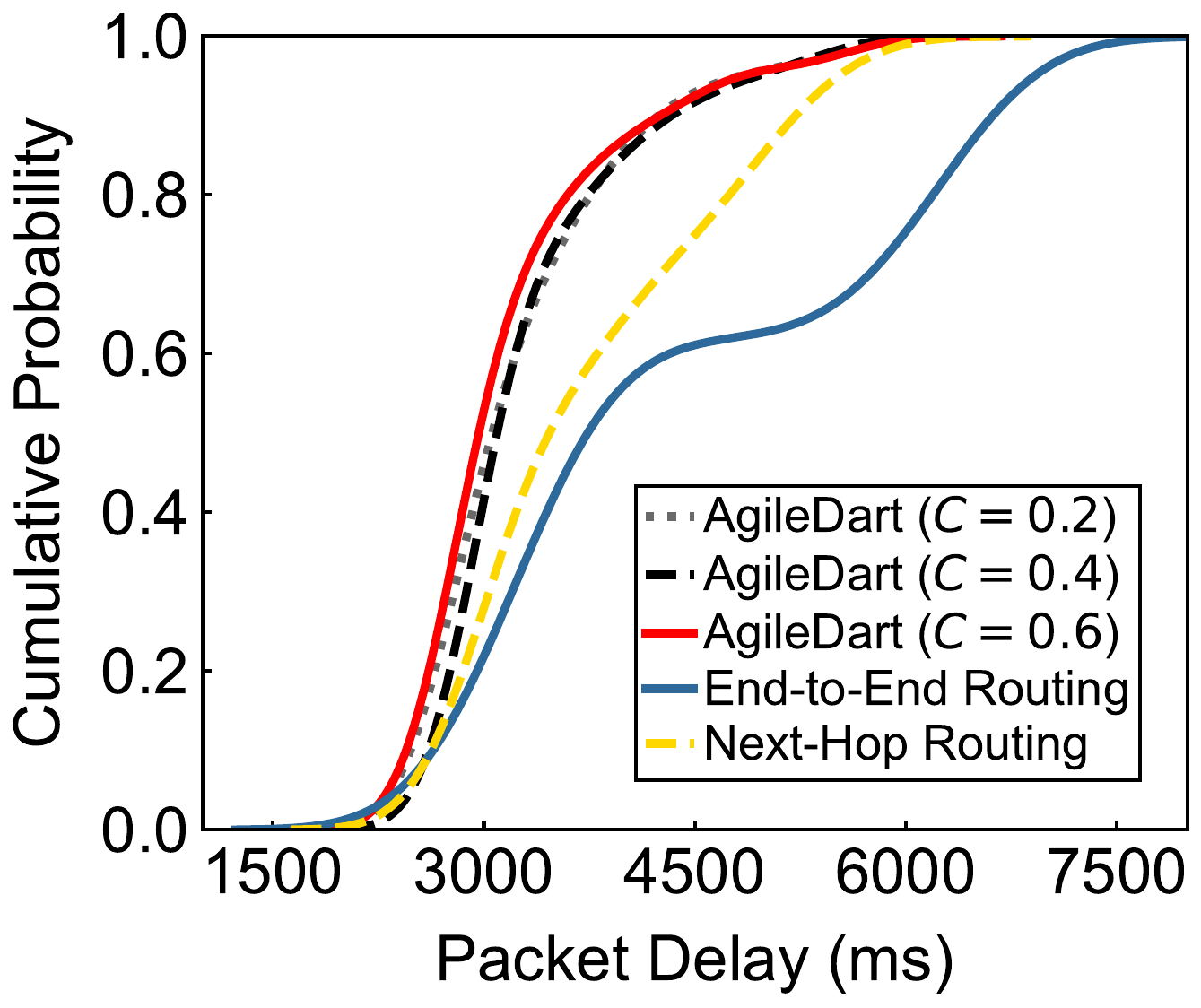} 
        \caption{CDF of packet delays of different algorithms' paths.}
        \label{fig:cdf_packet_delays}
    \end{minipage}\hspace{1.5em}
    \begin{minipage}{.27\linewidth}
        \vspace{-0.1in}
        \centering
        \captionsetup{width=\textwidth} %
        \includegraphics[width=\linewidth]{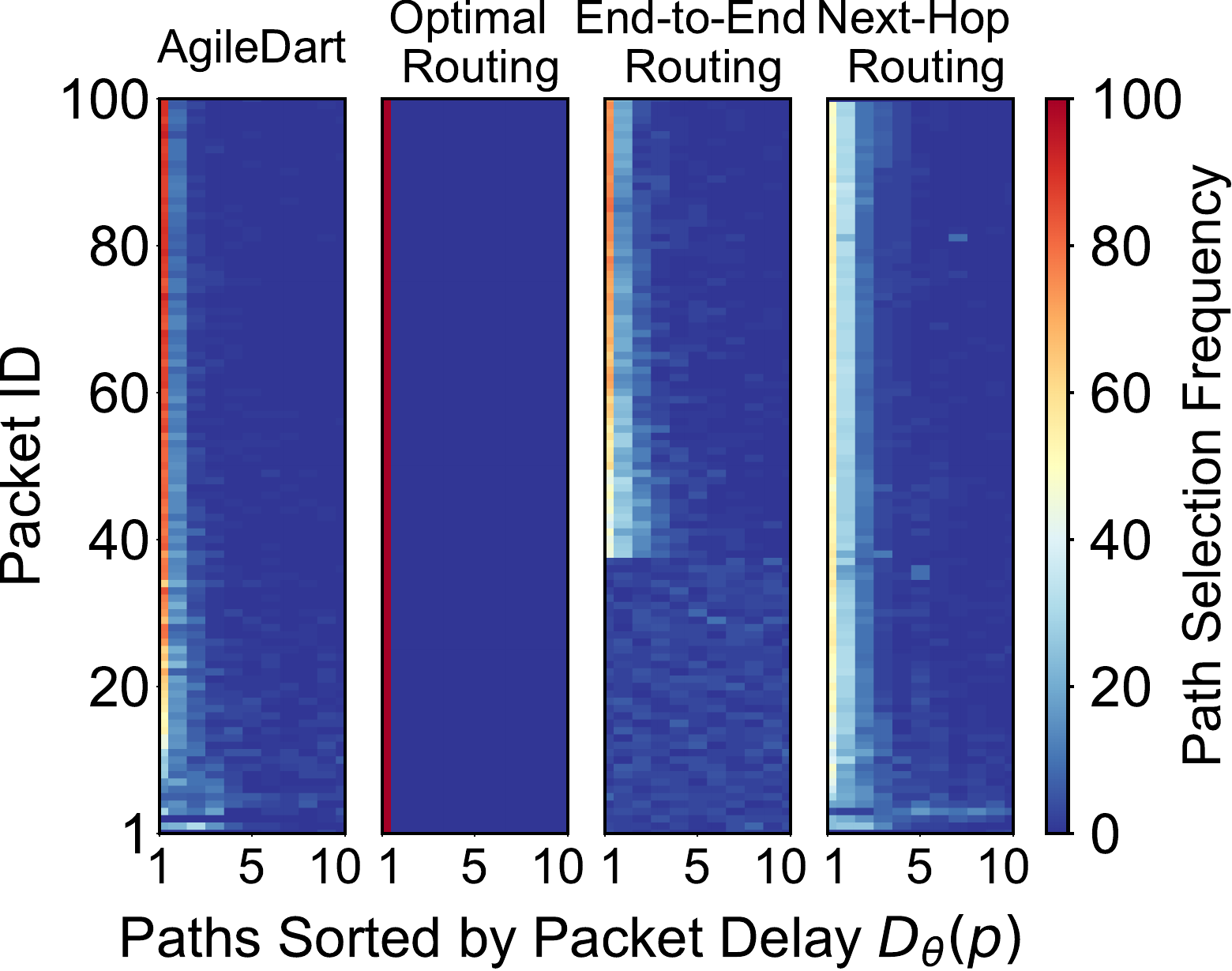}
        \caption{Path selection frequencies generated by different algorithms.}
        \label{fig:heatmap_selection_frequency}
    \end{minipage}\hspace{1.5em}
    \begin{minipage}{.25\linewidth}
        \centering
        \includegraphics[width=\linewidth]{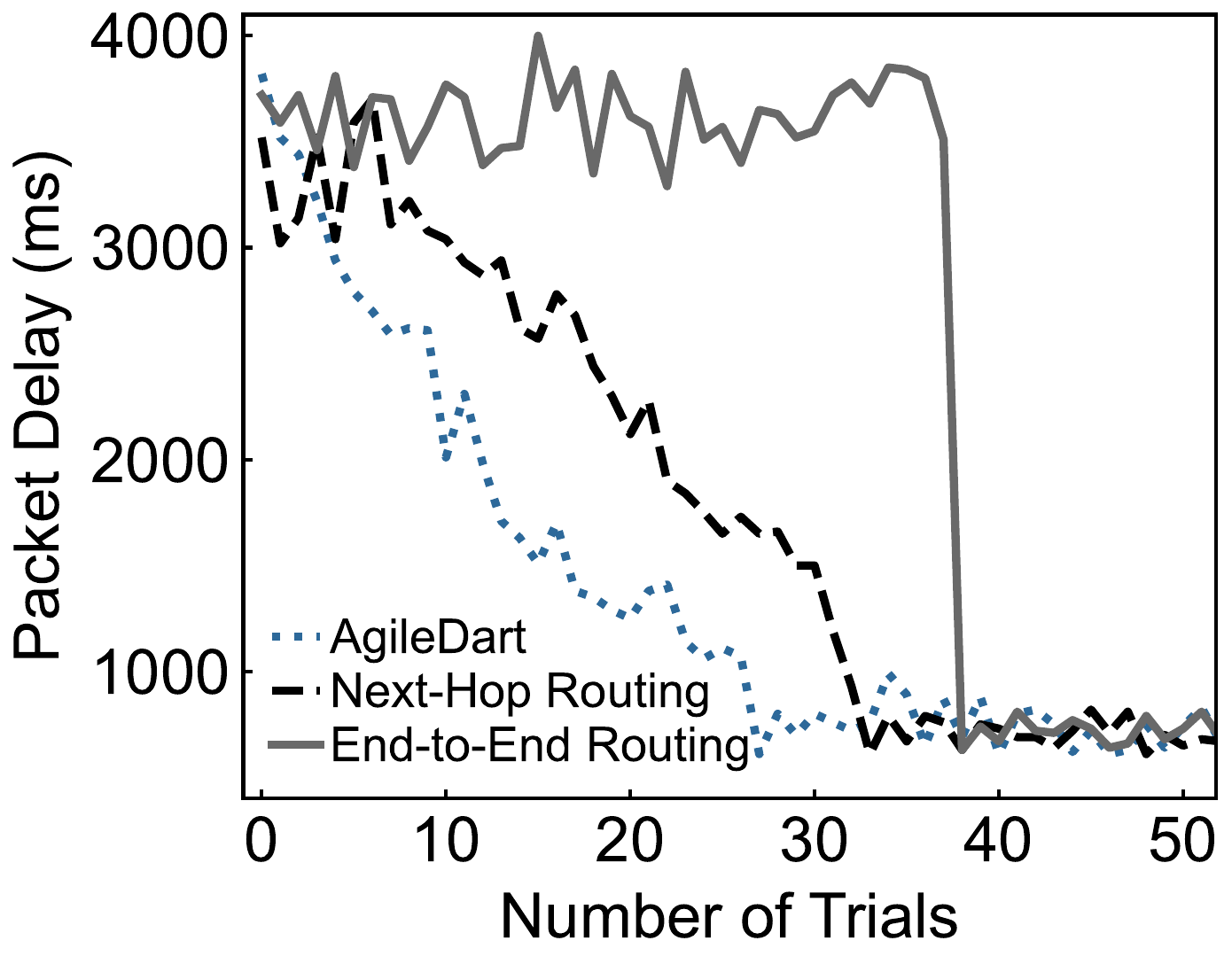} 
        \caption{Packet delays of paths at the different number of trials.}
        \label{fig:number_of_trials}
    \end{minipage}
\end{figure*}

\subsection{Elastic Scaling Analysis}
While scaling is a well-explored subject, our innovation lies in using the DHT leaf set to select the optimal candidate nodes for scaling up or scaling out. Therefore, our approach avoids the need for a central master for control and operates in a fully distributed manner. If many operators have bottlenecks at the same time, the system can adjust them all together.  

\textbf{\textit{Scale up \& scale out.}} 
In this experiment, we deploy three 4-stage topologies (\texttt{RemoveDuplicates}, \texttt{TopK}, \texttt{WordCount}). Figure~\ref{subfig:scale_up} shows the process of scaling up only. It begins from the moment a bottleneck is detected and continues until the system stabilizes. Figure~\ref{subfig:scale_out} shows the process of scaling up and subsequently scaling out. In this experiment, we apply pressure to the system by gradually increasing the number of instances (tasks) by 10 every 30 seconds until a bandwidth bottleneck occurs (at 60 seconds for the blue line and the black line, and at 90 seconds for the red line). This bottleneck can only be resolved by scaling up. The results show that the system stabilizes through the migration of instances to other nodes.

\textbf{\textit{Health score.}} Figure~\ref{subfig:scale_health_up} shows how the health score changes corresponding to Figure~\ref{subfig:scale_up}. Figure~\ref{subfig:scale_health_out} shows how the health score changes corresponding to Figure~\ref{subfig:scale_out}. It's worth noting that in cases where the objective of achieving a higher health score conflicts with the aim of improving throughput, it becomes necessary to strike a balance between the health score and system throughput. This balance can be achieved by adjusting the health score function to target a lower score.
\revised{More analysis on health scores can be found in 
Appendix~\ref{sec:health_score}.
}

    \begin{figure*}[t]
        \centering
        \subfloat[Different algorithms.]{%
            \includegraphics[width=.245\linewidth]{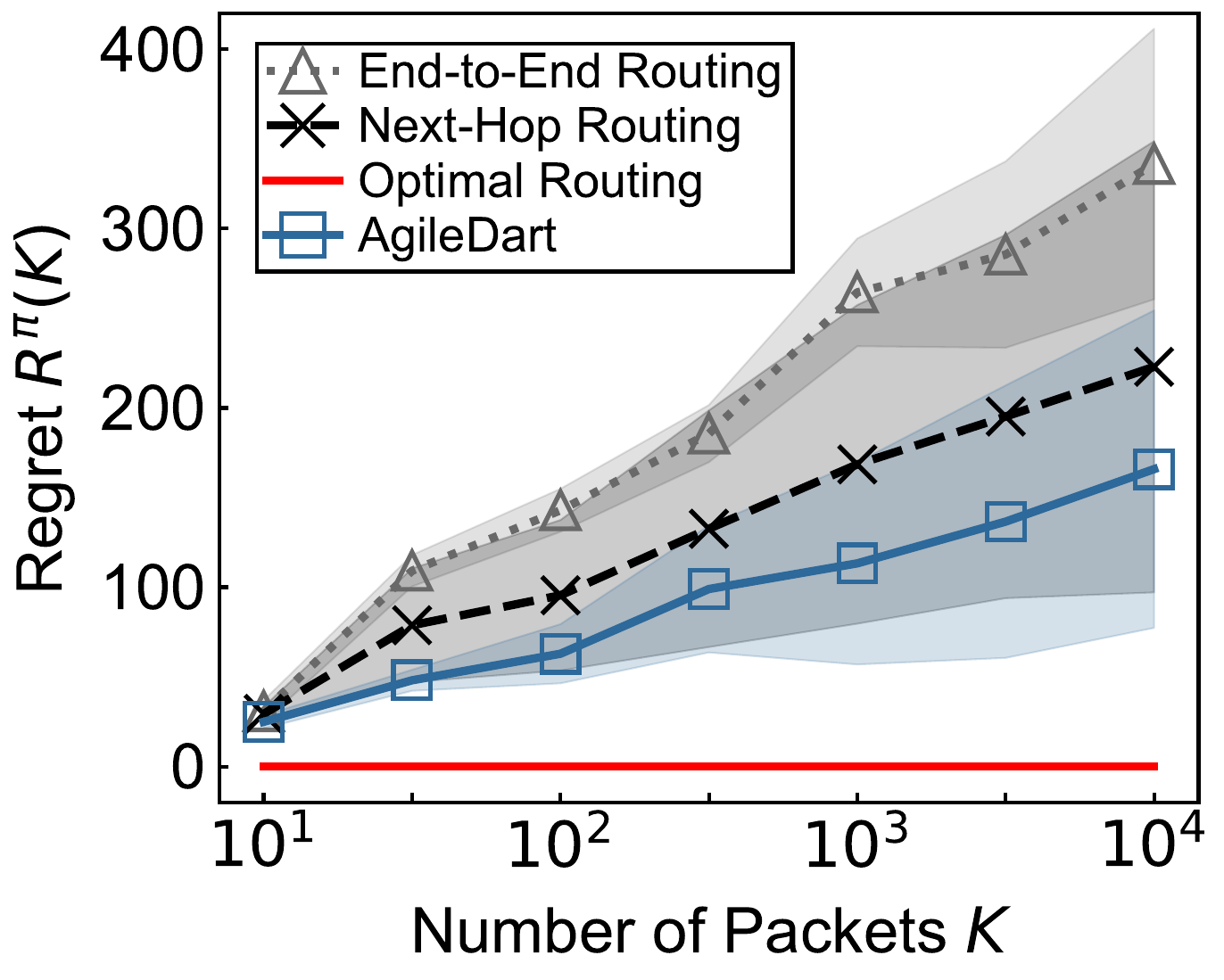}%
            \label{subfig:regret_four_algorithms}%
        }\hfill
                                \vspace{-0.05in}
        \subfloat[Different network sizes.]{%
            \includegraphics[width=.25\linewidth]
            {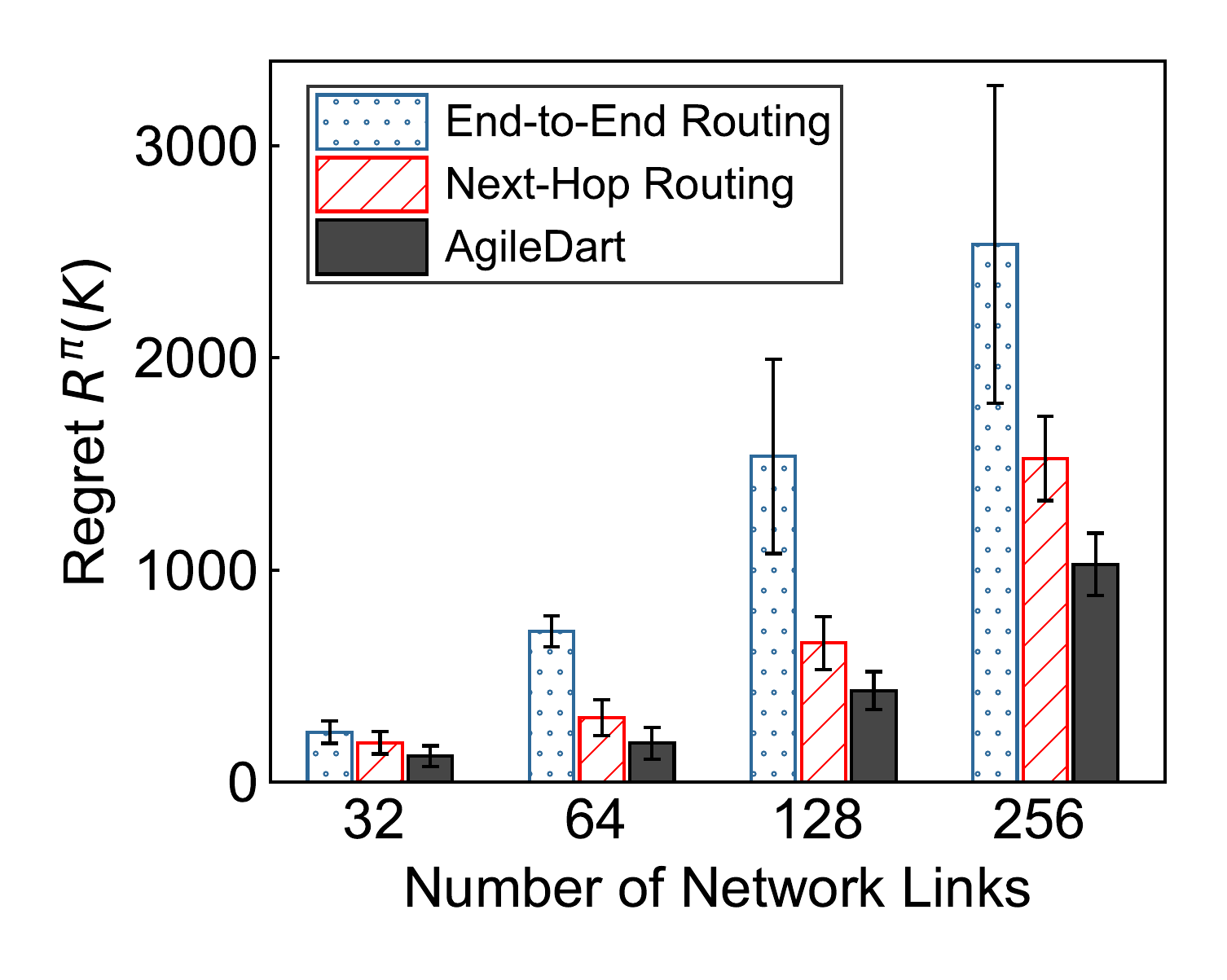}%
            \label{subfig:regret_comparison_network_size}%
        }\hfill
        \subfloat[Different hop counts.]{%
            \includegraphics[width=.24\linewidth]{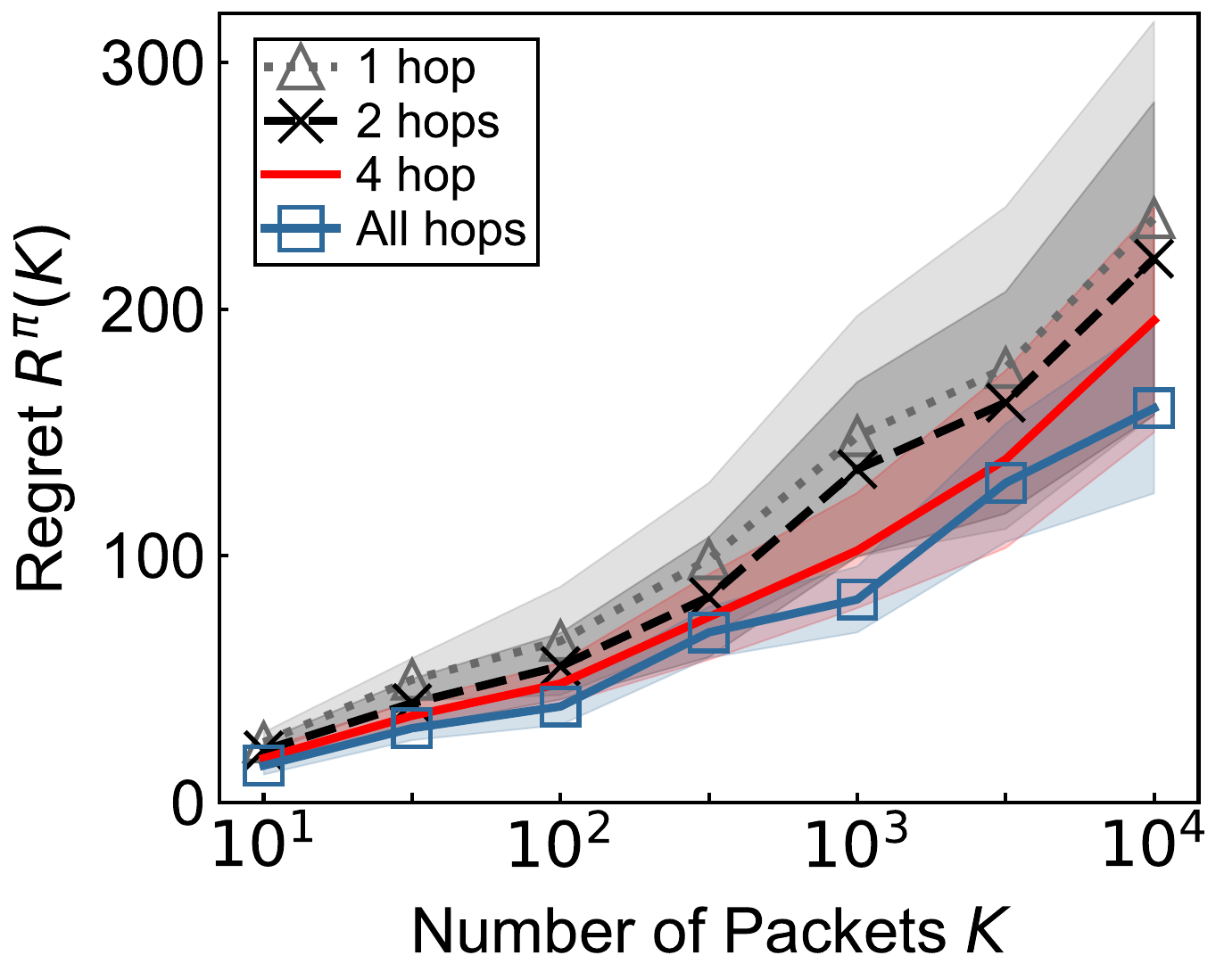}%
            \label{subfig:regret_comparison_different_hops}%
        }\hfill
        \subfloat[Different exploration factors.]{%
            \includegraphics[width=.24\linewidth]{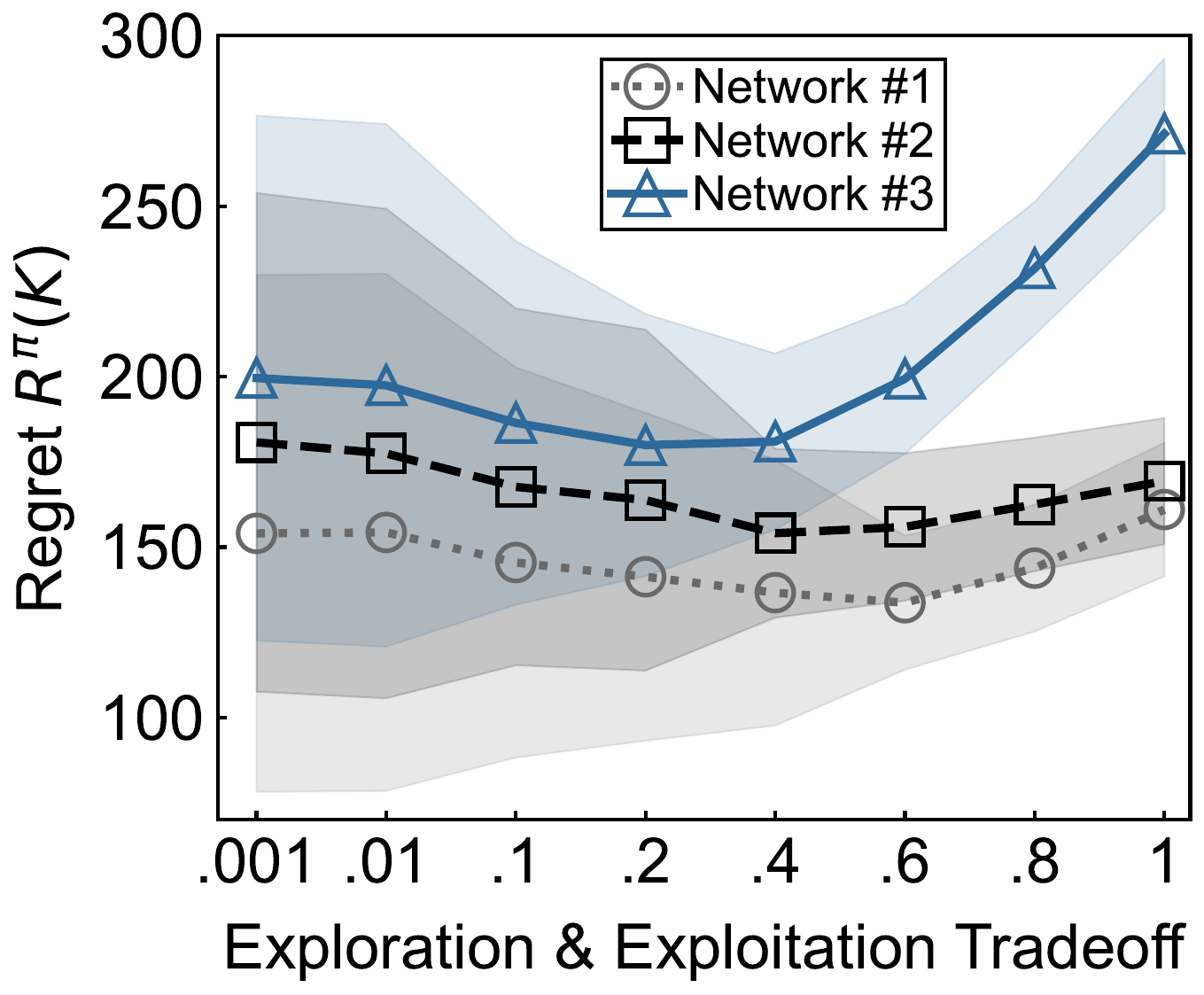}%
            \label{subfig:tradeoff_exploration_exploitation}%
        }
        \vspace{0.2in}
        \caption{Regret analysis on different path planning algorithms, network sizes, hops counts, and exploration factors.}
    \end{figure*}

\subsection{Data Shuffling Analysis}
We evaluate AgileDart's data shuffling model using different sources and sinks in a real-world road map at different scales.

\textbf{\textit{Applications in real-world road map.}}
We extracted a part of the road map in Sydney, Australia, from coordinates $-33.87463$ to $-33.88919$ in latitude and $151.21343$ to $151.21934$ in longitude (see Figure~\ref{subfig:different_source_sink_paths}). Each gray link represents a real-world road connecting two road nodes. We generated four applications with sources and sinks at different road nodes and the corresponding optimal source-to-sink paths, where the optimal source-to-sink path represents a path with the minimum packet delay among all possible paths.

\textbf{\textit{Path planning in real-world road map.}} Figure~\ref{subfig:same_source_sink_paths} presents the paths planned by 
AgileDart and its counterparts (end-to-end routing~\cite{combinatorial_network_optimization} and next-hop routing~\cite{adaptive_opportunistic_routing}) for an application after 50 trials 
(See 
Appendix~\ref{subsec:end2end_routing}
and 
Appendix~\ref{subsec:next_hop_routing}
for more details of end-to-end routing and next-hop routing, respectively). 
In particular, end-to-end routing selects paths by leveraging the lower confidence bound (LCB)~\cite{intro_mab}; next-hop routing selects the next hop based on empirical packet delays; optimal routing selects the optimal path all the time. We considered an IoT-based application that needs to send 100 \textit{KB/s} keyframes in the video data~\cite{iot_in_edge} from the source to the sink. We referred to~\cite{edge_network_speed, edge_network_bandwidth} to simulate network bandwidth over the road network. The expected packet delay of each link falls in the range from 50 \textit{ms} to 250 \textit{ms}, depending on the allocated bandwidth. The smaller the allocated bandwidth of a link, the longer its expected packet delay. The results show that optimal routing can find the shortest path with 5.9 \textit{second} of packet delay, and the packet delay of the path planned by AgileDart is 7.8 \textit{second}, which is closer to optimal routing than end-to-end routing and next-hop routing. 

\textbf{\textit{Packet delay distribution.}}
Figure~\ref{fig:cdf_packet_delays} shows the cumulative probability of packet delays of paths selected by different algorithms. The results show that the packet delays of 91\% of the paths planned by AgileDart are less than 4500 \textit{ms}, while only 78\% and 62\% of the paths planned by next-hop routing and end-to-end routing are less than 4500 \textit{ms} of packet delay, respectively.
This is because end-to-end routing and next-hop routing need to try more inferior paths than AgileDart before finding the optimal path, resulting in longer path delays.

\textbf{\textit{Path selection frequency.}} Figure~\ref{fig:heatmap_selection_frequency} shows the path selection frequencies generated by different algorithms, in which each color grid shows the frequencies of selecting the $x_{th}$ path for each packet. The $X$-axis represents the path in order from the optimal path to the $10_{th}$ best path. The $Y$-axis represents the sequential order of packets. The optimal routing always selects the optimal path for each packet. Next-hop routing sometimes finds the optimal path, but it selects some mediocre paths as well. End-to-end routing is the last to find the optimal path. Compared to them, AgileDart finds the optimal path faster and balances the exploration-exploitation tradeoff better by introducing the exploration factor into its cost function. 

\textbf{\textit{Number of trials.}} 
Figure~\ref{fig:number_of_trials} shows the packet delays of paths selected by different algorithms at the different number of trials. We can see that AgileDart finds the optimal path at the $26_{th}$ trial, which is faster than the $33_{rd}$ trial and the $38_{th}$ trial of next-hop routing and end-to-end routing, respectively. Moreover, AgileDart keeps trying different paths despite finding the optimal one. This is because the exploration factor guides AgileDart to explore some paths worth trying.
\revised{We conduct more convergence analysis in 
Appendix~\ref{sec:convergence}.
}

\subsection{Regret analysis}

\textbf{\textit{Path planning algorithm comparison.}}
Figure~\ref{subfig:regret_four_algorithms} shows the regret comparison of different algorithms. The results show that AgileDart achieves lower regret. This is because AgileDart considers the packet delays not only of the next hop but also of the subsequent path starting from the next hop to the sink, which avoids selecting paths with a low-delay first link but with a high overall delay.

Figure~\ref{subfig:regret_comparison_network_size} shows the regret of an exponentially increasing number of network links. We evaluate the regret of three algorithms using four networks with 32, 64, 128, and 256 links and 25, 36, 64, and 144 nodes, respectively. The results show that AgileDart can reach the lowest regret than its counterparts regardless of the number of network links. When the network size doubles, the number of source-to-sink paths increases dramatically, significantly increasing the number of trials for end-to-end routing to find the optimal path. Similar reasons apply to next-hop routing. Since it only considers the next hop, too many subsequent paths will also significantly increase its number of trials to find the optimal path.

\textbf{\textit{Number of hops.}} Figure~\ref{subfig:regret_comparison_different_hops} shows the regret comparison of different hop counts for subsequent paths leveraged by AgileDart. For “1 hop”, the minimum total empirical transmission cost with exploration adjustment $J_\tau(w)$ only evaluates the empirical transmission cost of the link between the next hop $w$ and the next hop of the next hop $w'$. Similarly, when “all hops” are considered, the expression $J_\tau(w)$ evaluates the cost of all the links from the next hop $w$ to the sink. The results show that considering all hops outperforms other options.

\textbf{\textit{Exploration factor.}} Figure~\ref{subfig:tradeoff_exploration_exploitation} shows the regret comparison of different exploration factors in different network sizes. We extractd three different networks from the road map in Sydney, Australia, with 16 nodes and 30 links, but with varying connectivity and bandwidth distribution. In particular, the ranges of the expected packet delays in network \#1, \#2, and \#3 are between 10 \textit{ms} and 100 \textit{ms}, 50 \textit{ms} and 100 \textit{ms}, and 100 \textit{ms} and 300 \textit{ms}, respectively. We use an exploration factor from 0.001 to 1 to plan paths from the same source to the same sink in different networks using AgileDart. When the exploration factor is set near 0.001, AgileDart primarily exploits known paths; conversely, setting it closer to 1 encourages the exploration of new paths. The results show that different networks require different exploration factors to achieve the lowest regret. For example, the exploration factor of 0.2 is the best fit to network \#3, where 0.4 works best for network \#2. This is because each network may have a unique topology or set of links that influence the optimal exploration factor. AgileDart provides flexibility to set an exploration factor to adapt to different network conditions.

\begin{figure}[t!]
    \centering
    \begin{subfigure}[b]{1\columnwidth}
    \centering
          \includegraphics[width=0.7\columnwidth]{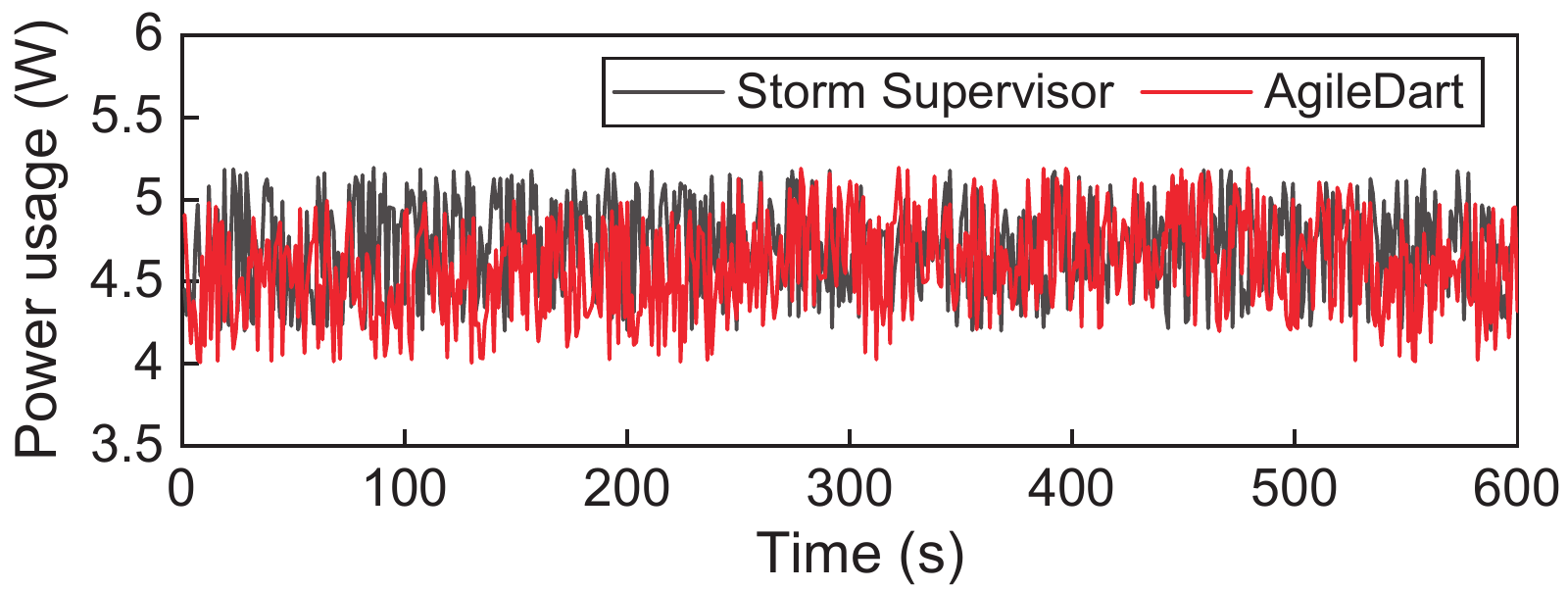}
          \vspace{-0.05in}
        \caption{Power overhead.}
        \label{subfig:overhead_power}
                    \vspace{0.25in}
    \end{subfigure}
        \vspace{0.25in}
    \begin{subfigure}[b]{1\columnwidth}
    \centering
        \includegraphics[width=0.7\columnwidth]{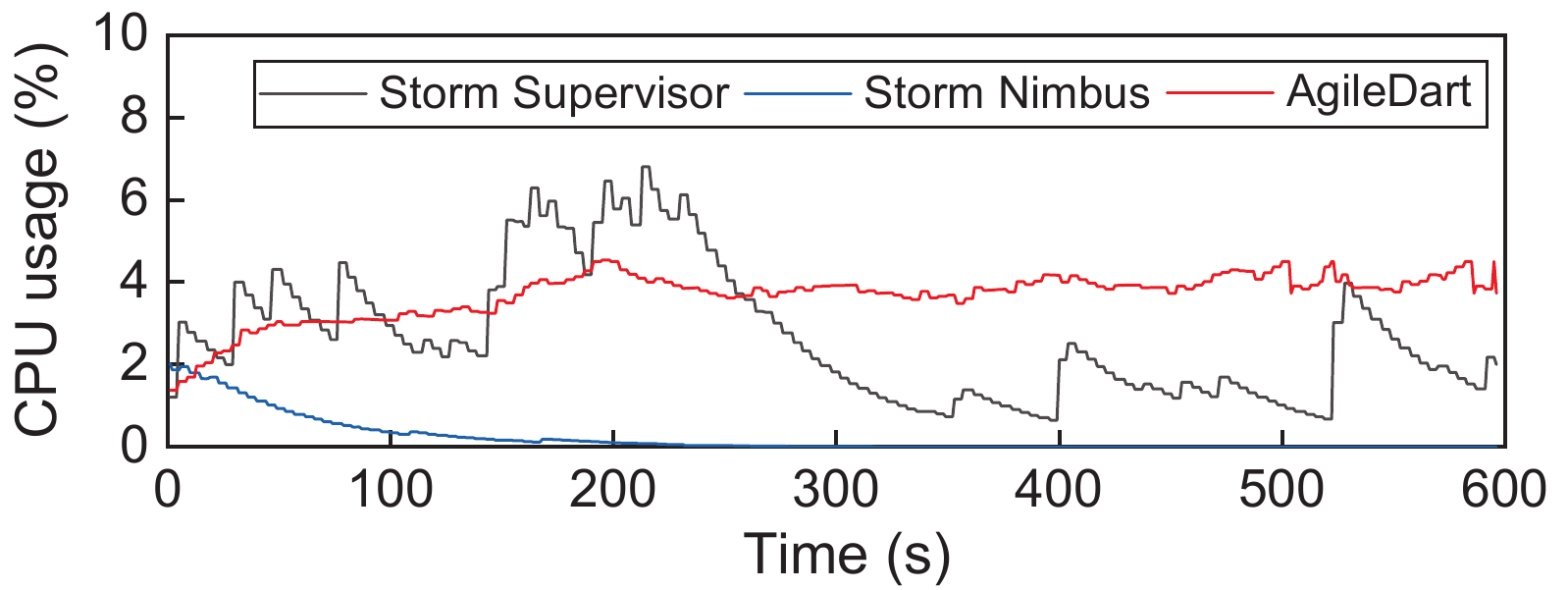}
          \vspace{-0.05in}
        \caption{CPU overhead.}
        \label{subfig:overhead_cpu}
    \end{subfigure}
        \vspace{0.25in}
    \begin{subfigure}[b]{1\columnwidth}
    \centering
        \includegraphics[width=0.7\columnwidth]{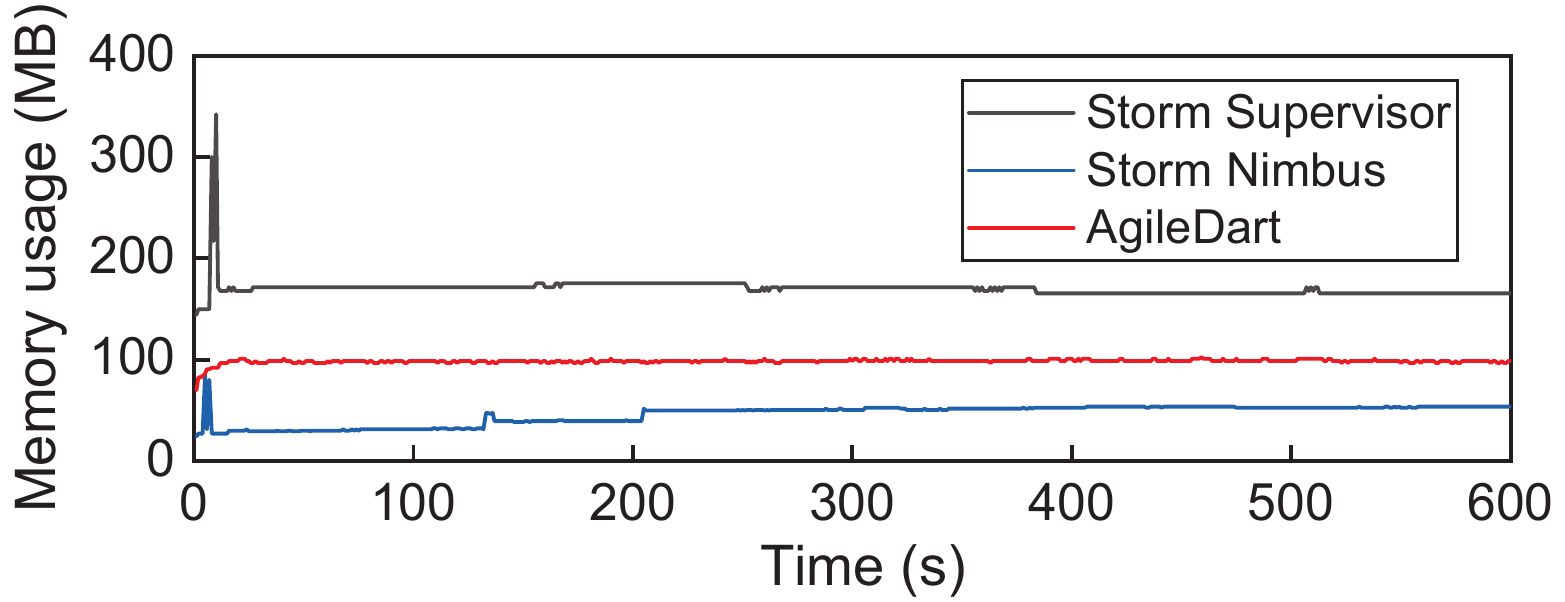}
          \vspace{-0.05in}
        \caption{Memory overhead.}
        \label{subsubfig:overhead_mem}
    \end{subfigure}
            \vspace{0.25in}
    \begin{subfigure}[b]{1\columnwidth}
    \centering
        \includegraphics[width=0.7\columnwidth]{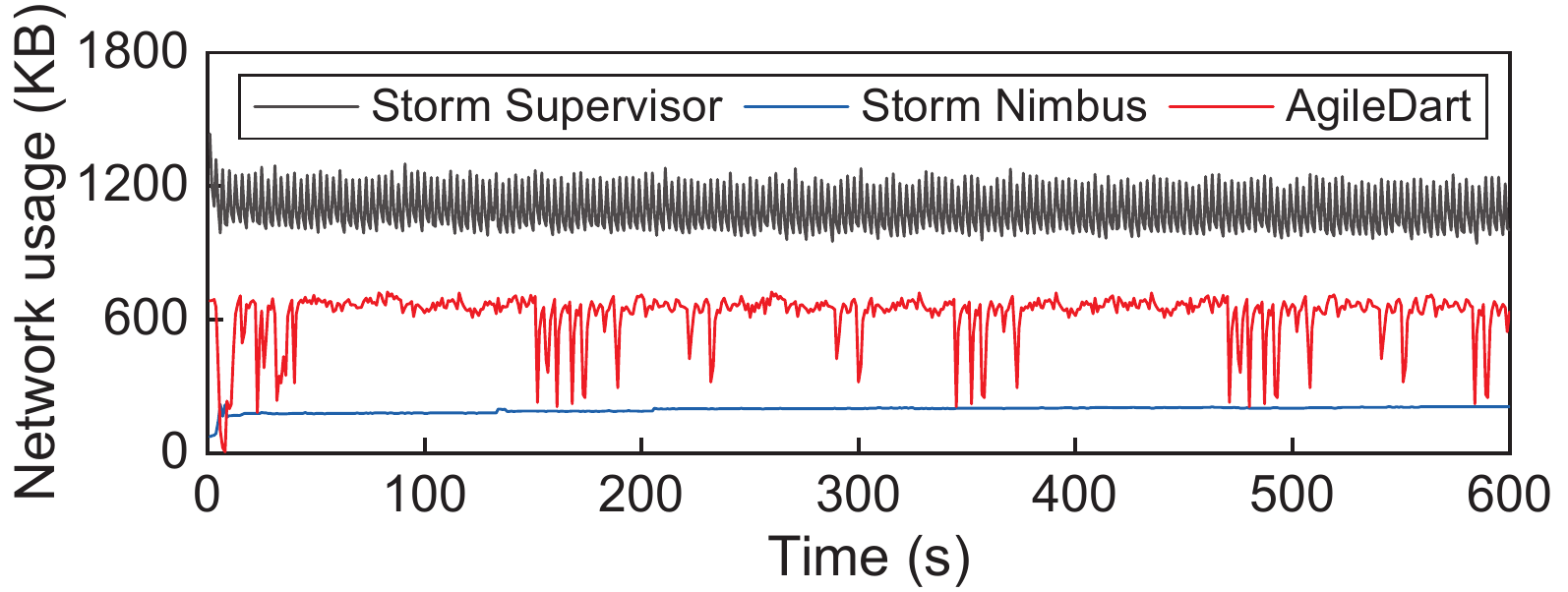}
          \vspace{-0.05in}
        \caption{Network overhead.}
        \label{subfig:overhead_network}
    \end{subfigure}
    \vspace{-0.3in}
    \caption{Overhead comparison of AgileDart and Storm.}

\end{figure}


\subsection{Overhead Analysis}
We evaluate AgileDart's runtime overhead in terms of the power overhead, CPU overhead, memory overhead, and network overhead. 

\textbf{\textit{Power overhead.}} IoT devices often rely on batteries or energy harvesters, and given their limited energy budgets, it's essential to ensure that performance gains don't come at an excessive power cost. To evaluate AgileDart's power usage, we use the MakerHawk USB Power Meter Tester~\cite{testor} to measure the power consumption of the Raspberry Pi 4. When plugged into a wall socket, the idle power usage is 3.35 watts. Figure~\ref{subfig:overhead_power} shows a comparison of the averaged power usage per device/node for AgileDart and Storm's supervisor while running the DEBS 2015 application. The results show that AgileDart exhibits lower power usage, with an average of 5.24 watts, compared to Storm's average of 5.41 watts. This demonstrates AgileDart's power efficiency.

\textbf{\textit{CPU overhead.}} Figure~\ref{subfig:overhead_cpu} shows a comparison of CPU overhead between AgileDart and Storm. The results show that AgileDart uses more CPU resources than Storm Nimbus and Storm supervisor. This increased CPU usage in AgileDart stems from its continuous monitoring of the health status of all operators, enabling auto-scaling decisions to adapt to workload and bandwidth variations in the edge environment. In contrast, Storm does not prioritize this aspect. 

\textbf{\textit{Memory overhead.}} Figure~\ref{subsubfig:overhead_mem} shows a comparison of the memory overhead between AgileDart and Storm. The results show that AgileDart maintains an average memory utilization of 98.23 MB, which is around 42.2\% less than Storm supervisor. This is because Storm's upstream bolt caches all data in memory before it finishes sending data to downstream blots. 

\textbf{\textit{Network overhead.}} Figure~\ref{subfig:overhead_network} shows the network traffic comparison between AgileDart and Storm. The results show that AgileDart uses 41.7\% less network traffic than Storm’s superior. This highlights AgileDart’s lightweight feature: AgileDart’s dataflow graph maintenance overhead (e.g., ping-pong messages used for overlay) is less than the Storm’s coordination overhead (e.g., Acks used for ZooKeeper).


\section{Related Work}
\label{sec:related}

Existing studies can be divided into five categories: \emph{cluster-based stream processing systems}, \emph{cloud-based IoT data processing systems}, \emph{edge-based data processing systems}, \emph{serverless-based stream processing systems}, and \emph{wide-area data analytics systems}. 

\textbf{\textit{Category 1: Cluster-based stream processing systems.}} Over the last decade, a bloom of industry stream processing systems have been proposed to support stream and batch processing on a compute cluster. Examples include Flink~\cite{ApacheFlink}, Samza~\cite{ApacheSamza}, Spark~\cite{ApacheSpark}, Storm~\cite{storm}, Millwheel~\cite{MillWheel}, Heron~\cite{Heron}, S4~\cite{S4}. These systems, however, are designed for low-latency intra-datacenter settings that have powerful computing resources and stable high-bandwidth connectivity, making them unsuitable for edge stream processing. Moreover, they mostly inherit MapReduce's ``single master/many workers'' architecture that relies on a monolithic scheduler for scheduling all tasks and handling failures and stragglers, suffering significant shortcomings due to the centralized bottleneck. \textcolor{
black}{SBONs~\cite{pietzuch2005evaluating} leverages distributed hash tables (DHTs) for service placement. However, it does not support DAG parsing, task scheduling, data shuffling, and elastic scaling, which are required for modern stream processing engines.} 

\textbf{\textit{Category 2: Cloud-based IoT data processing systems.}} In such a model, most of the data is sent to the cloud for analysis. Today many computationally-intensive IoT applications~\cite{nestcam, netatmo} leverage this model because the cloud environment can offer unlimited computational resources. Such solutions, however, cannot be applied to time-critical edge stream applications because (1) they cause long delays and strain the backhaul network bandwidth due to the high volumes of unaggregated data to be sent over the network;
and (2) offloading sensitive data (e.g., medical records) to third-party cloud providers may cause privacy issues.

\revised{\textbf{\textit{Category 3: Edge-based data processing systems.}} In such a model, data processing is performed at the edge without connectivity to a cloud backend~\cite{azureiot, greengrass, shen2015, shepherd, amnis, equality, beaver} or with cloud-edge cooperation~\cite{geoekuiper,hybrid_edge_spe,dscs}. This requires installing a hub device at the edge to collect data from other edge devices and perform data processing. 
GeoEkuiper~\cite{geoekuiper} abstracts spatial data operations into necessary SQL statements functions and leverages cloud-native messaging to improve throughput in resource-constrained edge devices. Shepherd proposes efficient dynamic placement of operators over the hierarchical structure between edge and cloud. Amnis~\cite{amnis} aims at reducing end-to-end latency and overall throughput by considering data locality and resource constraints when generating physical DAG workflows. 
EQUALITY~\cite{equality} proposes heuristic approaches to solving the trade-off between latency and overhead of data quality checks in edge stream processing. Beaver~\cite{beaver} aims to consider varying network bandwidths and heterogeneous computing resources for operator placement. 
These solutions, however, either require centralized coordination or are limited by the computational capabilities of the hub service and cannot support distributed data parallel processing across many devices, thus limiting throughput. It may also introduce a single point of failure once the hub device fails.

\textbf{\textit{Category 4: Serverless-based stream processing systems.}} In such a model, data processing is performed simultaneously on serverless and serverful platforms. Serverless is leveraged to handle event-driven workloads in a scalable and cost-effective manner, benefiting from the automatic scaling capabilities of serverless architectures that dynamically respond to demand fluctuations without manual intervention~\cite{sponge, enabling_stateful, stateful_function, performance_charac, beaver, storm_rts,splitserve,smartpick}.
Sponge~\cite{sponge} leverages serverless framework instances to offload unpredictable increases in workloads from existing VMs with low latency and cost. 
Smartpick~\cite{smartpick} exploits serverless and VMs together to realize composite benefits, i.e., agility from serverless and better performance with reduced cost from VMs.
Storm-RTS~\cite{storm_rts} proposes translating stream operations into function-as-a-service invocations to achieve transparent and predictable performance.  
This model complements AgileDart by offering bursty workload offloading and transparent and predictable performance, whereas AgileDart provides scalable system support.}

\textbf{\textit{Category 5: Wide-area data analytics systems.}} In such a model, data processing is performed collaboratively at both the edge and the cloud. Researchers have studied distributed “job placement” schedulers for wide-area data analytics systems using this model. They focus on determining where to distribute workloads across nodes in multiple geo-distributed sites. For example, many Storm and Spark-based systems (e.g., Flutter~\cite{flutter}, Iridium~\cite{iridium}, SAGE~\cite{SAGE}) and Hadoop systems (e.g., G-Hadoop~\cite{G-Hadoop2013}, G-MR~\cite{G-MR2014}, Nebula~\cite{Nebula2014}, Medusa~\cite{Medusa2016}) optimize the execution of wide-area data analytic jobs by assigning individual tasks to the site that has more data locality or moving data to the site that has more bandwidth. However, most of them ignore the dynamic nature of the wide-area environment. They make certain assumptions based on some theoretical models which do not always hold in practice. For example, Flutter~\cite{flutter}, Tetrium~\cite{hung2}, Iridium~\cite{iridium}, Clarinet~\cite{CLARINET}, and Geode~\cite{geode} formulate the task scheduling problem as a lexicographical integer linear programming (ILP) problem. 
They assume that all sites have infinite resources and the WAN bandwidth does not change. However, these idealized conditions are rarely the case in practice. 

Edgent~\cite{Edgent}, EdgeWise~\cite{EdgeWise}, and Frontier~\cite{2018frontier} are other stream processing engines tailored for the edge. They all point out the criticality of edge stream processing, but no effective solutions were proposed for scalable and adaptive edge stream processing. Edgent~\cite{Edgent} is designed for data processing at individual edge devices rather than full-fledged distributed stream processing. EdgeWise~\cite{EdgeWise} develops a congestion-aware scheduler to reduce backpressure, but it can not scale well due to the centralized bottleneck. Frontier~\cite{2018frontier} develops replicated dataflow graphs for fault-tolerance, but it ignores the edge dynamics and heterogeneity.

\section{Conclusion}

Existing stream processing engines were mostly designed for an over-provisioned cloud, making them a poor fit for the edge computing environment where resources are scarce and network links are unreliable and heterogeneous. In this paper, we present AgileDart, an agile and scalable edge stream processing engine, which enables fast stream processing of a large number of concurrently running low-latency edge applications’ queries at scale in the dynamic, heterogeneous edge environment. AgileDart presents two design innovations: First, it introduces a novel dynamic dataflow abstraction by leveraging distributed hash table (DHT) based peer-to-peer (P2P) overlay networks, which can automatically place, chain, and scale stream operators to reduce the query latencies, adapt to the workload variations, and recover from failures. Second, it introduces a novel bandit-based path planning model that can re-plan the data shuffling paths to adapt to unreliable and heterogeneous edge network links. 

We evaluate AgileDart on a real-world distributed network environment that has 100 Linux virtual machines (VMs) and Raspberry Pis. We port a full-stack standard IoT stream processing benchmark and real-world IoT stream applications with real-world datasets to AgileDart. We compare AgileDart with two stream processing engines (Storm and EdgeWise) and show significant improvements in scalability and adaptivity when processing a large number of real-world edge stream applications’ queries. \revised{We present discussions and future work in 
Appendix~\ref{sec:discussions}.
}

\bibliographystyle{IEEEtran}
\bibliography{References}

\vspace{-12mm}
\begin{IEEEbiography}[{\includegraphics[width=1in,height=1.25in,clip,keepaspectratio]{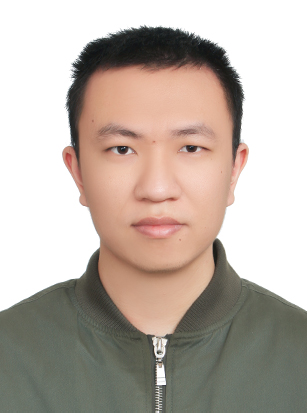}}]
{Cheng-Wei Ching}
received the MS degree in Computer Science and Information Engineering from National Chung Cheng University, Taiwan, in 2021. He is pursuing a PhD in Computer Science and Engineering at the University of California Santa Cruz, USA.
His research interests include distributed systems for machine learning, 
approximation algorithms and their applications, and mobile edge computing. 
\end{IEEEbiography}

\vspace{-12mm}

\begin{IEEEbiography}[{\includegraphics[width=1in,height=1.25in,clip,keepaspectratio]{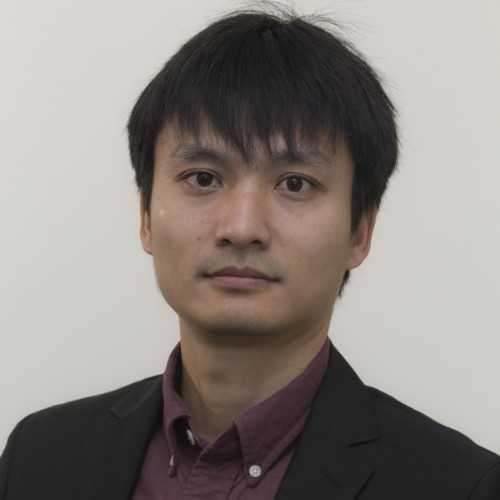}}]
{Xin Chen} received the BS degree in Computer Science from Shandong University, China, in 2009, the MS degree in Software Engineering from Tsinghua University, China, in 2012, and the PhD degree in Computer Science from the Georgia Institute of Technology, USA, in 2020. His research focuses on computer systems and machine learning systems, including solving path planning and collision detection problems.
\end{IEEEbiography}

\vspace{-12mm}

\begin{IEEEbiography}[{\includegraphics[width=1in,height=1.25in,clip,keepaspectratio]{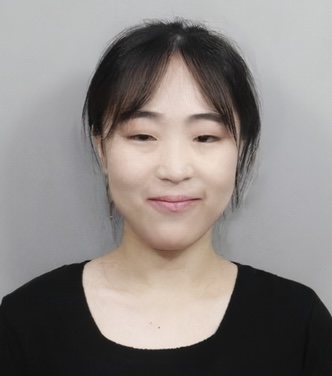}}]
{Chaeeun Kim} received both the BS and MS degrees in Computer Science from Kookmin University, South Korea, in 2020 and 2022, respectively. She is currently pursuing a PhD in Computer Science and Engineering at the University of California Santa Cruz, USA. Her research interests include stream processing systems, large-scale graph mining, distributed algorithms, and distributed systems.
\end{IEEEbiography}

\vspace{-12mm}

\begin{IEEEbiography}[{\includegraphics[width=1in,height=1.25in,clip,keepaspectratio]{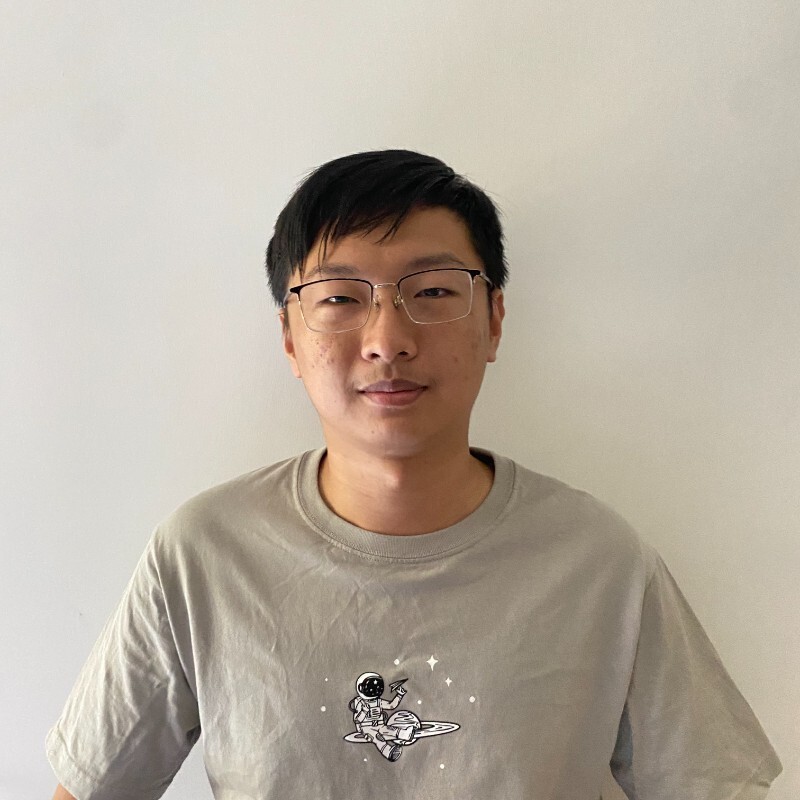}}]
{Tongze Wang} received both the BS and MS degrees in Computer Science and Engineering from the University of California Santa Cruz, USA, in 2021 and 2024, respectively. He continued to work as an Assistant Specialist at the University of California, Santa Cruz, to perform research in Computer Science after graduation. 
\end{IEEEbiography}

\vspace{-12mm}

\begin{IEEEbiography}[{\includegraphics[width=1in,height=1.25in,clip,keepaspectratio]{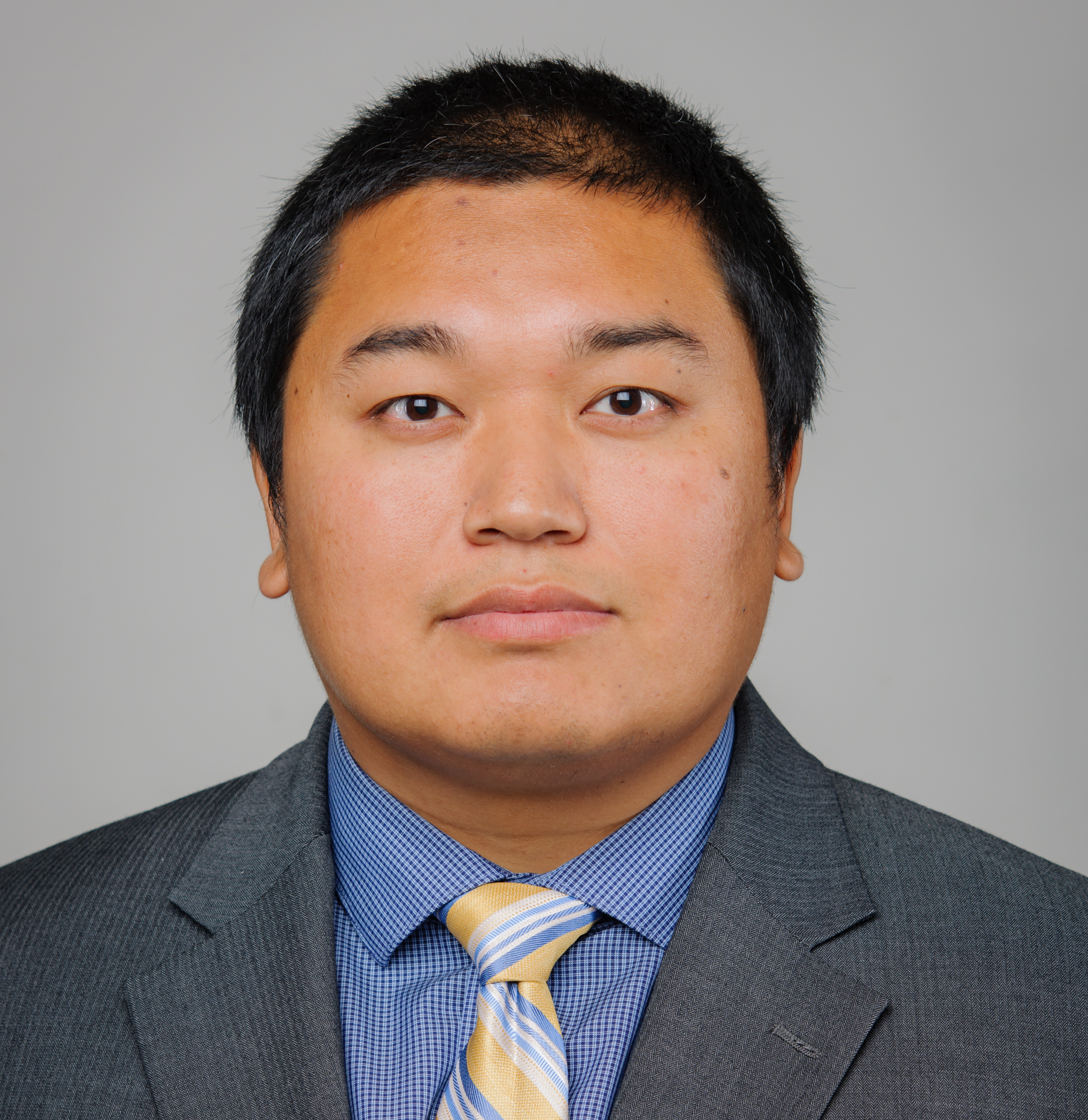}}]
{Dong Chen} received the BS degree in Computer Science from National University of Defense Technology, China, and the PhD degree from the University of Massachusetts Amherst, USA. His research interests include designing data-driven experimental computer systems with an emphasis on improving Cybersecurity, User Privacy and Sustainability of Cyber-Physical Systems (CPS) and the Internet of Things (IoT). He is currently an assistant professor at the Department of Computer Science, Colorado School of Mines.
\end{IEEEbiography}

\vspace{-12mm}

\begin{IEEEbiography}[{\includegraphics[width=1in,height=1.25in,clip,keepaspectratio]{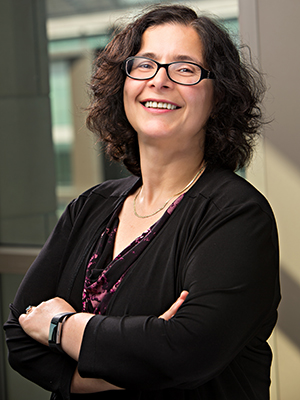}}]
{Dilma Da Silva} received the PhD degree in Computer Science from the Georgia Institute of Technology, USA, in 1997. She is a professor and holder of the Ford Motor Company Design Professorship II with the Department of Computer Science and Engineering, Texas A\&M University, USA. She is an ACM distinguished scientist. Her research interests include operating systems addresses the need for scalable and customizable system software. She is a member of the board of CRA-WP (Computer Research Association’s Committee on Widening the Participation in Computing) and a co-founder of the Latinas in Computing group.
\end{IEEEbiography}

\vspace{-12mm}

\begin{IEEEbiography}[{\includegraphics[width=1in,height=1.25in,clip,keepaspectratio]{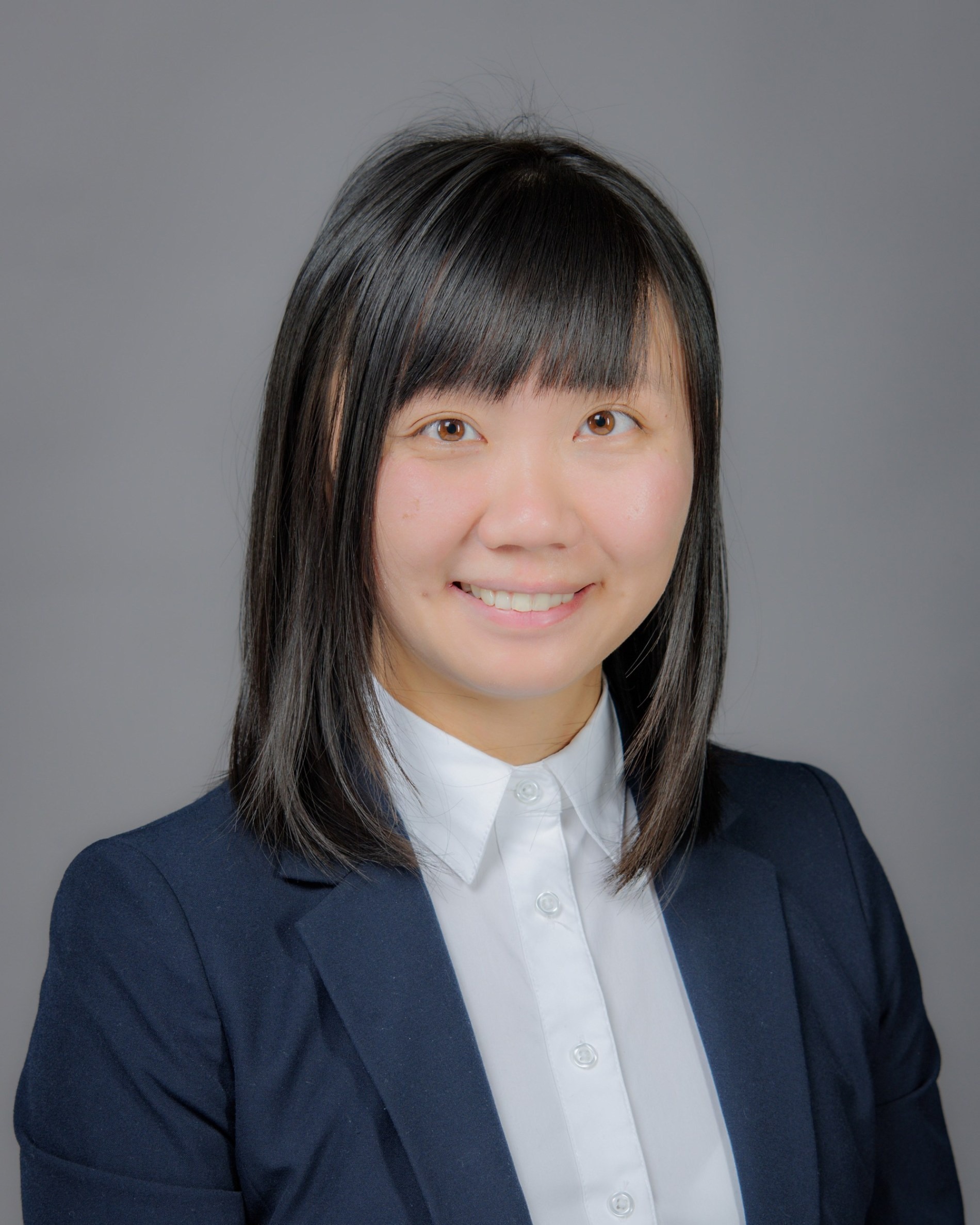}}]
{Liting Hu} received the BS degree in Computer Science from the Huazhong University of Science and Technology, China, in 2007, and the PhD degree in Computer Science from the Georgia Institute of Technology, USA, in 2016. She conducts experimental computer systems research in the areas of stream processing systems, cloud and edge computing, distributed systems, and systems virtualization. Currently she is an assistant professor at the University of California Santa Cruz.
\end{IEEEbiography}

\clearpage

\appendices

\renewcommand\thesubsection{\thesection.\arabic{subsection}}

\section{Edge Stream Processing}
\label{sec:appendix_background}

\subsection{Edge Stream Processing Query Model}
\label{subsec:query}

We follow a generic stream query model. A stream $s\in S$ is an infinite sequence of tuples $t\in T$. A tuple $t\in(\tau,p)$ has a payload $p$ and a timestamp $\tau\in N^+$ defined by event occurrence. A query is a directed acyclic dataflow graph, denoted as $Q=(V,E)$. Each vertex $v\in V$ corresponds to a stream operator $f_v$ that consumes input streams $I$ from its predecessor (upstream) vertices and produces output streams $O$ to its successor (downstream) vertices $(O=f_v(I))$. Each edge $e\in E$ represents a data flow between two vertices.

\begin{figure}[t]
  \centering
  \includegraphics[width=0.45\textwidth]{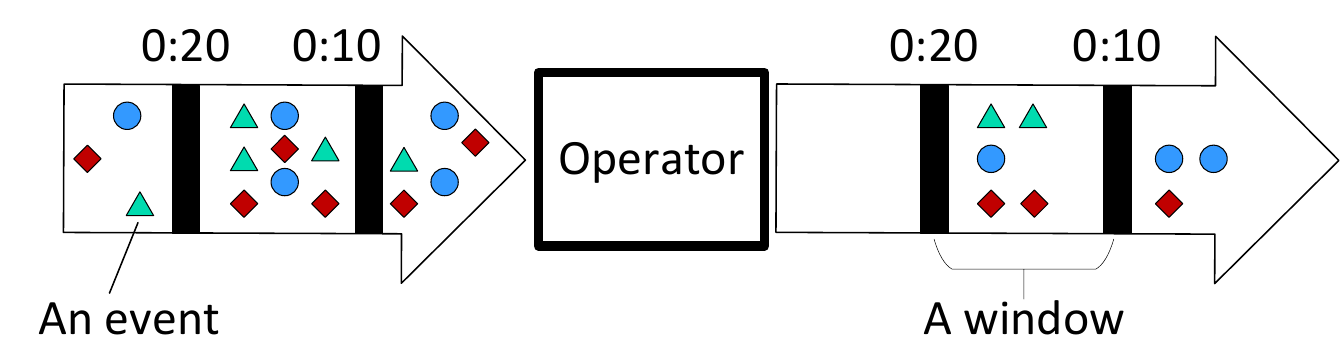}
  \caption{A stream of events flowing through a stream operator.}
  \label{fig:window_op}
\end{figure}

The source and the sink operators are specialized operators that receive input streams from sensors and output the results to actuators or the cloud. Operators execute on event-time scopes called \emph{windows}, as illustrated in Figure~\ref{fig:window_op}. The query latency is defined as the elapsed time starting from the moment the source operator receives the timestamp signaling the completion of the current window to the moment the sink operator externalizes the window results. We view the edge compute nodes are a distributed collection of hundreds of thousands of heterogeneous edge devices, routers, servers, and gateways, which are interconnected using a mix of WANs and LANs.

\subsection{Edge Stream Processing Execution Pipeline}
\label{subsec:pipeline}

\begin{figure*}[t]
 \centering 
 \includegraphics[scale=0.4]{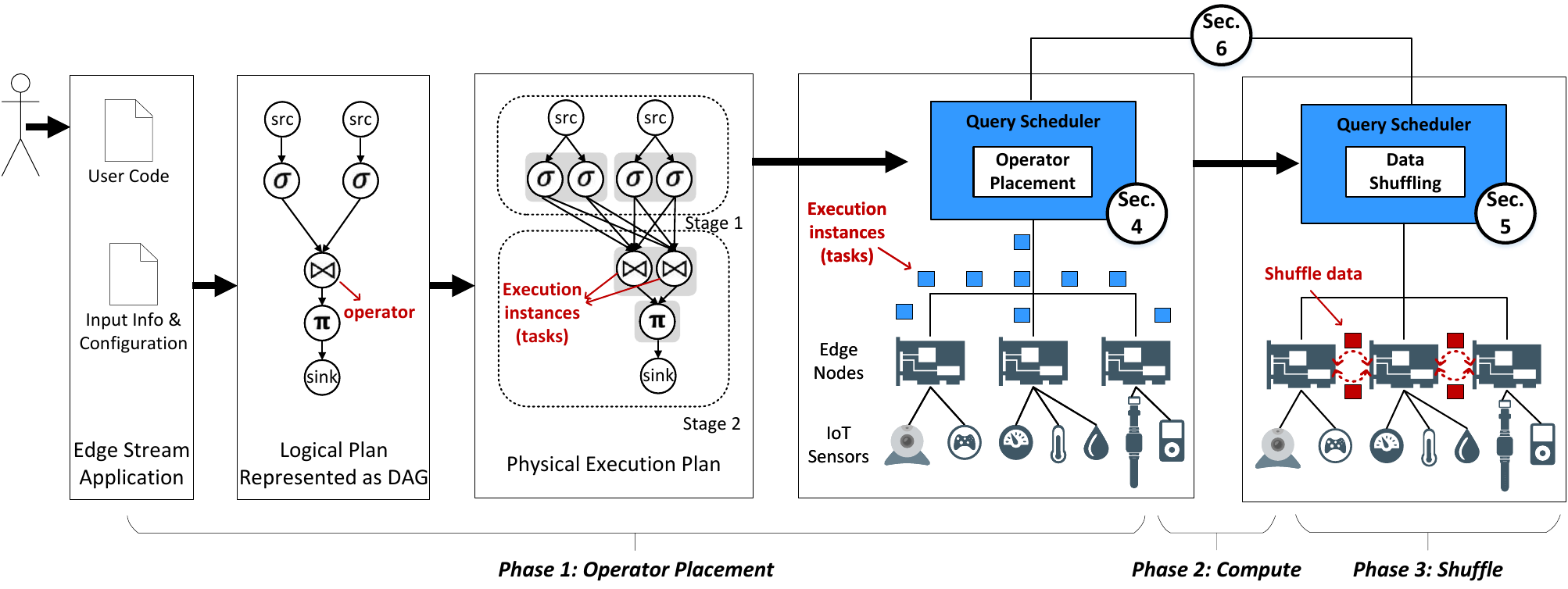}
 \caption{The execution pipeline of a typical edge stream application's queries.}
 \label{fig:pipeline}
\end{figure*}

Then let's have a brief overview of the execution pipeline behind edge stream processing applications. As shown in Figure~\ref{fig:pipeline}, processing typically occurs in a few key phases:

\textbf{Phase 1: Query parsing and optimization.} When an edge stream application is launched, its user code containing transformations and actions is first parsed into a logical execution plan represented using a DAG, where the vertices correspond to stream operators and the edges refer to data flows between operators. 

\textbf{Phase 2: Operator placement.} Afterward, the DAG is converted into a physical execution plan, which consists of several execution stages. Each stage can be further broken down into multiple execution instances (tasks) that run in parallel, as determined by the stage’s level of parallelism. \textit{This requires the system to place all operators’ instances on distributed edge nodes that can minimize the query latency and maximize the throughput}.

\textbf{Phase 3: Compute and shuffle.} Operator instances independently compute their local shard of data and shuffle the intermediate results from one stage to the next stage. Shuffle may create a lot of small, random I/O requests across machines, racks, or even sites, which may significantly delay the query processing. \textit{This requires the system to adapt to the bandwidth and workload variations, node joins and leaves, and failures and stragglers}.

    \begin{figure*}[h]
    \centering
        \subfloat[CPU cycle comparison.]{%
            \includegraphics[width=.3\linewidth]{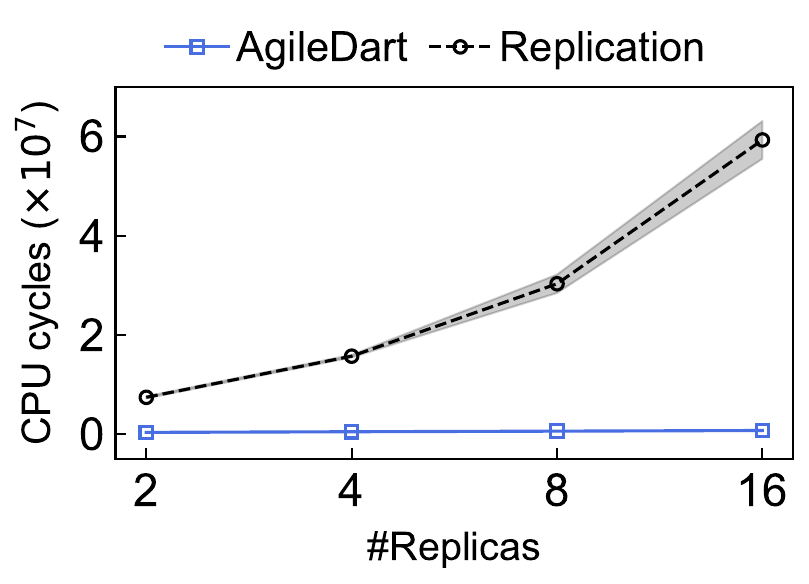}%
            \label{subfig:comp_overhead}%
        }\hspace{1.5em}
        \subfloat[Communication time comparison.]{%
            \includegraphics[width=.32\linewidth]{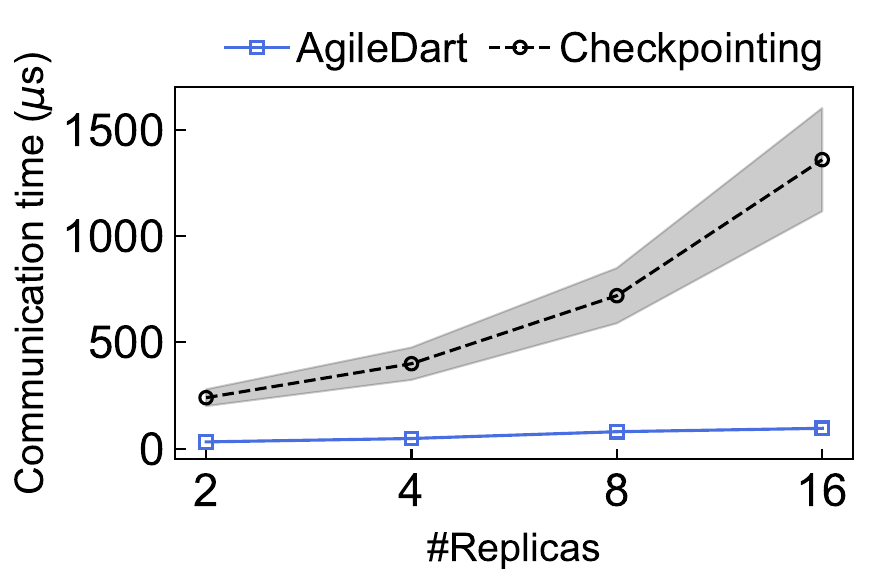}%
            \label{subfig:commu_overhead}%
        }
        \vspace{0.15in}
        \caption{Computation and communication overhead comparison.} 
    \end{figure*}

\section{Complexity Analysis}
\label{sec:time_complexity}

\revised{
We begin with $\omega_\tau(v,v')$. The KL divergence for two fractions, $\hat{\theta}_\tau(v,v')$, and $u$, is computed, which inherently has a constant time complexity of
$O(1)$. Given that the function $KL(\lambda,u)$ is strongly convex and increases for $\lambda\geq u$, identifying the minimum value of $u$ within $\omega_\tau(v,v')$ is equivalent to finding the root of a strictly convex and increasing function in one variable. This problem is efficiently addressed using a simple line search algorithm, resulting in a complexity of $O(|S|)$, where $|S|$ denotes the number of discrete samples of $u$. Hence, the complexity of $\omega_\tau(v,v')$ is $O(|S|)$. Subsequently, deriving $J_\tau(w)$ involves calculating $\omega_\tau(v,v')$ for each path in $|\mathcal{P}_w|$ and selecting the minimal value, resulting in a complexity of $O(|S||\mathcal{P}_w|)$. Furthermore, the computation of $C_\tau(v,v')$ for a node $v$ necessitates performing both $\omega_\tau(v,v')$ and $J_\tau(w)$ at most $|V|$ times each. This results in an overall complexity of $O(|V||S|) + O(|V||\mathcal{P}_w||S|)$.

In the context of the DHT-based consistent ring overlay, the maximum path length is at most $O(\log N)$, with each hop having at most $O(\log N)$ next-hop candidates. Consequently, this structural characteristic allows for the transformation of $|V|$ and $|\mathcal{P}_w|$ into $O(\log N)$ and $O(N^{\log \log N})$, respectively. Thus, the final overall complexity is $O(|S|(\log N)) + O(|S|(\log N) N^{\log \log N})$.
In a practical scenario with $|S| = 5$,  and $N= 10,000$, the term $|S|(\log N)+|S|(\log N) N^{\log \log N} \approx 5,040$, which is manageable even by resource-constrained edge nodes.

}

\section{Failure Recovery Overhead Comparison}
\label{sec:recovery_overhead}

\revised{

We ran an application with a predictive analysis task to evaluate the computation and communication overhead of Replication, Checkpointing, and AgileDart. For a fair comparison, we fixed the range of the number of replicas from 2 to 16. 
Specifically, the number of replicas for Replication represents the number of tasks run in parallel. For Checkpointing, it means the copies of states sent to the cloud-based HDFS. For AgileDart, it represents the number of encoded fragments, and we set $k=2$. We use CPU cycles to evaluate computation overhead and communication time to evaluate communication overhead. We referred to~\cite{edge_network_speed, edge_network_bandwidth} to simulate the data rate from edge networks to the cloud, whereas the data rate in a local network is 1Gbps.

\Cref{subfig:comp_overhead} shows the CPU cycle comparison of Replication and AgileDart. We can see that Replication requires far more CPU cycles than AgileDart, and the gap increases dramatically when the number of replicas doubles.
This is because running several predictive analysis tasks in parallel takes more CPU cycles than executing encoding and decoding states. 
\Cref{subfig:commu_overhead} compares the communication time of Checkpointing and AgileDart. Checkpointing takes more time to transmit states to the cloud-based HDFS. This is because AgileDart sends data fragments to neighborhood set nodes, which are physically close to the current node, so they can usually communicate with each other via a local network. In contrast, Checkpointing must send the states to the cloud-based HDFS over edge networks whose data rate may be 100 Mbps at most~\cite{edge_network_bandwidth,edge_network_speed}. Hence, even though AgileDart and Checkpointing back up a fixed-size state simultaneously, AgileDart can finish sooner.

}
\section{Health Score Analysis}
\label{sec:health_score}

    \begin{figure*}[h]
    \centering
        \subfloat[Number of instances over time with different capacities.]{%
            \includegraphics[width=.41\linewidth]{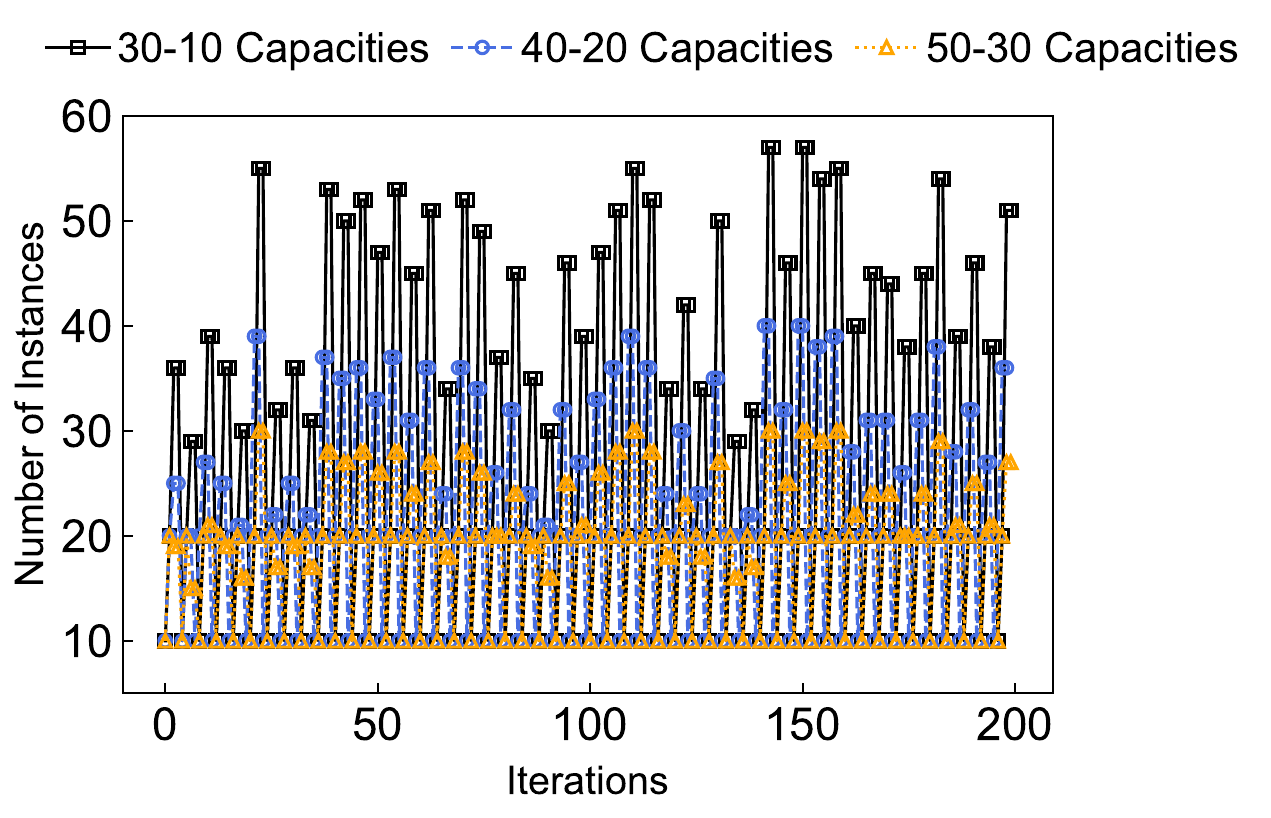}%
            \label{subfig:instance_dif_cap}%
        }\hspace{1.5em}
        \subfloat[Number of instances over time with different two initial steps.]{%
            \includegraphics[width=.41\linewidth]{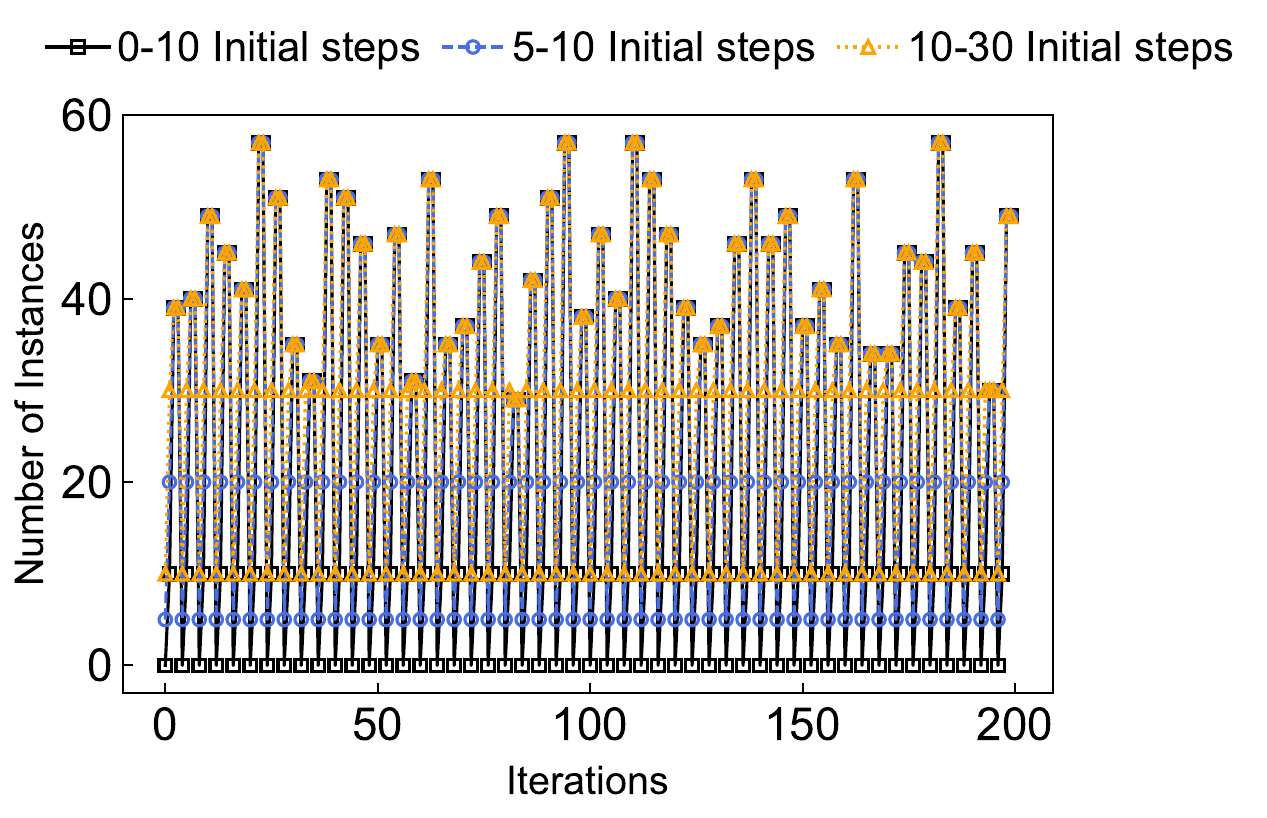}%
            \label{subfig:instance_dif_init}%
        }
        \vspace{0.15in}
        \caption{Number of instances to add using different parameters.} 
    \end{figure*}

\revised{

We generated 50 phases, each with different input rates (i.e., $R_{p_n}$) and queue sizes (i.e., $Q_{p_n}$). We set $\alpha = 0.5$ to consider input rates and queue sizes equally important.
\Cref{subfig:instance_dif_cap} shows the numbers of instances over 50 phases given different instance capacities. The label 30-10 Capacities represents that each instance can increase 30 input rates (i.e., $r=30$) and 10 queue sizes (i.e., $q=10$) for the system. The results show that the proposed heuristic approach helps find the optimal number of instances to satisfy the required input rate and queue size regardless of instance capacities. Also, the approach can adapt to different phases quickly, where each phase takes at most four iterations to meet the requirement. 

\Cref{subfig:instance_dif_init} shows the numbers of instances over 50 phases given different initial steps. The label 0-10 Initial steps represent that the first two steps in the Secant root-finding method (i.e., $x_0, x_1$) are 0 and 10, respectively. When a new phase comes, the system always creates 0 and 10 instances in the first two trials to see if they meet the requirements of the latest phase. 
The results show that different initial steps do not affect finding the optimal number of instances in various phases. At most, four steps are taken to find the optimal number. 

}
\section{Extended Version of Different Routing
Policies}
\label{sec:dif_routing_policy}

\subsection{End-to-end routing} 
\label{subsec:end2end_routing}

The pseudo-code of the end-to-end routing~\cite{combinatorial_network_optimization} is given in Algorithm~\ref{pseudo:end_to_end}. Whenever source node $s$ receives a packet at time slot $\tau \geq 1$, it first runs line~\ref{state:select_path_end_to_end} and line~\ref{state:minimum_lcb_end_to_end} to find path $p'$, where path $p'$ yields the minimum lower confidence bound (LCB)~\cite{intro_mab}. After sending the packet along path $p'$, source $s$ receives feedback from the nodes that path $p'$ travels through and updates the packet delay of path $p'$ and the number of visits to path $p'$ and each link in path $p'$ (line~\ref{state:update_feedback_end_to_end}). Table~\ref{tab:notation_end_to_end} summarizes the key variables in the end-to-end routing.

\begin{algorithm}[h]
    \caption{End-to-end routing ~\cite{combinatorial_network_optimization} for source node $s$}
    \label{pseudo:end_to_end}
    \begin{algorithmic}[1]
        \For{time slot $\tau \geq 1$ }
            \State Select path $p'\in \mathcal{P}$, where
            \label{state:select_path_end_to_end}
            \State \quad $p'\in \argmin_{p\in \mathcal{P}} LCB_\tau(p)$, where 
            \label{state:minimum_lcb_end_to_end}
            \Statex   
            $LCB_\tau(p) = \underbrace{\frac{\sum_{t=1}^\tau \textbf{1}_{(I_t = p)}\sum_{i\in p}D_{t}(i)}{N_\tau(p)}}_{\text{exploitation term}} -\underbrace{\sqrt{\frac{(L+1)\log (\tau)}{\sum_{i\in p}N_\tau(i)}}}_{\text{exploration term}}$.
            \State Collect feedback on path $p'$, and update  packet delay $\sum_{i\in p'}D_{\tau}(i)$, and 
        number of visits $N_\tau(p')$ and $N_\tau(i), \forall i \in p'$.
        \EndFor
        \label{state:update_feedback_end_to_end}        
    \end{algorithmic}
\end{algorithm}

The LCB of a path at each time slot consists of two terms: the \textit{exploitation term} and the \textit{exploration term}. These two terms balance the exploration-exploitation tradeoff.
For each time slot, a source node selects a path with the minimum LCB and uses the observed packet delay to update the LCB of the selected path.

We first introduce the necessary notations.
Let $D_{\tau}(i)$ denote the packet delay of link $i$ at time slot $\tau$. Let $N_\tau(p)$ and $N_\tau(i)$ denote the number of visits to path $p$ and the number of visits to link $i$ up to time slot $\tau$, respectively. The exploitation term of a path represents the empirical packet delay of the path up to time slot $\tau$, while the exploration term is the square root of the ratio between $\big((L+1)\log(\tau)\big)$ and $\sum_{i\in p}N_\tau(i)$, where $L > 0$ is a constant and $\sum_{i\in p}N_\tau(i)$ represents the summation of the number of visits to each link $i$ in path $p$ up to time slot $\tau$.

Thus, the LCB of link $p$ at time slot $\tau$ is defined as follows:
\begin{align}
    LCB_\tau(p) = \underbrace{\frac{\sum_{t=1}^\tau \textbf{1}_{(I_t = p)}\sum_{i\in p}D_{t}(i)}{N_\tau(p)}}_{\text{exploitation term}} -\underbrace{\sqrt{\frac{(L+1)\log \tau}{\sum_{i\in p}N_\tau(i)}}}_{\text{exploration term}},
    \nonumber
\end{align}
where $\textbf{1}_{(I_t = p)}$ denotes the indicator function that equals 1 if the condition $I_t = p$ is true and 0 otherwise. Here, $I_t$ represents the selected path at time slot $t$.
The path $p'$ with the minimum LCB is selected at time slot $\tau$, that is,
\begin{align}
    p' \in \argmin_{p\in \mathcal{P}} LCB_\tau(p),
    \nonumber
\end{align}
to route a packet from the source to the destination, and the observed packet delay is used to update the LCB of path $p'$.

\subsection{Next-hop routing} 
\label{subsec:next_hop_routing}

\begin{table}[t]
\setlength{\belowcaptionskip}{10pt} 
\renewcommand\arraystretch{1.25}
 \small
\begin{tabular}{cp{6.7cm}}
    \Xhline{2\arrayrulewidth}
    \rule{0pt}{1.1\normalbaselineskip}\textbf{Variable} & \textbf{Description} \vspace{0.05in}\\ \hline 
    \rule{0pt}{1.1\normalbaselineskip}
    $L$ & Constant in the end-to-end routing.\\
    $\mathcal{P}$    &    Set of all possible loop-free paths from source node $s$ to destination node $d$ in $G$. \\
    $p$    &  Path in $\mathcal{P}$.\\
    $s$    &  Source node.     \vspace{0.05in}     \\
    \hline
    \rule{0pt}{1.1\normalbaselineskip}
    $D_\tau(i)$ & Packet delay of link $i$ at time slot $\tau$.\\
    $I_t=p$ &  Selected path is $p$ at time slot $t$.\\
    $LCB_\tau(p)$ & Lower confidence bound of path $p$ at time slot $\tau$.\\
    $N_\tau(i)$ & Number of visits to each link $i$ in path $p$ up to time slot $\tau$.\\
    $N_\tau(p)$ & Number of visits to path $p$ up to time slot $\tau$.\\
    $\textbf{1}_{(I_t=p)}$ &  Indicator function that is equal to 1 if the condition $I_t = p$ is true and 0 otherwise.     \vspace{0.05in}     \\
    \Xhline{3\arrayrulewidth}
\end{tabular}
\caption{Key variables in the end-to-end routing algorithm.}
\label{tab:notation_end_to_end}
\end{table}

The pseudo-code of the next-hop routing~\cite{adaptive_opportunistic_routing} is given in Algorithm~\ref{pseudo:next_hop_routing_appendix}. Whenever node $v$ receives a packet at time slot $\tau \geq 1$, it first randomly picks a value $\epsilon$ between 0 and 1 (line~\ref{state:random_number_next_hop}). 
If the drawn value $\epsilon$ is smaller than $\Big(1-\frac{1}{N_\tau(v)}\Big)$ (line~\ref{state:if_statement_next_hop}), node $v$ selects link $(v,v')\in E$ with the minimum empirical packet delay (line~\ref{state:select_links_next_hop} and line~\ref{state:minimum_packet_delay_next_hop}); otherwise, node $v$ selects a link randomly (line~\ref{state:random_link}).
After sending the packet to the next hop $v'$ through link $(v,v')$, node $v$ receives feedback from node $v'$ and updates the empirical packet delay $\hat{D}_\tau(v,v')$ and number of visits to node $v$ and link $(v,v')$ (line~\ref{state:update_feedback_next_hop}). 
Table~\ref{tab:notation_next_hop} summarizes the key variables in the next-hop routing.

\begin{algorithm}[t]
    \caption{Next-hop routing~\cite{adaptive_opportunistic_routing} for node $v$}
    \label{pseudo:next_hop_routing_appendix}
    \begin{algorithmic}[1]
        \For{time slot $\tau \geq 1$ }
            \State Let $\epsilon$ be a random number between 0 and 1
            \label{state:random_number_next_hop}
            \If{$\epsilon \leq \Big(1- \frac{1}{N_\tau(v)}\Big) $}
            \label{state:if_statement_next_hop}
                \State Select link $(v,v')\in E$, where 
                \label{state:select_links_next_hop}
                \State \quad $v' \in \argmin_{w\in V:(v,w)\in E} \hat{D}_\tau(v,w)$, where
                \Statex \quad \quad \quad $\hat{D}_\tau(v,w) =  \frac{\sum_{t=1}^\tau \textbf{1}_{(I_t = (v,w))}D_{t}(v,w)}{N_\tau(v,w)}$.
                \label{state:minimum_packet_delay_next_hop}
            \Else
                \State Select link $(v,v')\in E$ randomly.
                \label{state:random_link}
            \EndIf
            \State Collect feedback on link $(v,v')$, and update empirical packet delay $\hat{D}_\tau(v,v')$, and number of visits $N_\tau(v)$ and $N_\tau(v,v')$.
            \label{state:update_feedback_next_hop}
        \EndFor
    \end{algorithmic}
\end{algorithm}

\begin{table}[h]
\setlength{\belowcaptionskip}{10pt} 
\renewcommand\arraystretch{1.25}
 \small
\begin{tabular}{cp{6.5cm}}
    \Xhline{3\arrayrulewidth}
    \rule{0pt}{1.1\normalbaselineskip}\textbf{Variable} & \textbf{Description} \vspace{0.05in}\\ \hline 
    \rule{0pt}{1.1\normalbaselineskip}
    $G$    &  Graph with a set of nodes $V$ and a set of links $E$.\\
    $V$    &  Set of nodes in $G$. \\
    $E$    &  Set of links in $G$.\\
    $\epsilon$  \vspace{0.05in}  &  Random value between 0 and 1. \\
    \hline
    \rule{0pt}{1.1\normalbaselineskip}
    $(v,v')$    &  Link from node $v$ to node $v'$.         \\
    $D_\tau(v,w)$& Packet delay of link $(v,w)$ at time slot $\tau$.\\
    $\hat{D}_\tau(v,w)$ & Empirical packet delay of link $(v,w)$ up to time slot $\tau$.\\
    $I_t=(v,w)$ & Selected link is $(v,w)$ at time slot $t$. \\
    $N_\tau(v)$ & Number of visits to node $v$ up to time slot $\tau$.\\
    $N_\tau(v,w)$ & Number of visits to link $(v,w)$ up to time slot $\tau$.\\
    $\textbf{1}_{(I_t=(v,w))}$ &  Indicator function that is equal to 1 if the condition $I_t=(v,w)$ is true and 0 otherwise. \vspace{0.05in}\\
    \Xhline{3\arrayrulewidth}
\end{tabular}%
\caption{Key variables in the next-hop routing algorithm.}
\label{tab:notation_next_hop}
\end{table}

The next-hop routing leverages an $\epsilon$-greedy algorithm~\cite{intro_mab} to execute exploration and exploitation, where $\epsilon$ is a random value between 0 and 1. At each time slot, $\epsilon$ is randomly drawn between 0 and 1. If $\epsilon$ is smaller than a given threshold, it executes exploitation; otherwise it executes exploitation. Instead of relying on the complete randomness of $\epsilon$, the next-hop routing sets the threshold as the number of visits to a node. When a node has been visited a sufficient number of times, the next-hop routing prefers exploitation over exploration and vice versa.

We first introduce the necessary notations. Let $D_\tau(v,w)$ denote the packet delay of link $(v,w)$ at time slot $\tau$. Let $N_\tau(v)$ and $N_\tau(v,w)$ denote the number of visits to node $v$ and to link $(v,w)$ up to time slot $\tau$, respectively. Then, the empirical packet delay $\hat{D}_\tau(v,w)$ of link $(v,w)$ at time slot $\tau$ is defined as 
\begin{align*}
    \hat{D}_\tau(v,w) = \frac{\sum_{t=1}^\tau \textbf{1}_{(I_t = (v,w))}D_{t}(v,w)}{N_\tau(v,w)},
\end{align*}
where $\textbf{1}_{(I_t = (v,w))}$ denotes the indicator function that equals 1 if the condition $I_t = (v,w)$ is true and 0 otherwise. Here, $I_t$ represents the selected link at time slot $t$. When $\epsilon$ is smaller than $\Big(1- \frac{1}{N_\tau(v)}\Big)$, the link with minimum empirical packet delay would be selected, that is,
\begin{align*}
    (v,v')\in E, \text{ and } v' \in \argmin_{w\in V:(v,w)\in E} \hat{D}_\tau(v,w).
\end{align*}
Then, the observed packet delay is used to update $\hat{D}_\tau(v,v')$.

\section{Convergence Analysis}
\label{sec:convergence}

\captionsetup[subfigure]{labelformat=empty}
\begin{figure*}[h]
    \centering
    \subfloat[(a) Network with 25 nodes.]{%
        \includegraphics[width=.32\linewidth]{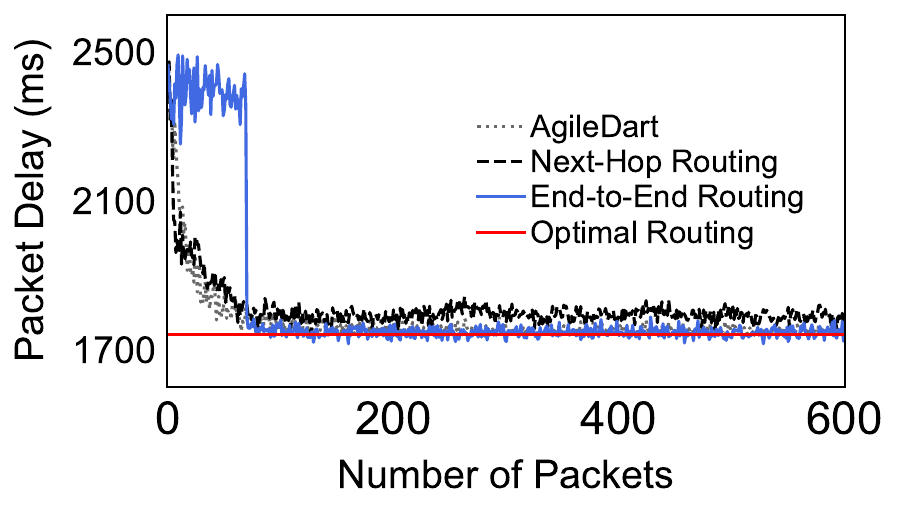}%
        \label{subfig:convergence_25_nodes}%
    }
                            \vspace{-0.05in}
    \subfloat[(b) Network with 36 nodes.]{%
        \includegraphics[width=.32\linewidth]        {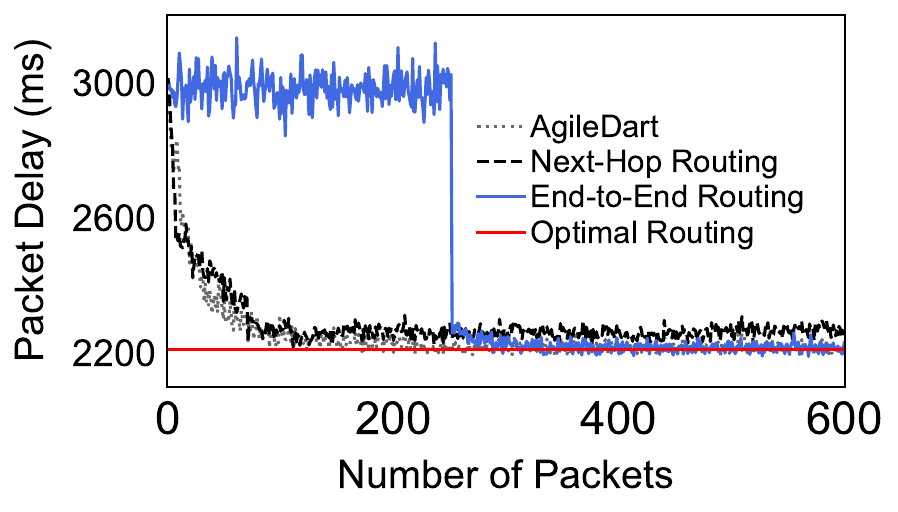}%
        \label{subfig:convergence_36_nodes}%
    }
    \subfloat[(c) Network with 64 nodes.]{%
        \includegraphics[width=.32\linewidth]{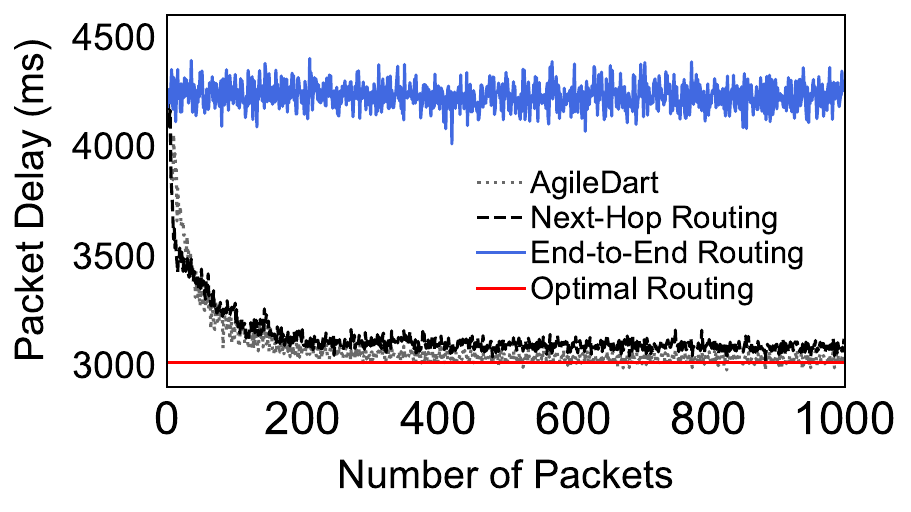}%
        \label{subfig:convergence_64_nodes}%
    }
    \vspace{0.2in}
    \caption{Convergence analysis on different path planning algorithms using different network sizes.}
\end{figure*}

\revised{

We evaluate the convergence of three algorithms using four networks with 32, 64, and 128 links and 25, 36, and 64 nodes, respectively. Each link is given a delay randomly generated from the range between 100 \textit{ms} to 500 \textit{ms}.
AgileDart, next-hop routing, and end-to-end routing aim to find the optimal path from source to destination within 1,000 packets, whereas optimal routing always selects the optimal paths.

The results of the network with 25 nodes, the network with 36 nodes, and the network with 64 nodes are presented in~\Cref{subfig:convergence_25_nodes},~\Cref{subfig:convergence_36_nodes}, and~\Cref{subfig:convergence_64_nodes}, respectively. The results of each algorithm are averaged 1,000 times to offset the randomness of delays.
We can see that AgileDart achieves optimal routing within the fewest number of packets regardless of network size. In contrast, next-hop routing eventually reaches the sub-optimal paths. This is because next-hop routing considers the optimal next hop but ignores the quality of subsequent paths. End-to-end routing reaches optimal routing in networks with only 25 and 36 nodes. This is because end-to-end routing needs to try each possible path from the source to the destination, and hence, when the network size expands, end-to-end requires more than 1,000 trials to reach optimal routing.

}
\section{Discussions and Future Work}
\label{sec:discussions}

\revised{
\textit{\textbf{Deployment complexity.}} 
We provide three proofs of the lightweight, practical deployment of AgileDart.

First, on the implementation side, AgileDart is built only on top of two software stacks: Apache Flume~\cite{flume} (v.1.9.0) and Pastry~\cite{freepastry} (v.2.1). In particular, Apache Flume is similar to Kafka~\cite{ApacheKafka} that provides distributed services for collecting and aggregating large amounts of streaming event data. On the other hand, Pastry is an overlay network and routing framework based on DHTs. We implemented AgileDart's innovative features on Pastry alone, like autonomous operator placement in the dynamic dataflow abstraction, the scaling mechanism, topic-based trees, and the failure recovery mechanism. As long as a physical node can run Pasty, it can enjoy all of AgileDart's features without other complex dependencies. In addition, since AgileDart relies only on Pastry for implementation, any modifications can be made to AgileDart easily.

Secondly, on the deployment side, we chose Apache Flume~\cite{flume} and Pastry~\cite{freepastry} to develop AgileDart because both frameworks are developed in Java. Java is famous for its cross-platform compatibility, allowing one to seamlessly deploy AgileDart across various environments without requiring substantial modifications. We have shown that AgileDart worked well seamlessly in a heterogeneous computing setting with 10 Raspberry Pis hosting source operators and 100 Linux VMs hosting internal and sink operators in Section VII.

Lastly, we have released AgileDart's preliminary version source code on GitHub\footnote{\url{https://github.com/ElvesLab/DART}}. Users can download \texttt{.jar} files to run AgileDart, and developers can download the Java source code and easily add new suitable features for their deployment needs.

\textit{\textbf{Resource management and task scheduling.}} 
A P2P node in AgileDart can be seen as a \textit{logical} node. We can map a physical edge node with rich resources to more P2P nodes in AgileDart, whereas mapping resource-constrained edge nodes to fewer P2P nodes. For example, suppose three physical edge nodes with 2, 4, and 8 CPU cores, respectively. Then, the nodes with 4 and 8 CPU cores can serve as 2 and 3 logical P2P nodes in the DHT-based P2P overlay network, respectively. One can follow a similar way for different resources such as memory, network bandwidths, or GPUs. 

In the DHT-based consistent ring overlay, the node mapped to two P2P nodes is assigned two NodeIds. Although these two NodeIDs may not be numerically close, the proximity metrics (e.g., hop counts, RTT, cross-site link congestion levels) in the dynamic dataflow abstraction indicate that these two P2P nodes are physically close. Moreover, since the operators are distributed evenly across the overlay network due to the hash, the nodes mapped to two P2P nodes have more chances of being operators than those mapped to a P2P node. For the elastic scaling mechanism and failure recovery mechanism, since each P2P node operates individually and autonomously, the nodes mapped to multiple P2P nodes follow the same procedures. Lastly, since P2P nodes in the overlay network represent \textit{resources}, a physical node mapped to several P2P nodes can manage its resources by shutting down or spawning P2P nodes at an application level. Similarly, the source operator can follow the dynamic dataflow abstraction to route a task with resource requirements to P2P nodes that meet the requirements.

\textit{\textbf{More aspects of Quality of Service (QoS).}} The computing entities in edge stream processing systems are usually edge nodes or IoT devices, where they are typically resource-limited and battery-powered. 
Furthermore, they transmit input data, intermediate data, and processing states over edge networks, which are generally unpredictable and stochastically varying due to unreliable links and random access protocols (e.g., in wireless networks), mobility  (e.g., in mobile ad-hoc networks), randomness of demand (e.g., workload surges) in arbitrary (and dynamic) geographical edge locations. 
As such, other aspects of QoS, like network congestion, data loss, and energy consumption, are critical to edge stream processing systems. Although AgileDart's primary focus is shortening the end-to-end processing latency of edge stream processing systems by improving scalability and adaptivity, and we also show that AgileDart's power and resource usage is less than Storm's, it would be interesting to explore jointly optimizing power and resource usage with unreliable network connections while keeping low latency.

}

\end{document}